\documentclass[11pt, oneside]{article}   	
\usepackage{jheppub}
\usepackage{amsmath}
\usepackage{amssymb}
\usepackage{amsfonts}
\usepackage{amsthm}
\usepackage{amsbsy}
\usepackage{array}
\usepackage{mathtools}
\usepackage{subcaption}
\usepackage{graphicx}
\usepackage[usenames,dvipsnames]{xcolor}

\usepackage{tikz}
\usetikzlibrary{decorations}
\usetikzlibrary{trees}
\usetikzlibrary{decorations.pathmorphing}
\usetikzlibrary{decorations.markings}
\usetikzlibrary{external}
\usetikzlibrary{intersections}
\usetikzlibrary{shapes,arrows}
\usetikzlibrary{arrows.meta}
\usetikzlibrary{calc}
\usetikzlibrary{shapes.misc}
\usetikzlibrary{decorations.text}
\usetikzlibrary{backgrounds}
\usetikzlibrary{fadings}
\usetikzlibrary{tikzmark,calc,arrows,shapes,decorations.pathreplacing}

\colorlet{mred}{black!10!red}
\colorlet{mgreen}{black!30!green}
\colorlet{mblue}{black!30!blue}
\colorlet{morange}{blue!70!red}
\colorlet{mgrey}{black!60!white}

    \usepackage{tikz,environ}
\usetikzlibrary{snakes}
\usetikzlibrary{decorations}
\usetikzlibrary{trees}
\usetikzlibrary{decorations.pathmorphing}
\usetikzlibrary{decorations.markings}
\usetikzlibrary{external}
\usetikzlibrary{intersections}
\usetikzlibrary{shapes,arrows}
\usetikzlibrary{arrows.meta}
\usetikzlibrary{calc}
\usetikzlibrary{shapes.misc}
\usetikzlibrary{decorations.text}
\usetikzlibrary{backgrounds}
\usetikzlibrary{fadings}
\usetikzlibrary{tikzmark,calc,arrows,shapes,decorations.pathreplacing}

\tikzset{
	cross/.style={cross out, draw=black, minimum size=2*(#1-\pgflinewidth), inner sep=0pt, outer sep=0pt, mred
    },
	graviton/.style={decorate,line width=0.3mm, decoration={snake,amplitude=0.3mm, segment length=1.5mm}},
	photon/.style={decorate, decoration={snake}, draw=red},
	scalar/.style={postaction={decorate},
	},
	massive/.style={postaction={decorate},
		line width=0.75mm,
	},
    scalar/.style={postaction={decorate},
		line width=0.5mm,
	},
	massless/.style={postaction={decorate},
	},
	masslessWithDot/.style={postaction={decorate},
		decoration={
			markings,
			mark=at position 0.5 with {\fill circle (2pt);}}
	},
	massiveWithDot/.style={postaction={decorate},
		line width=0.5mm,
		decoration={
			markings,
			mark=at position 0.5 with {\fill circle (2pt);}}
	},
	massiveWithArrow/.style={postaction={decorate},
		line width=0.75mm,
		decoration={
			markings,
			mark=at position 0.5 with {\arrow{latex}}}
	},
	massiveLin/.style={postaction={decorate},
		double,
		thick,
		fill=white
	},
	massivePhi/.style={postaction={decorate},
		line width=0.75mm,
		dashed
	},
	masslessPhi/.style={postaction={decorate},
		dashed
	},
	unitaryCut/.style={postaction={draw,densely dashed,blue,thin},
		line width = 0.2cm,white
	},
	gluon/.style={decorate, draw=magenta,
		decoration={coil,amplitude=4pt, segment length=5pt}},
	partial ellipse/.style args={#1:#2:#3}{
		insert path={+ (#1:#3) arc (#1:#2:#3)}
	},
	cross/.style={cross out, draw=black, minimum size=2*(#1-\pgflinewidth), inner sep=0pt, outer sep=0pt},
	branchCut/.style={postaction={decorate},
		snake=zigzag,
		decoration = {snake=zigzag,segment length = 2mm, amplitude = 2mm}	
	}
	cross/.default={1pt}
}

\colorlet{mred}{black!30!red}
\colorlet{mgreen}{black!30!green}
\colorlet{mblue}{black!30!blue}
\colorlet{morange}{blue!70!red}

\newcommand{\boxgrav}{
\begin{tikzpicture}[thick,scale=1.5]
		\coordinate (e1) at (-1,0.25);
			\coordinate (e2) at (1,0.25);
			\coordinate (e3) at (1,-1.25);
			\coordinate (e4) at (-1,-1.25);	
			
			\coordinate (v1) at (-0.5,0);
			\coordinate (v2) at (0.5,0);
			\coordinate (v3) at (0.5,-1);
			\coordinate (v4) at (-0.5,-1);
			
			\draw[scalar] (e1) -- (v1);
			\draw[scalar] (v1) -- (v2);
			\draw[scalar] (e4) -- (v4);
			\draw[scalar] (v2) -- (e2);
			\draw[scalar] (v4) -- (v3);
			\draw[scalar] (v3) -- (e3);
			
			\draw[graviton] (v1) -- (v4);
			\draw[graviton] (v2) -- (v3);
			\draw[scalar] (v3) -- (e3);

		\end{tikzpicture}
}

\newcommand{\trigrav}{
\begin{tikzpicture}[thick,scale=1.5]
		\coordinate (e1) at (-1,0.25);
			\coordinate (e2) at (1,0.25);
			\coordinate (e3) at (1,-1.25);
			\coordinate (e4) at (-1,-1.25);	
			
			\coordinate (v1) at (-0.5,0);
			\coordinate (v2) at (0.5,0);
			\coordinate (v3) at (0,-0.5);
			\coordinate (v4) at (0,-1);
			
			\draw[scalar] (e1) -- (v1);
			\draw[scalar] (v1) -- (v2);
            \draw[scalar] (v2) -- (e2);
			\draw[scalar] (e4) -- (v4);
			\draw[scalar] (v4) -- (e3);

            \draw[graviton] (v4) -- (v3);
			\draw[graviton] (v1) -- (v3);
			\draw[graviton] (v2) -- (v3);
		
		\end{tikzpicture}
}

\newcommand{\bubgrav}{
\begin{tikzpicture}[thick,scale=1.5]
		\coordinate (e1) at (-1,0.25);
			\coordinate (e2) at (1,0.25);
			\coordinate (e3) at (1,-1.25);
			\coordinate (e4) at (-1,-1.25);	
			
			\coordinate (v1) at (0,0);
			\coordinate (v2) at (0,-0.25);
			\coordinate (v3) at (0,-0.75);
			\coordinate (v4) at (0,-1);
			
			\draw[scalar] (e1) -- (v1);
            \draw[scalar] (v1) -- (e2);
			\draw[scalar] (e4) -- (v4);
			\draw[scalar] (v4) -- (e3);

            \draw[graviton] (v4) -- (v3);
			\draw[graviton] (v1) -- (v2);
			\draw[graviton] (0.015,-0.5) circle (0.25);

		\end{tikzpicture}
}

\newcommand{\gravitoncut}[2]{
\begin{tikzpicture}
        \coordinate (e1) at (-2.25,-1);	
	    	\coordinate (e2) at (-2.25, 1);
	    	\coordinate (e3) at ( 2.25, 1);
	    	\coordinate (e4) at ( 2.25,-1);
    
	    	\coordinate (v1) at (-1.25, 0);
	    	\coordinate (v2) at ( 1.25, 0);
	    	
	    	\coordinate (m1) at (0, 1.00);
	    	\coordinate (m2) at (0,-1.00);
    
	    	\draw[scalar] (e1) -- (v1);
	    	\draw[scalar] (e2) -- (v1);
	    	\draw[scalar] (e3) -- (v2);
	    	\draw[scalar] (e4) -- (v2);
    
	    	\draw[graviton, name path=rung1] (v1) to[out=90, in=90](v2);
	    	\draw[graviton, name path=rung2] (v1) to[out=-90, in=-90](v2);
	    	\draw[line width=0.25mm, dashed, name path=rung4, white] (0, 1) to (0, -1);
	    	
            \draw[fill=lightgray, opacity=1] (v1) circle (0.65);
            \draw[fill=lightgray, opacity=1] (v2) circle (0.65);
	    	\node  at (v1) {$\mathcal{M}$};
	    	\node  at (v2) {$\mathcal{M}^*$};
    
	    	\fill[name intersections={of=rung1 and rung4, name=i,total=\t},white] foreach \s in {1,...,\t}{(i-\s) circle (5pt)};
	    	\fill[name intersections={of=rung2 and rung4, name=i,total=\t},white] foreach \s in {1,...,\t}{(i-\s) circle (5pt)};
    
	    	\draw[line width=0.3mm, dashed, name path=rung4, blue] (m1) to (m2);
    
	        \node[below left]  at (e1) {$p_1$};
	        \node[above left]  at (e2) {$p_2$};
	        \node[above right] at (e3) {$p_3$};
	        \node[below right] at (e4) {$p_4$};
	    	\node[above, fill=white, opacity=1] at (m1) {#1}; 
	    	\node[below, fill=white, opacity=1] at (m2) {#2}; 
\end{tikzpicture}}

\newcommand{\scalarcut}[2]{
\begin{tikzpicture}
        \coordinate (e1) at (-2.25,-1);	
	    	\coordinate (e2) at (-2.25, 1);
	    	\coordinate (e3) at ( 2.25, 1);
	    	\coordinate (e4) at ( 2.25,-1);
    
	    	\coordinate (v1) at (-1.25, 0);
	    	\coordinate (v2) at ( 1.25, 0);
	    	
	    	\coordinate (m1) at (0, 1.00);
	    	\coordinate (m2) at (0,-1.00);
    
	    	\draw[scalar] (e1) -- (v1);
	    	\draw[scalar] (e2) -- (v1);
	    	\draw[scalar] (e3) -- (v2);
	    	\draw[scalar] (e4) -- (v2);
    
	    	\draw[scalar, name path=rung1] (v1) to[out=90, in=90](v2);
	    	\draw[scalar, name path=rung2] (v1) to[out=-90, in=-90](v2);
	    	\draw[line width=0.25mm, dashed, name path=rung4, white] (0, 1) to (0, -1);
	    	
            \draw[fill=lightgray, opacity=1] (v1) circle (0.65);
            \draw[fill=lightgray, opacity=1] (v2) circle (0.65);
	    	\node  at (v1) {$\mathcal{M}$};
	    	\node  at (v2) {$\mathcal{M}^*$};
    
	    	\fill[name intersections={of=rung1 and rung4, name=i,total=\t},white] foreach \s in {1,...,\t}{(i-\s) circle (5pt)};
	    	\fill[name intersections={of=rung2 and rung4, name=i,total=\t},white] foreach \s in {1,...,\t}{(i-\s) circle (5pt)};
    
	    	\draw[line width=0.3mm, dashed, name path=rung4, blue] (m1) to (m2);
    
	        \node[below left]  at (e1) {$p_1$};
	        \node[above left]  at (e2) {$p_2$};
	        \node[above right] at (e3) {$p_3$};
	        \node[below right] at (e4) {$p_4$};
	    	\node[above, fill=white, opacity=1] at (m1) {#1}; 
	    	\node[below, fill=white, opacity=1] at (m2) {#2}; 
\end{tikzpicture}}

\newcommand \ibub {\tikz \draw (0,0) ellipse (3pt and 1.5pt);}
\newcommand \itri {\vartriangleright}
\newcommand \ibox {\square}

\def\be#1\ee{\begin{align}#1\end{align}}
\newcommand \nn {\nonumber}

\newcommand   \g  {\gamma}
\newcommand   \e  {\epsilon}
\newcommand   \f  {\phi}
\newcommand   \lr  [3] {\left #1 #2 \right #3}
\newcommand   \p   [1] {\lr({#1})}

\newcommand   \s  {\phi}

\newcommand   \gr  {h}

\makeatletter
\def\@fpheader{\ }
\makeatother

\title{Graviton loops and negativity}
\author{Cyuan-Han Chang$^{1,2}$ and Julio Parra-Martinez$^3$}
\affiliation{$^{1}$Walter Burke Institute for Theoretical Physics, Caltech, Pasadena, California 91125, USA}
\affiliation{$^{2}$Kadanoff Center for Theoretical Physics \& Enrico Fermi Institute, University of Chicago, Chicago, Illinois 60637, USA}
 \affiliation{$^3$Institut des Hautes Études Scientifiques, 
 91440 Bures-sur-Yvette, France}
\emailAdd{cchang10@uchicago.edu}
\emailAdd{julio@ihes.fr}

\date{}
\abstract{We revisit dispersive bounds on Wilson coefficients of scalar effective field theories (EFT) coupled to gravity in various spacetime dimensions, by computing the contributions from graviton loops to the corresponding sum rules at low energies. Fixed-momentum-transfer dispersion relations are often ill-behaved due to forward singularities arising from loop-level graviton exchange, making naive positivity bounds derived from them unreliable. Instead, we perform a careful analysis using crossing-symmetric dispersion relations, and compute the one-loop corrections to the bounds on EFT coefficients.
We find that including the graviton loops generically allows for negativity of Wilson coefficients by an amount suppressed by powers of Newton's constant, $G$. The exception are the few couplings that dominate over (or are degenerate with) the graviton loops at low energies. In $D=4$, we observe that assuming that the eikonal formula captures the correct forward behavior of the amplitude at all orders in $G$, and for energies of the order of the EFT cutoff, yields bounds free of logarithmic infrared divergences.}

\begin{document}

\maketitle
\pagenumbering{roman}
\setcounter{page}{2}
\newpage
\pagenumbering{arabic}
\setcounter{page}{1}

\section{Introduction}

Our theories of fundamental physics are described using the language of effective field theory (EFT), whereby any observable is parameterized at low energies as an expansion in small energies and momenta compared to some ultraviolet scale, $\Lambda$. Locality, relativistic causality and unitarity, together with various symmetries, are the basic principles which make EFTs, in fact, effective. For instance, locality dictates that various terms in the EFT expansion correspond to terms in a local action, function of a set of low-energy degrees of freedom; and dimensional analysis organizes this expansion as a series of derivative corrections. Each of these terms comes with free so-called Wilson coefficients, which must be determined experimentally or by matching to a more fundamental description.  

This paper concerns the allowed values of Wilson coefficients in a theory with dynamical gravity, and beyond the strict weak-coupling limit. Well-known dispersive arguments show that the aforementioned principles impose non-trivial constraints on the space of these coefficients \cite{Pham:1985cr, Adams:2006sv}. For instance, forward dispersion relations for scattering amplitudes imply that many Wilson coefficients must be positive in unitary theories. By utilizing crossing symmetry at low energies, one can also derive two-sided bounds on EFT coefficients that are consistent with dimensional analysis \cite{Caron-Huot:2020cmc,Tolley:2020gtv,Arkani-Hamed:2020blm,Bellazzini:2020cot, Sinha:2020win}. This type of analysis has recently led to numerous results constraining the space of consistent EFTs, see e.g., \cite{Caron-Huot:2021rmr, Caron-Huot:2022ugt, Caron-Huot:2022jli,Caron-Huot:2024tsk,Li:2023qzs,Bellazzini:2017fep,Bellazzini:2021oaj,Riembau:2022yse,Bellazzini:2023nqj,Fernandez:2022kzi,Albert:2022oes,Albert:2023jtd,Albert:2023seb,Albert:2024yap,deRham:2017avq,deRham:2017zjm,deRham:2018qqo,Alberte:2019xfh,Alberte:2020bdz,Alberte:2021dnj,Haldar:2021rri,Raman:2021pkf,Zahed:2021fkp,Chowdhury:2021ynh,AccettulliHuber:2020oou,Henriksson:2021ymi,Henriksson:2022oeu,Arkani-Hamed:2021ajd, Chiang:2021ziz,Chiang:2022ltp,Zhang:2021eeo,Trott:2020ebl,Tokuda:2020mlf,Noumi:2022wwf,Wang:2020jxr,Wang:2020xlt,Li:2021lpe,Hong:2023zgm,Xu:2023lpq,Chen:2023bhu,Ma:2023vgc,Hong:2024fbl,Xu:2024iao,Wan:2024eto,Dong:2024omo,Guerrieri:2020bto,Guerrieri:2021ivu,Guerrieri:2022sod,EliasMiro:2021nul,Haring:2022sdp,EliasMiro:2022xaa, EliasMiro:2023fqi,Haring:2023zwu,Guerrieri:2024jkn,Heller:2023jtd,Delacretaz:2025ifh}.

In a theory with dynamical gravitons, however, forward singularities due to graviton exchange preclude most of such arguments. 
This problem was recently circumvented in \cite{Caron-Huot:2021rmr, Caron-Huot:2022ugt} by considering scattering with transverse wavefunctions localized at small impact parameters $b\sim 1/\Lambda$, which tames the forward contributions in dispersion relations. (See \cite{Beadle:2024hqg} for an alternative approach). The result is that, in the strict weak coupling limit, the leading Wilson coefficients can be negative by an amount proportional to Newton's constant, $G$ (times an infrared logarithm, $\log(\Lambda/\mu_{\rm IR})$, in dimension $D=4$).

The main result of this work is that the loss of the requirement positivity due to graviton exchange is generic for almost all Wilson coefficients when including loop corrections\footnote{A related point was recently made in Ref.~\cite{Kaplan:2024qtf} refining familiar superluminality arguments \cite{Adams:2006sv}.}, with the exception of a few that dominate over the loops at low energies.
The negativity is however computable and still small, i.e., suppressed by $G$.
 
Concretely, our setup concerns the low-energy EFT expansion of the 2-to-2 scattering amplitude of a single relativistic real scalar $\phi$.\footnote{For simplicity, we assume that the scalar has a shift symmetry $\phi\to\phi+b$, which forbids the constant terms in the EFT expansion.} Consider first the scattering amplitude without gravity,
\be
{\cal M}(s,t) = g_2\, (s^2+ t^2+ u^2) + g_3\, s t u + g_4\, (s^2+ t^2+ u^2)^2 + \cdots + \text{(loops)},
\ee
written in terms of the familiar Mandelstam variables $s= (p_1+p_2)^2$, $t= (p_2+p_3)^2$ and $u= (p_1+p_3)^2$, and a priori unknown Wilson coefficients $g_i$. Terms in this amplitude are in correspondence with operators (up to field redefinitions) in the effective action 
\begin{equation}
  S^\phi =  \int d^Dx 
 \left( \frac12 (\partial_\mu\phi)^2 + \frac{g_2}{2}(\partial\phi)^4 - \frac{g_3}{3} (\partial\phi)^2(\partial\partial\phi)^2 + 4g_4 (\partial\partial\phi)^4  + \cdots \right) \,.
\end{equation}

A famous dispersion relation (reviewed below) \cite{Pham:1985cr, Adams:2006sv} yields the following sum rule for the forward amplitude at $t=0$,
\be
g_{2} = \int_{\Lambda^2}^\infty \frac{ds}{s^3}\, \text{Im} {\cal M}(s,t=0) = \int_{\Lambda^2}^\infty \frac{ds}{s^2}\, \sigma(s) \, \,,
\ee
owing to the optical theorem, where $\sigma$ is the total cross section, which is positive due to unitarity. The sum rule then implies a positivity constraint on the leading Wilson coefficient
\be
g_{2} \ge 0.
\ee
Similar arguments using forward dispersion relations show that in a weakly coupled theory one must have $g_{2n}\ge 0$ \cite{Adams:2006sv}.

Next, consider the theory coupled to gravity
\begin{equation}
  S = \frac{1}{16\pi G} \int d^Dx \sqrt{g}
 \left( \frac12 (\nabla_\mu\phi)^2 + \frac{g_2}{2}(\nabla\phi)^4 - \frac{g_3}{3} (\nabla\phi)^2(\nabla\nabla\phi)^2 + \cdots \right)+ S_{\rm grav}\,,
\end{equation}
where the gravitational effective action is\footnote{Note that the shift symmetry, $\phi \to \phi + b$, forbids interactions of the form $\phi^n R$, etc.}
\begin{equation}
 S_{\rm grav} = \frac{1}{16\pi G} \int d^Dx \sqrt{g}
 \left( R + \frac{\alpha_2}{4} C^2+ \frac{\alpha_4}{12} C^3 - \frac{\beta_1^{(3)}}{2} (\nabla\phi)^2 C^2 +  \cdots \right) \,,\label{eq: schematically S_grav}
\end{equation}
and $C$ is the Weyl tensor. We focus on the weak coupling limit $G\Lambda^{D-2}\ll 1$, so that gravitational loop corrections can be controlled perturbatively.
The corresponding amplitude at low energies now receives contributions from graviton exchange
\be\label{eq:amp_scalarwithG}
{\cal M}(s,t) =& 8\pi G\left(\frac{st}{u} + \frac{su}{t} + \frac{tu}{s}\right) + g_2\, (s^2+ t^2+ u^2) + g_3\, s t u 
+ \cdots + \text{(loops)}\,.
\ee
Due to this, the bound for $g_2$ gets modified to \cite{Caron-Huot:2021rmr}
\be \label{eq:g2D}
g_{2} \ge - {\cal O}(1) GM^{-2}\,,  \qquad (D>4)\,,
\ee
where $M\leq \Lambda$ is the low-energy scale at which we perform scattering experiments and the ${\cal O}(1)$ is a computable number that depends on the dimension, $D$. A similar analysis shows that the more irrelevant couplings $g_{2n}$ with $n>1$ must remain positive.

\begin{figure}[t]
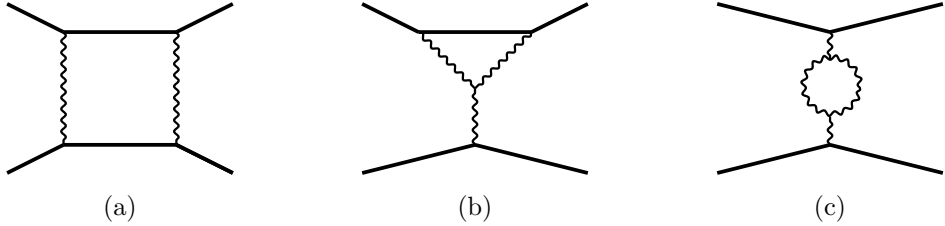

    \centering
    \begin{subfigure}[b]{0.3\textwidth}
         \centering
        \raisebox{-32pt}{\scalebox{1}{\boxgrav}}
         \caption{
         }
         \label{fig:box}
     \end{subfigure}
     \hspace{-0.2cm}
         \begin{subfigure}[b]{0.3\textwidth}
         \centering
        \raisebox{-32pt}{\scalebox{1}{\trigrav}}
         \caption{
         }
         \label{fig:tri}
     \end{subfigure}
     \hspace{-0.2cm}
              \begin{subfigure}[b]{0.3\textwidth}
         \centering
        \raisebox{-32pt}{\scalebox{1}{\bubgrav}}
         \caption{
         }
         \label{fig:bub}
     \end{subfigure}
    \caption{Sample graviton loop diagrams contributing to the one-loop four-scalar amplitude. Scalars are denoted by straight lines and gravitons by wavy lines.}
    \label{fig:gravloops}
\end{figure}

Several references have considered the effects of low-energy loops on such dispersive bounds \cite{Bellazzini:2020cot, Bellazzini:2021oaj,Li:2022aby, Caron-Huot:2024lbf, Ye:2024rzr, Peng:2025klv}, but none studied the contributions from loops of gravitons in detail. In this work we find that, in even dimensions, $D=2n$, including the leading graviton loops such as those in Figure~\ref{fig:gravloops} generically yields bounds of the form
\be
g_{2n} \ge - {\cal O}(1) G^2M^{D-4n}\,,  \qquad (n>1, D>4)\,,
\ee
for all Wilson coefficients except the ones that dominate over the graviton loops at low energies. For instance, we find that in $D=6$ the subleading coefficient $g_4$ satisfies the bound 
\begin{equation}
    g_4 \ge - 0.057 G^2/M^2 \,,  \qquad (D=6)\,,
\end{equation}
whereas in $D> 8$ it is still required to be positive. Similarly all other $g_{2n}$'s are allowed to be negative for $4n>D$. In the dimension where the coefficient is degenerate with graviton loops, the sign can go either way and one must compute the bound to determine it.  The conclusion is that, in a fixed even dimension,\footnote{The structure of loop contributions in odd dimensions is quite different, so we leave their analysis for future work.} \emph{the majority of Wilson coefficients are allowed to be negative by a small amount}.

Thus, the naive analysis from forward dispersion relations fails beyond the tree approximation. This is due to additional forward singularities from graviton loops, which are among various technical difficulties encountered using fixed-$t$ dispersion relations to obtain bounds on Wilson coefficients. In this work, we perform our analysis using crossing-symmetric sum-rules \cite{Sinha:2020win, Auberson:1972prg, Mahoux:1974ej}, which are better behaved in the presence of loops and allow us to obtain reliable bound such as the one quoted above.

In $D=4$, \eqref{eq:g2D} is modified to include an infrared logarithm which arises from the smearing of the $1/t$ tree-level graviton pole against transverse wavefunctions \cite{Caron-Huot:2021rmr}
\be \label{eq:g2D4}
g_{2} \ge - 25\times 8\pi G/M^{2} \log(0.3 M b_{\rm max})\,,  \qquad (D=4)\,,
\ee
where $b_{\rm max}\sim 1/\sqrt{t_{min}}$ is an IR cutoff in impact parameter space introduced by the transverse wavefunction. Indeed, graviton loop corrections generate higher powers of these infrared logarithms which become large as one takes $t_{min}\to 0$ (or $b_{\rm max}\to \infty$).

Finally, we observe that by assuming that the behavior of the amplitude in the forward limit $t\to 0$ is well captured by the familiar eikonal formula, one can resum such infrared logarithms in $D=4$ to all orders in $G$ and obtain finite bounds, e.g., of the form $g_2 \ge - {\cal O}(1)/M^4$. This builds upon the recent observations that the eikonal amplitude saturates spin-2 dispersive sum rules \cite{Caron-Huot:2021rmr,Caron-Huot:2021enk} and can generate the graviton pole \cite{Haring:2024wyz}. These bounds are independent of $G$, so strictly speaking they go beyond the weak coupling limit and they are best interpreted as bounds on non-perturbative scattering observables that reduce to the EFT couplings at low energies.

This paper is organized as follows: in Section~\ref{sec:fixed-t_tree} we review fixed-$t$ dispersive sum rules, explain how they are used to obtain bounds on Wilson coefficient at tree-level, and comment on various issues that arise when trying to include graviton loops in this analysis. In Section~\ref{sec:fixed-a_tree} we review the crossing-symmetric dispersion relations and use them to derive tree-level bounds on Wilson coefficients, which we compare to the bounds obtained from fixed-$t$ dispersion relations. Then, in Section~\ref{sec:fixed-a_loop} we explain how to include graviton loop corrections in the crossing-symmetric dispersive sum rules and corresponding bounds, and we illustrate the loss of the requirement positivity in various examples in $D\ge 6$. In Section~\ref{sec:4Dbounds} we perform the analysis in $D=4$. We illustrate the appearance of infrared logarithms from loops, and we explain how to resum them to obtain finite bounds using the eikonal formula. Finally in Section~\ref{sec:conclusion} we summarize our findings and provide an outlook. We also include several appendices: In Appendix~\ref{app:one-loop_amp} we present the one-loop corrections to the four-point scalar amplitude, including graviton loops in EFT; in Appendix~\ref{app:dispersive_int} we provide the details of a technical computation of the dispersive integrals for the graviton box diagram; and in Appendix~\ref{app:numerics} we explain the details of our numerical optimization setup for computing our bounds.

\paragraph{Note added:} During the process of completing this manuscript we were made aware of overlapping results with the upcoming work \cite{Beadle2025}. We thank the authors for sharing their draft with us and coordinating submission.

\section{Fixed-$t$ dispersive bounds with gravity}\label{sec:fixed-t_tree}

Let us start by reviewing the basic setup to compute bounds using dispersion relations at fixed $t$.
Our assumptions are exactly as in \cite{Caron-Huot:2021rmr}, namely 
\begin{itemize}
    \item Our theory can be described at low energies as a weakly coupled EFT with cutoff $\Lambda$ such that $G\Lambda^{D-2}\ll 1$ and $g_2\Lambda^D\ll1$. We keep the combination $\lambda=g_2\Lambda^2/G$ fixed and generically $O(1)$.\footnote{This ratio controls the strength of gravity, with $\lambda = \infty$ being a non-gravitational theory of scalars, and $\lambda \ll 1$ a theory dominated by gravitational interactions.}

    \item Causality, as embodied in the analyticity and crossing properties of the scattering amplitude.

    \item Rotational invariance and unitarity, which result in a partial wave expansion 
    \be
    {\cal M}(s,t) = \sum_J n^{(D)}_J\, a_J(s)\, {\cal P}_J\left(1+\tfrac{2t}{s}\right)
    \ee
    with positive spectral density
    \be
     0\le \rho_J(s) =s^{\frac{D-4}{2}} \text{Im}\, a_J(s) \le 2.
    \ee
    Here, ${\cal P}_J$ are the usual Gegenbauer polynomials
    \be\label{eq:Gegenbauer_def}
     {\cal P}_J\left(x\right) = {}_2 F_1(-J,J+D-3;\tfrac{D-1}{2};\tfrac{1-x}{2}),
    \ee
    and the normalization factor $n^{(D)}_J$ is given by
    \be
    n^{(D)}_J = \frac{(4\pi)^{\frac{D}{2}} (2J+D-3) \Gamma(J+D-3)}{\pi \Gamma\left(\tfrac{D-2}{2}\right)\Gamma(J-2)}.
    \ee
\end{itemize}
Traditionally, one further assumes that the amplitude in the Regge limit satisfies
\be\label{eq:Regge_assumption}
\lim_{|s|\to \infty} \frac{{\cal M}(s,t)}{s^2} = 0\,.
\ee
Recently, it was argued in \cite{Haring:2022cyf, Caron-Huot:2022ugt} that this follows from the assumptions listed above when the amplitudes are smeared over $t$. This will indeed be the case after applying the class of functionals that we use in this work. Additionally, it was also shown in \cite{Caron-Huot:2021enk} that the flat space limit of the rigorous CFT dispersion relations agree with the twice-subtracted flat space dispersion relations, which follow from \eqref{eq:Regge_assumption}. 

\subsection{Review: Fixed-$t$ dispersive sum rules}

\begin{figure}[t]
	\centering
    \begin{tikzpicture}
	\draw[ mgrey] (0,-2) -- (0,2);
	\draw[ mgrey] (-3,0) -- (3,0);
	\node (a) at (2.5,1.5) {$s$};
	\draw (a.north west) -- (a.south west) -- (a.south east);	
	
	\draw (0,-0.1) node[cross=2pt] {};
	\draw[decoration={snake, amplitude=0.6}, decorate, thick,mred] (3,-0.1) -- (0,-0.1) ;
		
	\draw (0.75,0.1) node[cross=2pt] {};
	\draw[decoration={snake, amplitude=0.6}, decorate, thick,mred] (-3,0.1) -- (0.75,0.1) ;
	
	\draw[thick,mblue] (2.25,-0.2) arc (0:-180:2.25) ;
	\draw[thick,mblue] (2.25,+0.2) arc (0:+180:2.25) ;
	
	\draw (0,   0) node[xshift=-0.2cm, yshift=-0.4cm] {\scriptsize $0$};
	\draw (0.75,0) node[xshift=+0cm, yshift=+0.4cm] {\scriptsize $-t$};
	\draw (2.25,0) node[xshift=-0.3cm, yshift=+0.3cm] {\scriptsize $M^2$};
	\draw (-2.25,0) node[xshift=0.6cm, yshift=+0.3cm] {\scriptsize $-M^2-t$};
	
	\draw[thick,mblue] (2.25,+0.2) -- (3,+0.2);
        \draw[thick,mblue] (-3,+0.2) -- (-2.25,+0.2);
	\draw[thick,mblue] (2.25,-0.2) -- (3,-0.2);
        \draw[thick,mblue] (-3,-0.2) -- (-2.25,-0.2);
	
	\end{tikzpicture}
\caption{Contour that gives rise to the sum rule. The contour consists of a low-energy arc $|s+t/2|=M^2+t/2$, and a high-energy part along the branch cuts on the real axis.
}
    \label{fig:contour-fixed-t}
\end{figure}
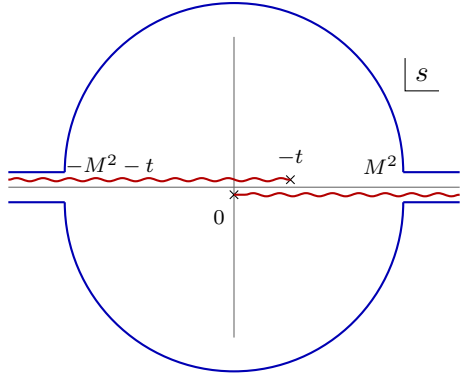

Together, the assumptions above imply that the amplitude satisfies the twice-subtracted fixed-$t$ dispersion relation,
\be
\frac{{\cal M}(s,t)}{s(s+t)} 
= \int_0^\infty \frac{ds'}{\pi}\left(\frac{1}{s'-s} + \frac{1}{s'+s+t}\right)
\text{Im} \left[  
\frac{{\cal M}(s',t)}{s'(s'+t)}
\right],
\ee
which is valid for $t<0$ and $s$ not on the real axis. This dispersion relation can be shown to arise from a deformation of the contour in Figure~\ref{fig:contour-fixed-t}.
We will not review here its derivation, which can be found e.g., in Ref.~\cite{Caron-Huot:2020cmc}.

The first step to deriving dispersive bounds is to separate the dispersion relation above into low- and high-energy contributions, using a scale $M\leq \Lambda$. The low-energy contributions to the dispersion integral can be written as the integral over arcs $|s+t/2|=M^2+t/2$ (see Figure~\ref{fig:contour-fixed-t}). The integral can be computed using the EFT,
\be 
B(s,t)|_{\rm low} &= 
 \int_0^{M^2} \frac{ds'}{\pi}\left(\frac{1}{s'-s} + \frac{1}{s'+s+t}\right) \text{Im} \left[  
\frac{{\cal M}(s',t)}{s'(s'+t)}
\right] - \frac{{\cal M}(s,t)}{s(s+t)} \nn \\
 &= 
 \int_{\rm arc} \frac{ds'}{4\pi i}\left(\frac{1}{s'-s} + \frac{1}{s'+s+t}\right) \frac{{\cal M}(s',t)}{s'(s'+t)} - \frac{{\cal M}(s,t)}{s(s+t)}.
\ee

 The high-energy contributions can be written using the partial wave expansion as
\begin{align}
B(s,t)|_{\rm high} &= 
 \int_{M^2}^\infty \frac{ds'}{\pi}\left(\frac{1}{s'-s} + \frac{1}{s'+s+t}\right) \frac{\text{Im} {\cal M}(s',t)}{s'(s'+t)} \\
 &= \bigg \langle \frac{(2 s' + t)}{(s' - s) (s' + s + t)} \frac{{\cal P}_J\left(1+\tfrac{2t}{s'} \right)}{(s'+t)} \bigg\rangle,
 \end{align}
where we have defined the positive average
\be\label{eq:highenergysum_definition}
\big\langle (\cdots) \big \rangle = \sum_J  n^{(D)}_J \int_{M^2}^\infty \frac{ds'}{\pi} s'^{\frac{2-D}{2}}  \rho_J(s')\,  (\cdots)\,.
\ee
Expanding at small $s$ and $ s+t$,\footnote{Each term can be conveniently extracted  by computing the contour integral
\[
\oint_\infty \frac{ds}{4\pi i} \left(\frac{1}{s} + \frac{1}{s+t}\right) \frac{B(s,t)}{(s(s+t))^{k/2-1}} .
\]}
\begin{equation}
   B(s,t) = \sum_{k \text{ even}} B_{k}(t)  (s(s+t))^{\frac{k}{2}-1}, 
\end{equation}
one gets a family of sum rules labeled by the so-called Regge spin $k$
\be
 - B_{k}(t)|_{\rm low} = 
 B_{k}(t)|_{\rm high}
 \quad (k \;\; \text{even})\,.
\ee

These sum rules express a combination of EFT Wilson coefficients on the LHS as a high-energy average over known functions with unknown but positive coefficients. 
For example, the $k=2$ and $k=4$ sum rules for the amplitude \eqref{eq:amp_scalarwithG} computed at tree level are
\begin{align}
- B_{2}(t)|_{\rm low} &= \frac{8\pi G}{-t} +   2 g_2 -  g_3 \,t + 8 g_4 \,t^2  - 2 g_5 \,t^3  + \cdots, \\
- B_{4}(t)|_{\rm low} &= 4 g_4 - 2 g_5 \, t  +  (24 g_6 + g_6' ) \, t^2 - 8  g_7 \, t^3 + \cdots,
\end{align}
and the corresponding high-energy contributions are
\begin{align}
    {B}_{2}(t)|_{\rm high} &=\left\langle \frac{(2m^2+t){\cal P}_J(1+\tfrac{2t}{m^2})}{m^2(m^2+t)^2}\right\rangle, \\
    {B}_{4}(t)|_{\rm high} &=\left\langle \frac{(2m^2+t){\cal P}_J(1+\tfrac{2t}{m^2})}{m^4(m^2+t)^3}\right\rangle \,.
\end{align}

\subsection{Tree-level bounds}
Armed with the dispersive sum rules, we can derive bound on the Wilson coefficients by applying a functional which is positive when acting on a linear combination of the high-energy combinations
\be
\sum_{k} F_k[{B}_{k}(t)|_{\rm high}] \ge0.
\ee
Then by applying the same functional to the low-energy contributions one obtains a bound on the EFT Wilson coefficients. In practice, the functional is chosen to give an optimal bound when acting on the low energy contributions.

An often-used class of functionals are linear combinations of derivatives at the forward limit $t=0$,
\be
F_k[\,\cdot\,] = \sum_{n}^{N_{k, \rm max}} \alpha_{k,n}\, \partial_t^n \cdot \big|_{t=0}.
\label{eq:forw-funct}
\ee
For instance, the forward limit of an arbitrary spin sum rule is, at high energies,
\be
{B}_{k}(0)|_{\rm high} = \left\langle \frac{2}{m^{2k}}\right\rangle,
\ee
and at low energies
\be
-{B}_{k}(0)|_{\rm low} =  2^{\frac{k}{2}} g_{k}\qquad (k > 2\, \text{ even})
\ee
which yields the bounds
\be
g_{k} \ge 0 \qquad (k > 2, \text{ even})
\ee
For $k=2$, the forward limit is ill-defined due to the $1/t$ pole from graviton exchange. We will return to this point below.

A key observation made in Ref.~\cite{Caron-Huot:2020cmc} is that crossing at low energies implies that multiple sum rules are sensitive to the same Wilson coefficient. For example, the term
\be
{\cal M} \supset g_6 (s^2+t^2+u^2)^3 = 8g_6 (s^6 + 6 s^4 t^2 + 6 s^2 t^4 + \cdots)
\ee
will feature in the $B_6, B_4$ and $B_2$ sum rules. This observation enables the existence of so-called {\it improved sum rules}, and {\it null constraints}, which are special linear combinations of the sum rules that truncate, or all together vanish, in the EFT expansion at tree level. The leading improved sum rules are\footnote{Here, the forward derivatives of the amplitude must be taken after the integral, e.g.,
    \be
     \oint \frac{ds}{4\pi i s}  \frac{t^3 \partial_{t} {\cal M}(s,0)}{s^2 (s^2-t^2)}  =  \lim_{t^\ast\to 0} \frac{\partial}{\partial t^\ast}   \oint \frac{ds}{4\pi i s}   \frac{t^3  {\cal M}(s,t^\ast)}{s^2 (s^2-t^2)} 
    \ee}
    \begin{align}
        B_2^{\rm imp}(t) &= B_2(t) - 2 t^2 B_4(0)  + t^3 \partial_{t}B_4(0) + 3 t^4 B_6(0) + t^{5} \partial_{t}B_6(0)+\cdots \nonumber \\
        &= \oint \frac{ds}{4\pi i s} \left[ \frac{{\cal M}(s,t)}{s \left(s+t\right)} - \frac{t^3 \partial_{t} {\cal M}(s,0)}{s^2 (s^2-t^2)} + \frac{ \left(2 s^2-t^2\right) t^2 {\cal M}(s,0)}{s^3 (s+t) (s^2-t^2)}\right] \,, \label{eq:B2imp_def} \\
        B_4^{\rm imp}(t) &= B_4(t) - 2 t^3 \partial_tB_6(0) + \frac12 t^4 \partial^2_tB_6(0) + 2 t^4 B_8(0) + \cdots \nonumber \\
        &=\oint \frac{ds}{4\pi i s} \left[ \frac{{\cal M}(s,t)}{s^2 \left(s+t\right)^2} -\frac{t^4 \partial_{t}^2 {\cal M}(s,0)}{2 s^4 \left(s^2-t^2\right)} +\frac{ \left(2 s^2- st-2 t^2\right)t^3\partial_{t} {\cal M}(s,0)}{s^5 \left(s+t\right) (s^2-t^2)}  \right.\nonumber\\ 
        &\hspace{5cm} \left.+ \frac{\left(3 s^3 -4  st^2 - 2 t^3\right) t^2 {\cal M}(s,0)}{s^5 \left(s+t\right)^2 \left(s^2-t^2\right)} \right] \,,
    \end{align}
whose tree-level EFT contributions indeed truncate at low energies,
\begin{align}
        -B_{2}^{\rm imp}(t)\big|_{\rm low}^{\rm tree} &=  \frac{8\pi G}{-t} + 2 g_2 - g_3 t , \\
        -B_4^{\rm imp}(t)\big|_{\rm low}^{\rm tree} &= 4 g_4 - 2 g_5 t + g_6' t^2 \,.
\end{align}
Similarly, the lowest spin null constraint is
    \begin{align}
        X_4(t) &=  B_4(t) - B_4(0) - t \partial_tB_4(0)  - \frac12 t^2 \partial^2_tB_4(0)  - 2 t^3 \partial_tB_6(0)   - \frac12 t^4 \partial^2_tB_6(0)  + \cdots \nonumber \\
        &=\oint \frac{ds}{4\pi i s} \left[ \frac{{\cal M}(s,t)}{s^2 \left(s+t\right)^2} - \frac{t^2 \partial_t^2 {\cal M}(s,0)}{2 s^2 \left(s^2-t^2\right)} - \frac{ \left(s^2-st-t^2\right) t\partial_{t} {\cal M}(s,0)}{s^3 \left(s+t\right) \left(s^2-t^2\right)} \right.\nonumber\\ 
        &\hspace{5cm} \left.- \frac{\left(s^2+ st+t^2\right) \left(s^2-
   st-t^2\right) {\cal M}(s,0)}{s^4 \left(s+t\right)^2 \left(s^2-t^2\right)}\right],
    \end{align}
and 
\be
X_4(t)\big|_{\rm low}^{\rm tree}  = 0\, .
\ee
Note that $X_4(t)$ can be obtained from the improved spin-$4$ sum rule using $X_4(t) = B_4^{\rm imp}(t) - B_4^{\rm imp}(0) - t \partial_t B_4^{\rm imp}(0) - \frac{t^2}{2}\partial^2_t B_4^{\rm imp}(0)$.

The existence of improved sum rules and null constraints greatly simplifies the optimization problem as one can consider combinations
\be
\sum_{k} F_k[{B}^{\rm imp}_{k}(t)|_{\rm high}] + \sum_{k} H_k[{X}_{k}(t)|_{\rm high}] \ge 0
\ee
which at low energies, ignoring loops, just give
\be
-\sum_{k} F_k[{B}^{\rm imp}_{k}(t)|^{\rm tree}_{\rm low}] \ge 0\,.
\ee
Thus, in practice the null constraints facilitate the task of finding positive functionals. Furthermore, since the improved sum rules $B_k^{\rm imp}$ truncate at $O(t^k)$ at low energies one can also consider functionals that annihilate the low energy part beyond a given $k$. For instance in the class of functionals in Eq.~\eqref{eq:forw-funct} on can consider functionals with $\alpha_{k,n} = 0$ for $n<k$, which effectively generates additional null constraints.

Another useful class of functionals which sidestep the issues with singular forward limits are the so-called impact-parameter functionals. These are defined by integrating the sum rules over the momentum transfer $t=-p^2$ with compact support for $p \in [0, M]$
\be
  {\cal F}_k[\,\cdot\,] = \int_0^M dp f_k(p) \cdot \big|_{t=-p^2},
\ee
where $f_k(p)$ is a polynomial 
\be
f_k(p) = p^r \sum_{n} \beta_{k,n} p^n,
\ee
whose coefficients are fixed by the optimization problem. This typically results in functionals which are peaked at small impact parameter $b\sim 1/p \sim 1/M $. For examples of tree-level bounds from impact-parameter functionals see e.g., \cite{Caron-Huot:2021rmr, Caron-Huot:2022jli,Caron-Huot:2022ugt}.

\subsection{Issues with fixed-$t$ dispersion relations at loop level}\label{sec:fixed-t_loop}

Several technical issues arise when considering the contribution of low-energy loops in fixed-$t$ dispersion relations. Just like the familiar $1/t$ singularity at tree level, these issues are specific to theories with a massless spin-$2$ tree-level exchange. For this reason, they were not encountered in Refs.~\cite{Bellazzini:2020cot,Bellazzini:2021oaj} which also considered loop corrections to dispersive fixed-$t$ sum rules without gravity. 

First, at small $t$, additional forward singularities arise from graviton exchanges at loop level. For instance, the one-loop contribution to the amplitude from minimal coupling in even dimension, $D$, yields a logarithmic dependence 
\be
{\cal M}^{\rm 1-loop} \sim & \,(8\pi G)^2 s^3 (-t)^{\frac{D-6}{2}} \log(-t),
\ee
which arises from the double graviton exchange in in Figure~\ref{fig:box}.
This behavior remains after performing the dispersive integral, and hence the sum rules $B_k(t)\big|_{\rm low}$ for all Regge spin $k$ have logarithmic divergences starting at $(-t)^{\frac{D-6}{2}}$. Note that the spin-$2$ improved sum rule $B_2^{\rm imp}(t)$ defined in \eqref{eq:B2imp_def} requires $B_k(0)$ and $\partial_t B_k(0)$, and $B_4^{\rm imp}(t)$ requires even higher forward derivatives. As a result, the improved sum rules and null constraints are generically ill-defined due to the $\log(-t)$ divergences, unless the spacetime dimension $D$ is large enough.\footnote{For example, $B_2^{\rm imp}(t)$ is well-defined in $D\geq 10$, and $B_4^{\rm imp}(t)$ is well-defined in $D\geq 12$.}

When using impact-parameter functionals, a further subtlety arises at large $t$. Namely that the appearance of logarithms in the amplitude, e.g., ${\cal M} \supset s\log(-s)$ can result in singularities in the dispersion relation at $t = -M^2$ from the subtractions,
\be
\int_{\rm arc} \frac{ds}{2\pi i}\frac{1}{s} \frac{{\cal M}(s,t)}{[s(s+t)]^{\frac{k}{2}}} \sim \frac{1}{(M^2+t)^{\frac{k}{2}}}  + \cdots \,.
\ee
These can invalidate the use of functionals which do not vanish rapidly enough as $t\to -M^2$, but that otherwise had no issues at tree level. 

Interestingly, superconvergent sum rules for spinning external states avoid these issues, as they are automatically improved and do not require any subtractions. However, these sum rules do not exist for scalar external states.

We give the explicit expressions for the one-loop contribution to fixed-$t$ sum rules in Appendix \ref{app:dispersive_int} and the ancillary file \texttt{GravitonLoop.m}. One can check that they indeed have the above features at $t=0$ and $t=-M^2$. Several strategies can be used to deal with these issues. For example, one might restrict the class of functional such that they truncate the EFT expansion and avoid singularities at small and large impact parameters. Another option is to consider partially-improved sum rules, as in Ref.~\cite{Beadle:2024hqg}. In this work, instead, we will abandon fixed-$t$ dispersion relations altogether, and use instead a different family of dispersion relations.

\section{Crossing-symmetric dispersive bounds with gravity}\label{sec:fixed-a_tree}

The issues with fixed-$t$ dispersion relations described above motivate the use of crossing-symmetric dispersion relations \cite{Sinha:2020win, Auberson:1972prg, Mahoux:1974ej}, which instead fix the crossing-symmetric variable,
\be \label{eq:adef}
a = \frac{stu}{st+tu+su}\,.
\ee
At small $t$, one has $a \to t$, so the behavior of the fixed-$t$ and fixed-$a$ sum rules are very similar. Manifest crossing symmetry, however, results in the automatic improvement of the sum rules. 
Furthermore, for physical kinematics $|a|$ cannot exceed $M^2/3$, and thus the aforementioned issues at large $t$ are also ameliorated.

\subsection{Review: crossing-symmetric dispersive sum rules}
We now give a review of the crossing-symmetric dispersion relations \cite{Sinha:2020win, Auberson:1972prg, Mahoux:1974ej}. We will focus on the case where the amplitude is completely crossing-symmetric. The main idea is to introduce fully crossing-symmetric variables,
\be
x=st+su+tu=-\frac{1}{2}(s^2+t^2+u^2)\,,\quad y=stu\,,\quad a=\frac{y}{x}\,,
\ee
and express the amplitude in terms of $x$ and $a$. For example, for the tree-level four-scalar amplitude given in \eqref{eq:amp_scalarwithG}, we have
\be\label{eq:Mtree_inxa}
{\cal M}_{\f\f\f\f}^{\textrm{tree}} &= 8\pi G\frac{x}{a} -2g_2 x+g_3 x a+4 g_4 x^2-2g_5 x^2 a+g_6' x^2 a^2 \nn \\
&-8g_6 x^3+4g_7 x^3 a-2g'_8x^3 a^2+g'_9x^3 a^3+ 16g_8x^4-8g_9 x^4 a + \cdots\,,
\ee
where the first term comes from the graviton exchange, and the others are contact interactions, which we have included up to $g_9, g'_9$ (see \eqref{eq:M4s_tree}). Note that for the contact interactions, the power of $a$ cannot be bigger than the power of $x$. This comes from locality of the amplitude and will be important in a moment.

In terms of the $x,a$ variables, the authors of \cite{Sinha:2020win} derived a twice-subtracted dispersion relation. (See also \cite{Li:2023qzs} for a detailed discussion with massless loops.) The dispersion relation is given by\footnote{Strictly speaking, the crossing-symmetric dispersion relation requires a stronger analyticity assumption than the fixed-$t$ dispersion relation. In particular, we will assume that the amplitude is analytic in the complex $s,t$ plane in some region $\mathcal{D} \subset \mathbb{C}^2$, and $\mathcal{D}$ contains the region $-\frac{M^2}{3}<a<0$.}
\be\label{eq:crossing_symmetric_relation}
{\cal M}(x,a) =& \alpha_0(a)+\int_0^{\infty} \frac{ds'}{\pi}\mathrm{Im} {\cal M}(s',t)\left(\frac{s}{s'(s'-s)}+\frac{t}{s'(s'-t)}+\frac{u}{s'(s'-u)}\right) \nn \\
=&\alpha_0(a)+x\int_0^{\infty} \frac{ds'}{\pi}\mathrm{Im} \frac{{\cal M}(s',\tau(s',a))}{(s')^3}\frac{(2s'-3a)(s')^2}{x(a-s')-(s')^3},
\ee
where $\alpha_0(a)$ is a subtraction function, and $\tau(s',a)$ is the inverse function from $a$ to  $t$,
\be\label{eq:tau_def}
\tau(s',a) = -\frac{s'}{2}\left(1-\sqrt{\frac{s'+3a}{s'-a}}\right).
\ee
Note that for fixed real and positive $s'$, one should consider $s'>a >-\frac{s'}{3}$ in order for $\tau(s',a)$ to be real, and this gives $\tau > -\frac{s'}{2}$. The region $\tau < -\frac{s'}{2}$ is covered by the other solution $-\frac{s'}{2}\left(1+\sqrt{\frac{s'+3a}{s'-a}}\right)$. One can obtain it either by going through the branch cut of \eqref{eq:tau_def} starting at $a=-\frac{s'}{3}$, or by simply performing a $t \leftrightarrow u$ crossing. In other words, Eq.~\eqref{eq:tau_def} illustrates that $t=\tau(s',a)$ defines a two-sheeted Riemann surface which is the double cover of the complex $a$ plane, and the two sheets are related by $t \leftrightarrow u$ crossing. Since crossing symmetry is built-in, we expect it suffices to consider $\tau > -\frac{s'}{2}$ for crossing-symmetric dispersion relation.

Similar to the fixed-$t$ dispersion relations, one can separate the integral on right-hand side of \eqref{eq:crossing_symmetric_relation} into a part below an energy scale $M$ and another part above $M$. In this section where we only discuss tree-level EFT contributions, we choose $M$ to be the equal to the cutoff of the EFT, $\Lambda$.

We can take derivatives with respect to $x$ around $x=0$ to obtain dispersive sum rules. By applying $\frac{(-1)^{k/2}}{(k/2)!}\partial^{k/2}_x|_{x=0}$ to \eqref{eq:crossing_symmetric_relation}, we obtain a family of sum rules parametrized by $a$ and an even integer $k$,
\be\label{eq:fixeda_sumrules}
-{\cal C}^{\textrm{low}}_{k}(a)={\cal C}^{\textrm{high}}_{k}(a)\,,
\ee
where $k=2,4,6,\ldots$ is the Regge spin of the sum rule. The right-hand side contains all the contributions above the EFT cutoff. Using the partial wave decomposition of ${\cal M}$, we can write it as
\be\label{eq:fixeda_highenergy_def}
{\cal C}^{\textrm{high}}_{k}(a) =
\left\langle \frac{{\cal P}_J(\sqrt{\tfrac{m^2+3a}{m^2-a}})}{m^{3k}}(m^2-a)^{\frac{k}{2}-1}(2m^2-3a) \right\rangle \equiv \left\langle {\cal C}_{k,a}[m^2,J] \right\rangle\,,
\ee
where we have used the $\left\langle \ldots \right\rangle$ notation defined by \eqref{eq:highenergysum_definition}, and ${\cal P}_J$ is the Gegenbauer polynomial \eqref{eq:Gegenbauer_def}.

On the other hand, ${\cal C}^{\textrm{low}}_{k}(a)$ is a computable quantity from the EFT. It is given by
\be\label{eq:fixeda_lowenergy_def}
-{\cal C}^{\textrm{low}}_{k}(a)\!=\!\frac{(-1)^{k/2}}{(k/2)!}\partial^{k/2}_x{\cal M}(x,a)\Big|_{x=0} \!\!-\! \int_0^{M^2}\!\frac{ds'}{\pi}\mathrm{Im}{\cal M}(s',\tau(s',a))\frac{(s'\!-\!a)^{\frac{k}{2}-1}(2s'\!-\!3a)}{(s')^{\frac{3k}{2}+1}}\,.
\ee
Later, we will see that the first term captures the contribution from the tree-level EFT amplitude, and the second term corresponds to the loop contribution of ${\cal M}_{\textrm{EFT}}$.

In the limit $t/s \to 0$, we have $a \to t$. So, in some sense $a$ can be thought of as the crossing-symmetric version of the Mandelstam variable $t$. Because of this, in what follows we will often refer to \eqref{eq:fixeda_sumrules} as fixed-$a$ sum rules, in contrast with the fixed-$t$ sum rules considered in section \ref{sec:fixed-t_tree}.

\subsection{Tree-level crossing-symmetric sum rules}

Now, we consider crossing-symmetric sum rules in the tree-level case, and derive tree-level bounds from these sum rules. Several bounds from forward limit of the crossing-symmetric sum rules have been considered in \cite{deRham:2022gfe}, but here we will be considering bounds from the impact parameter space functional method, which is more suitable for  studying theories with gravity and/or low-energy loops.

For concreteness, we consider the spin-2, spin-4, and spin-6 fixed-$a$ sum rules (i.e., $k=2,4,6$). Their high energy contributions are given by \eqref{eq:fixeda_highenergy_def}. On the other hand, the contributions from the low-energy tree-level EFT amplitude given by \eqref{eq:Mtree_inxa} can be written as
\be\label{eq:fixeda_EFT_spin246}
-{\cal C}^{\textrm{low;tree}}_{2}(a) &=-\frac{8\pi G}{a} +2 g_2 -g_3 a\,,\nn \\
-{\cal C}^{\textrm{low;tree}}_{4}(a) &=4g_4 -2g_5 a+g'_6a^2\,, \nn \\
-{\cal C}^{\textrm{low;tree}}_{6}(a) &=8g_6-4g_7 a+2g'_8a^2-g'_9a^3\,.
\ee
As mentioned in the previous subsection, locality of the amplitude implies that the power of $a$ cannot be greater than the power of $x$ in the tree-level amplitude. Therefore, the low-energy part of all the fixed-spin crossing-symmetric sum rules involve only a finite number of Wilson coefficients, unlike the fixed-$t$ sum rules which have infinitely many Wilson coefficients. So, fixed-$a$ sum rules are automatically improved, and this is one of the advantages of using them.

Another important family of sum rules for deriving dispersive bounds are the null constraints, which follow from crossing symmetry of the low-energy amplitude. It is straightforward to obtain them from the fixed-$a$ sum rules. For example, we can define
\be
\tilde{X}_{4}(a) = {\cal C}_{4}(a)-{\cal C}_{4}(0)-{\cal C}'_{4}(0)a - \frac{1}{2}{\cal C}''_{4}(0)a^2\,.
\ee
The low-energy contribution of $\tilde{X}_{4}(a)$ vanishes at tree-level, and hence gives a family of null constraints. For instance, $\tilde{X}_{4}'''(0)$ is equivalent to the null constraint from the $g_7$ coupling.\footnote{The relation between $\tilde{X}_{4}(a)$ and the usual fixed-$t$ null constraints $X_{4}(t)$ given in \cite{Caron-Huot:2020cmc} is
\be
\tilde{X}_{4}(a=t) = 2t^3 X_{4}(t) + \sum_{n=1}^{\infty} \frac{2t^{n+2}}{n!}\partial^{n-1}_t (t^{3n}\partial_t (t X_{2n+4}(t)))\,.
\ee
} However, as we see above, fixed-$t$ null constraints obtained in this way require taking some forward limits, which become ill-defined due to the graviton loops. It turns out this is also true for the fixed-$a$ null constraints. It is thus more appropriate to obtain them by applying suitable functionals to ${\cal C}_{k}(a)$ that annihilate the low energy contributions.

Since $a$ in Eq.~\eqref{eq:adef} plays a similar role to $t$, we will use functionals that integrate the sum rule ${\cal C}_{k}(-p^2)$ against some function $f(p)$. In the $s$-channel physical region $s\in [0,M^2],t\in [-s,0]$, the range of $a$ is $-\frac{M^2}{3}<a<0$. Therefore, the functionals will act as $\int_0^{\frac{M}{\sqrt{3}}}dp\ f(p)(\ldots)$. Moreover, we can further restrict to the space of functionals that remove unwanted low-energy contributions. For example, to get the null constraints, let us consider
\be
X_{k;h_k}\equiv \int_0^{\frac{M}{\sqrt{3}}}dp\ h_k(p){\cal C}_{k}(-p^2)\,,
\ee
with the restriction
\be\label{eq:hk_nullconstraints}
\int_0^{\frac{M}{\sqrt{3}}}dp\ h_k(p) p^{2n} =0,\qquad n=0,1,\ldots,\tfrac{k}{2}\,.
\ee
From \eqref{eq:fixeda_EFT_spin246}, one can see that all $X_{k;h_k}$ sum rules have vanishing low-energy contribution at tree-level. Crucially, this definition makes sense even with graviton loops (for appropriate $h_k$'s). Therefore, we will use $X_{k;h_k}$ as our null constraints.

\subsection{Tree-level bounds from crossing-symmetric sum rules}\label{sec:fixed-a_tree_bounds}

We are now ready to derive dispersive bounds using crossing-symmetric sum rules. We will focus on the tree-level bound here, and consider the 1-loop corrections to the bound in the next section. As a concrete example, we consider bounds on the $g_2$ and $g_3$ couplings. The idea is to look for functions $f(p),h_k(p)$ such that 
\be\label{eq:pols_fixeda}
\int_0^{\frac{M}{\sqrt{3}}}dp\ f(p) {\cal C}_{2,a=-p^2}[m^2,J] + \sum_{k=4,6,\ldots} X_{k;h_k}[m^2,J] \geq 0\,,\quad \forall m>M, J=0,2,4,\cdots\,,
\ee
with the additional constraints on $h_k(p)$ given by \eqref{eq:hk_nullconstraints}. For any $f(p),h_k(p)$ satisfying the above conditions, we get a bound
\be
\int_0^{\frac{M}{\sqrt{3}}}dp\ f(p)\left(\frac{8\pi G}{p^2}+2g_2+g_3 p^2\right) \geq 0\,.
\ee
For instance, to find a lower bound on $g_3$, one can solve the optimization problem
\be\label{eq:optimization_tree}
\textrm{maximize}\ \int_0^{\frac{M}{\sqrt{3}}}dp\ f(p)\left(-\frac{8\pi G}{p^2}-2g_2\right)\,,\qquad \int_0^{\frac{M}{\sqrt{3}}}dp\ f(p) p^2=1\,.
\ee

For the numerical computation, we consider the functions $f(p)$ and $h_{k}(p)$ given by
\be\label{eq:functional_pexpansion}
f(p)=\sum_{n=0}^{N_2}a_n p^{n_0+n}\,,\quad h_k(p)=\sum_{m=0}^{N_k}b_{k,m}p^m\,.
\ee
As shown in \cite{Caron-Huot:2021rmr}, in order to satisfy \eqref{eq:pols_fixeda} at $m\to \infty$ with fixed impact parameter $b=\frac{2J}{m}$, one has to choose $n_0$ appropriately depending on the spacetime dimension. For $D=6,8,10,\ldots$, one can choose $n_0=\frac{3}{2}$, and for $D=7,9,11,\ldots$, one can choose $n_0=2$.\footnote{Another way is to introduce a $(1-\sqrt{3}p)^{\#}$ factor \cite{Caron-Huot:2021enk} (the additional $\sqrt{3}$ is because we are using fixed-$a$ sum rules). We will discuss this more in section \ref{sec:4Dbounds}.} The additional constraints on $h_k(p)$ \eqref{eq:hk_nullconstraints} can be written as $\frac{k}{2}+1$ linear equations of the coefficients $b_{k,m}$'s. With this choice of basis for the functions $f(p),h_k(p)$, the optimization problem \eqref{eq:optimization_tree} with positivity constraints \eqref{eq:pols_fixeda} can be solved numerically using the semidefinite programming solver {\tt SDPB} \cite{Simmons-Duffin:2015qma,Landry:2019qug}. We give more details on the numerical implementation in Appendix \ref{app:numerics}.

A substantial part of this calculation is similar to the one using the fixed-$t$ sum rules done in \cite{Caron-Huot:2021rmr}, but there are two main differences. Firstly, the functional we use only integrate over $p$ from $0$ to $\tfrac{M}{\sqrt{3}}$ instead of from $0$ to $M$, where $M$ is the size of the low-energy contour. This means we are using a smaller space of functionals, and therefore we expect the bounds to be weaker than the ones from fixed-$t$ sum rules. This is the price we have to pay when using the crossing-symmetric sum rules. 

The second difference is that the functions entering the positivity conditions \eqref{eq:pols_fixeda}, such as ${\cal C}_{2,-p^2}[m^2,J]$, are different in the fixed-$a$ and fixed-$t$ case. To impose positivity for all values of $m>M$ and $J$, one should consider both the condition at finite $m$ and $J$, and in the impact parameter space where one takes $m \to \infty$ with the impact parameter $b=\frac{2J}{m}$ kept fixed. For the finite $m$ and $J$ region, one has to plug in the expressions for high-energy contributions to the fixed-$a$ sum rules as given in \eqref{eq:fixeda_highenergy_def}, and then follow the same discretization and refinement procedure described in \cite{Caron-Huot:2021rmr}. On the other hand, the positivity conditions in the impact parameter space are in fact identical for the fixed-$a$ and fixed-$t$ sum rules. This follows from the fact that $a\to t$ at large $s$, and hence there is no distinction between fixed-$a$ and fixed-$t$ after taking the $m\to \infty$ limit.

\begin{figure}[t]
\centering
\includegraphics[width=12cm]{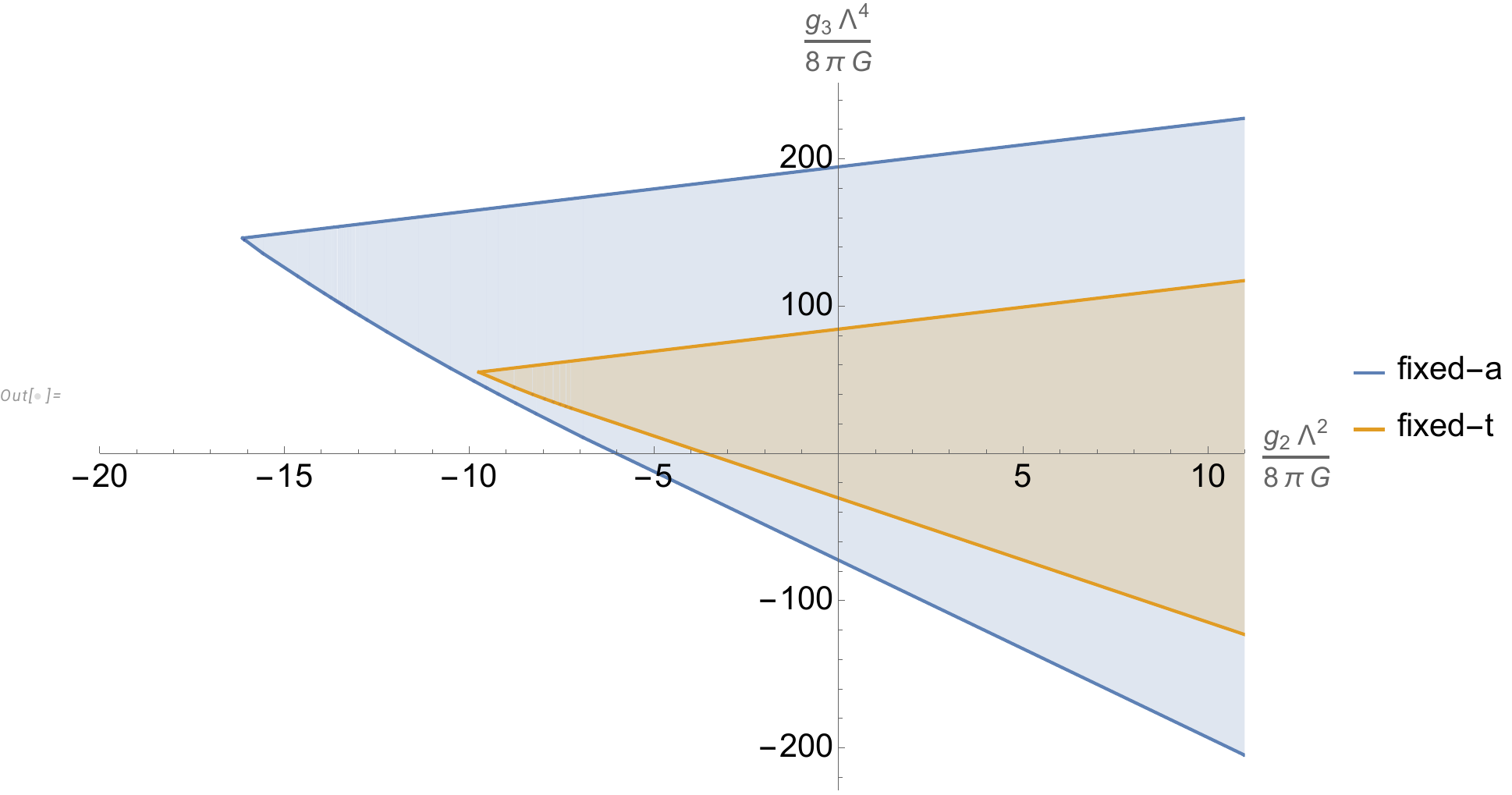}
\caption{A comparison of the tree-level bounds on $g_2,g_3$ in $D=6$ from the crossing-symmetric (fixed-$a$) sum rules and the bounds from the fixed-$t$ sum rules obtained in \cite{Caron-Huot:2021rmr}. We include the spin-2, spin-4, spin-6 sum rules, and the bounds are computed using a 17-dimensional space of functionals (see Appendix \ref{app:numerics} for more details).}
\label{fig:g2-g3-tree}
\end{figure}

In Figure \ref{fig:g2-g3-tree}, we show the tree-level bounds for the dimensionless couplings $\frac{g_2 \Lambda^2}{8\pi G},\frac{g_3 \Lambda^4}{8\pi G}$ in $D=6$ obtained from fixed-$a$ and fixed-$t$ sum rules. To obtain the bound, we include the contribution from the ${\cal C}_{2}(a)$ sum rule and the null constraints $X_{k;h_k}$ with $k=4,6$. We use \eqref{eq:functional_pexpansion} with $\{N_2,N_4,N_6\}=\{7,9,8\}$. After imposing the constraints \eqref{eq:hk_nullconstraints}, this gives a 17-dimensional space of functionals. For comparison, we also plot the bounds from the fixed-$t$ sum rules obtained in \cite{Caron-Huot:2021rmr} (with the same dimension of the functional space). As explained above, the fixed-$a$ bound is weaker than the fixed-$t$ bound because it uses a smaller functional space. However, the fixed-$t$ bounds become divergent after including the graviton loops, while the fixed-$a$ ones have finite 1-loop corrections which we will compute in the next section.

The method can also be adjusted to compute bounds on the other Wilson coefficients such as $g_4,g_6$. More concretely, one can again consider the spin-2,4,6 sum rules, and integrate them against functions $f(p), h_4(p),h_6(p)$. However, we relax the constraints \eqref{eq:hk_nullconstraints} and just impose
\be
&\int_0^{\frac{M}{\sqrt{3}}}dp\ h_4(p) p^{2} =\int_0^{\frac{M}{\sqrt{3}}}dp\ h_4(p) p^{4}=0,\nn \\
&\int_0^{\frac{M}{\sqrt{3}}}dp\ h_6(p) p^{2}=\int_0^{\frac{M}{\sqrt{3}}}dp\ h_6(p) p^{4}=\int_0^{\frac{M}{\sqrt{3}}}dp\ h_6(p) p^{6}=0.
\ee
In other words, we choose to not remove all the low energy contributions from the spin-4 and spin-6 sum rules. This will lead to a bound given by
\be
\int_0^{\frac{M}{\sqrt{3}}}dp\ f(p)\left(\frac{8\pi G}{p^2}+2g_2+g_3 p^2\right) + 4g_4\int_0^{\frac{M}{\sqrt{3}}}dp\ h_4(p) + 8g_6\int_0^{\frac{M}{\sqrt{3}}}dp\ h_6(p) \geq 0.
\ee
Similarly, by simply making different choices for the constraints we impose on $f(p),h_k(p)$, we can derive bounds involving different Wilson coefficients.

\begin{figure}[t]
\centering
\begin{subfigure}{.48\textwidth}
  \centering
  \includegraphics[width=\linewidth]{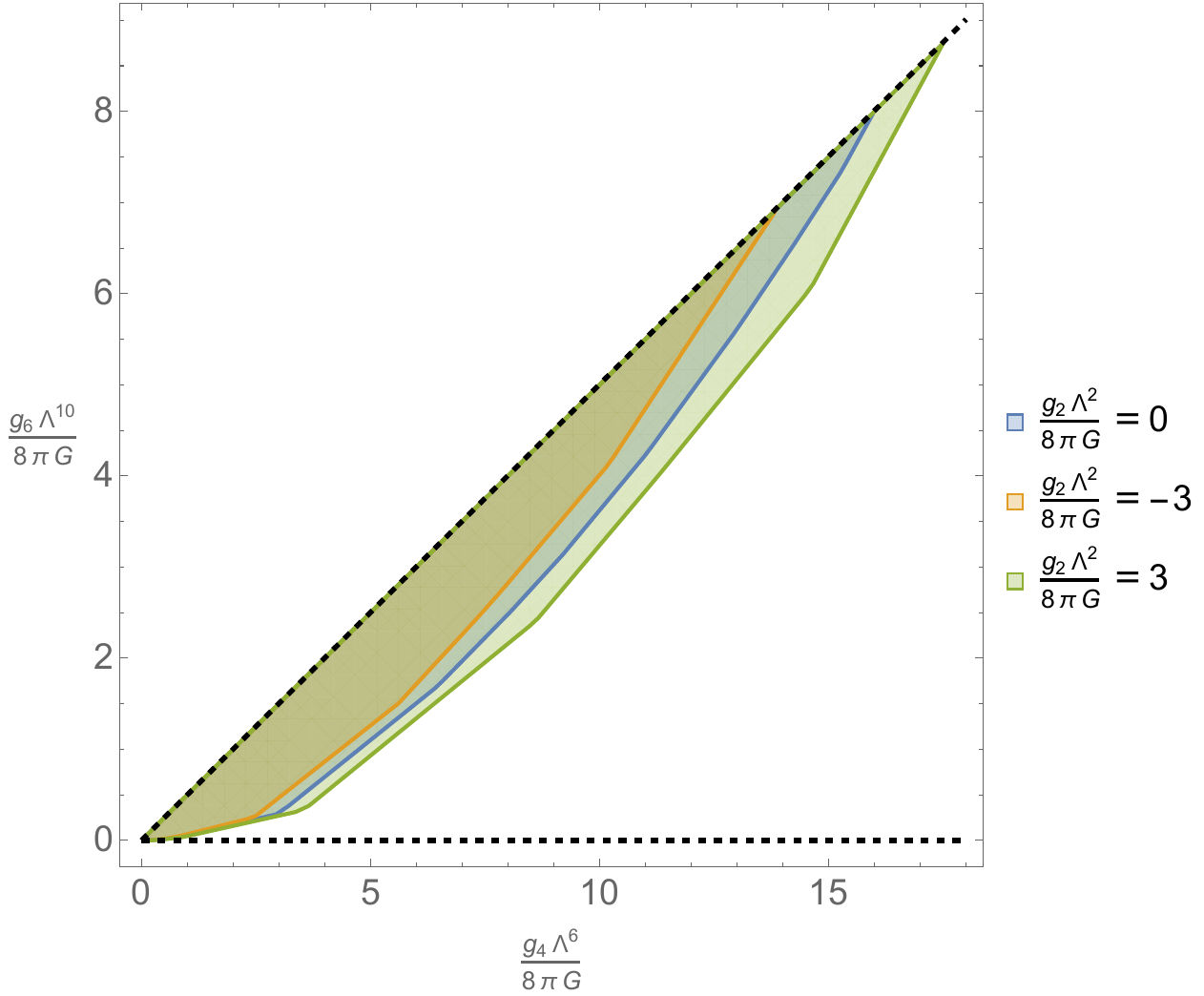}
\end{subfigure}%
\begin{subfigure}{.5\textwidth}
  \centering
  \includegraphics[width=.8\linewidth]{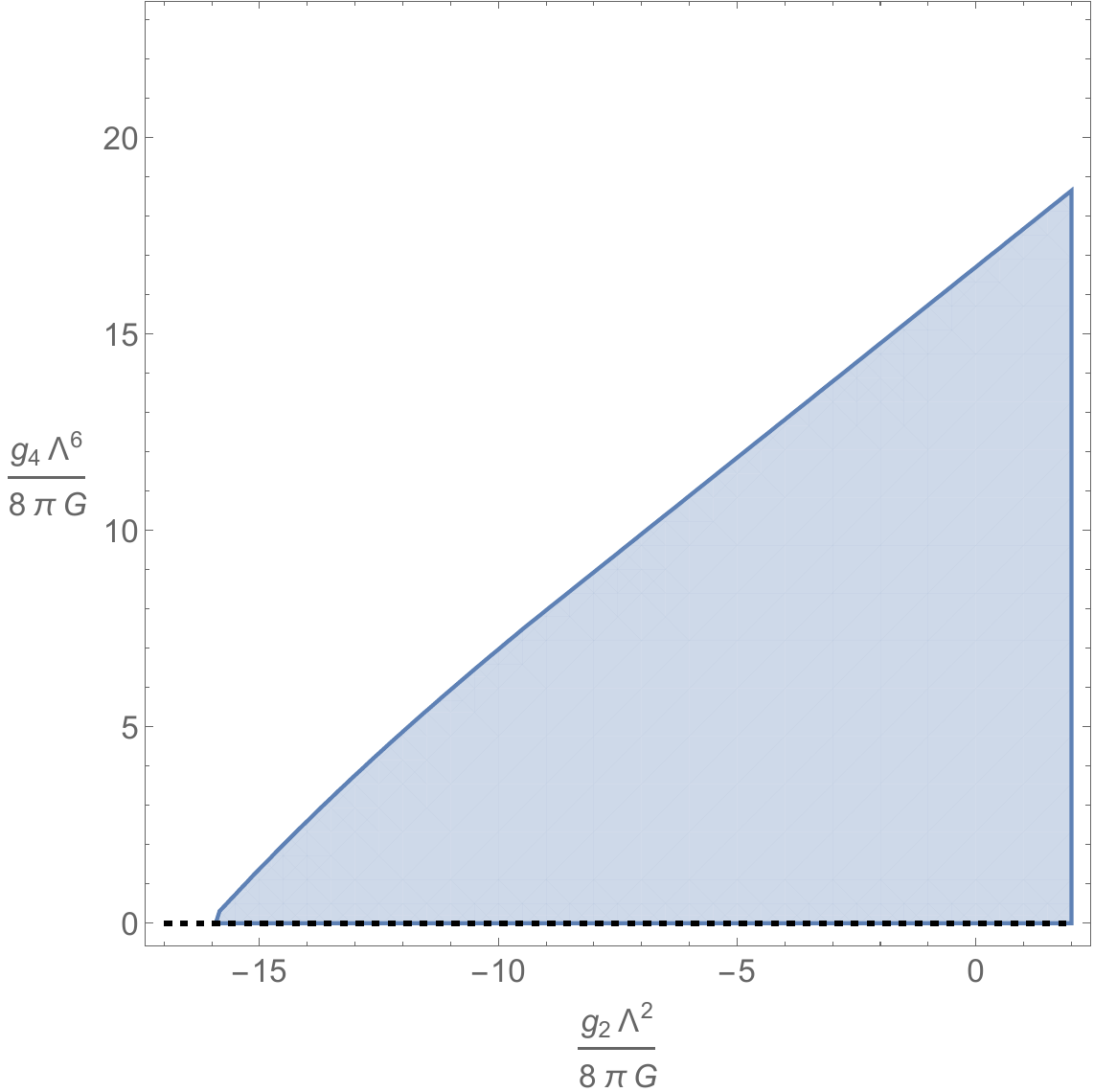}
\end{subfigure}
\caption{Tree-level bounds on higher couplings using fixed-$a$ sum rules ${\cal C}_{k,a}$ with $k=2,4,6,8$. \textbf{Left:} Allowed region of $\frac{g_4 \Lambda^{6}}{8\pi G}$ and $\frac{g_6 \Lambda^{10}}{8\pi G}$ with three different values of $\frac{g_2 \Lambda^{2}}{8\pi G}\in \{-3,0,3\}$. The two black dashed lines are $\frac{g_6}{g_4} \leq \frac{1}{2 \Lambda^4}$ and $g_6 \geq 0$. The space of functionals is $24$-dimensional. \textbf{Right:} Allowed region of $\frac{g_2 \Lambda^{2}}{8\pi G}$ and $\frac{g_4 \Lambda^{6}}{8\pi G}$. The black dashed line is the positivity bound $g_4\geq 0$. The space of functionals is $23$-dimensional. Note that $\frac{g_2 \Lambda^2}{8\pi G}$ has no upper bound, whereas $\frac{g_4 \Lambda^6}{8\pi G}$ has two-sided bounds (for fixed $g_2$).}
\label{fig:g4-g6_g2-g4}
\end{figure}

In Figure \ref{fig:g4-g6_g2-g4}, we show two examples of the bounds on higher couplings using the approach described above. Additionally, we also include a $k=8$ sum rule $X_{8,h_8}$ as a null constraint. On the left, we keep only $G, g_2,g_4,g_6$ at low energy, and plot the bounds on $g_4$ and $g_6$ (rescaled by $G$) for three different values of $g_2$. The allowed regions have two tangent lines given by $\frac{g_6}{g_4} \leq \frac{1}{2\Lambda^4}$ and the positivity bound $g_6 \geq 0$, which can be obtained from the forward limit. On the right, we keep only $G,g_2,g_4$ at low energy and plot the allowed region for $g_2,g_4$. The allowed region is also saturated by the positivity bound $g_4 \geq 0$.

\section{Graviton loop corrections to crossing-symmetric dispersive bounds}\label{sec:fixed-a_loop}

In this section, we consider how the dispersive sum rules are modified by the inclusion of loops in the effective field theory, including loops of gravitons. Let us begin by making some general comments about the effects of loops in  dispersive sum rules.

The dispersive sum rules are a relationship between integrals over the low energy arcs which are assumed to be calculable in the EFT, and high-energy terms integrals over the unknown, but positive, UV spectral density. At strict weak coupling, i.e., at tree-level, the EFT amplitude is meromorphic, so the computation of the low-energy arcs reduces to a computation of residues, which is insensitive to the location of the arc. Thus the only dependence on the EFT cutoff scale arises from the Wilson coefficients. At loop level, however, various subtleties arise.

First of all, beyond tree-level the EFT amplitude will contain UV divergences hence it is renormalization-scheme dependent. We choose to work in dimensional regularization with renormalization scale $\mu$, and in a scheme where the counterterms cancel not only the UV divergences but the full polynomial part of the loop amplitude, so that e.g., the renormalized $g_3$ coupling is the coefficient of the $stu$ term in the full amplitude.\footnote{For $g_3$ in $D=6$, the difference between our scheme and the $\overline{MS}$ scheme is given by
\be
g_3^{\overline{MS}} = g_3^{here}-(8\pi G)^2\frac{221849}{7526400\pi^4}\,.
\ee
} Furthermore, in what follows we always choose $\mu=M$, and therefore the bounds will be in terms of $g_i(M)$, the renormalized couplings at the scale $M$.

Second, when writing the low-energy arcs in terms of the EFT couplings, one often encounters the problem that the arcs contain infinitely many couplings. At tree-level, this EFT series can be truncated by using the improved sum rules or the crossing-symmetric sum rules. However, at one-loop, neither of this approach can truncate the EFT series. Therefore, to be able to truncate this infinite series, one inevitably needs extra assumptions on the behavior of the irrelevant EFT couplings $g_k$'s \cite{Bellazzini:2020cot}. 

Schematically, the one-loop contributions to the low-energy arcs, after applying the impact parameter functional, are of the form
\be
M^{\#}\p{\# G^2 + \# G g_2 M^{2}+ \# G g_3 M^{4} + \# g_2^2 M^4 + \# G g_4 M^6 + \# g_2 g_3 M^6 + \ldots},
\ee
where $M$ is the size of the arc, and the power of the overall factor $M^{\#}$ depends on the spacetime dimension and the Regge spin of the sum rule. We can rewrite this series as
\be\label{eq:arc_series_example}
G^2 M^{\#}&\left(\# + \# \frac{g_2\Lambda^2}{G} \p{\frac{M}{\Lambda}}^{2}+ \# \frac{g_3 \Lambda^4}{G} \p{\frac{M}{\Lambda}}^{4} + \# \p{\frac{g_2\Lambda^2}{G}}^2 \p{\frac{M}{\Lambda}}^{4} \right. \nn \\
&\quad \left.+ \# \frac{g_4\Lambda^6}{G} \p{\frac{M}{\Lambda}}^{6} + \# \frac{g_2\Lambda^2}{G} \frac{g_3 \Lambda^4}{G} \p{\frac{M}{\Lambda}}^6 + \ldots\right).
\ee
If we choose the size of the arc $M$ to be the EFT cutoff $\Lambda$, then \eqref{eq:arc_series_example} becomes an infinite sum of $O(1)$ numbers. This breakdown of the EFT expansion is simply the fact that we are pushing the arcs all the way to the scale where the EFT becomes invalid. On the other hand, if we choose $M<\Lambda$, then the first few terms in \eqref{eq:arc_series_example} become more important thanks to the $\p{M/\Lambda}^{\#}$ factor from the EFT counting. This allows us to suppress the higher order terms for the one-loop part of the arcs. 

It was shown in \cite{Caron-Huot:2020cmc,Caron-Huot:2021rmr} that the dimensionless ratio of Wilson coefficients, $g_k \Lambda^{2(k-2)}/g_2$ for EFT without gravity and $g_k \Lambda^{2(k-1)}/G$ for EFT with gravity, all satisfy two-sided bounds at tree-level in terms of the EFT cut-off $\Lambda$ for small values of $k$.\footnote{The only exception is $g_2 \Lambda^{2}/G$, which only has lower bound. As mentioned earlier, we assume it is $O(1)$ so that we have a controlled loop expansion.} Exploring the asymptotics of these tree-level bounds at large $k$ is an interesting problem on its own, but it is beyond the scope of this work. Thus, in this paper we assume that all $g_k \Lambda^{2(k-1)}/G$'s are $O(1)$ numbers.\footnote{Most of our results should remain true as long as $g_k \Lambda^{2(k-1)}/G$'s do not grow exponentially, so that the EFT series can still be suppressed by the $\p{M/\Lambda}^{\#}$ factor.}

Finally, we make a comment on the more practical side that concerns the computation of the low energy contributions to the sum rules. There are two strategies that one can follow: The first one is the obvious one, i.e., compute the one-loop EFT amplitude and input it into the sum rule. Alternatively, one can directly compute the imaginary part of the one-loop amplitude using the optical theorem and use it directly for the sum rule. As it turns out, this second strategy fails for the fixed-$t$ dispersion relations, as the imaginary part of the amplitude cannot reproduce terms which are analytic in $s$. The basic reason is that, in general $$\text{Im} \frac{{\cal M}(s,t)}{s(s+t)} \neq  \frac{1}{s(s+t)}\text{Im} {\cal M}(s,t)\,,$$ where the difference is the contributions from the poles at $s=0,-t$, which capture the analytic terms in the amplitude. Fortunately, the fixed-$a$ dispersion relations do not suffer from such issue, since all terms of the amplitude are non-analytic in $s'$.

In what follows we will begin explaining how to compute the imaginary part of the one-loop amplitude, which enters the crossing-symmetric dispersion relations. Then we comment on how from it we can infer the full amplitude using a version of the unitarity method \cite{Bern:1994zx, Bern:1994cg}, and we present the contributions from loops in the EFT to the dispersive sum rules. Finally, we explain how to use such loop-corrected sum rules to compute bounds on Wilson coefficients and we present several examples.

\subsection{One-loop EFT amplitude from its imaginary part}\label{sec:1loop_amplitude}

\begin{figure}
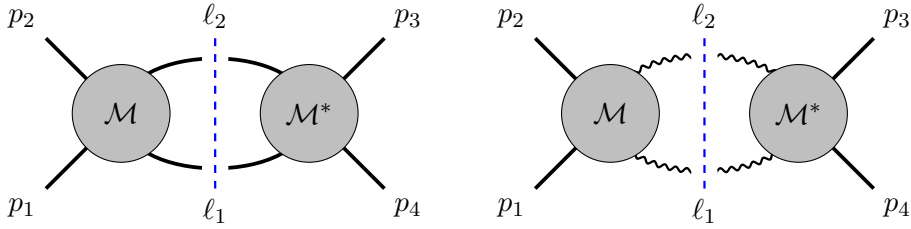

    \centering
    \raisebox{-32pt}{\scalebox{1}{\scalarcut{$\ell_2$}{$\ell_1$}}}  \hspace{0.5cm}\raisebox{-32pt}{\scalebox{1}{\gravitoncut{$\ell_2$}{$\ell_1$}}}
    \caption{Unitarity cuts contributing to the discontinuity of the four-point scalar amplitude in the $s$-channel. The diagram on the left corresponds to scalar particles crossing the cut, whereas the one on the right corresponds to gravitons crossing the cut.}
    \label{fig:cuts}
\end{figure}

The imaginary part of the one-loop EFT amplitude can be simply computed using the optical theorem, which relates it to unitarity cuts, given in terms of the product of tree-level amplitudes summed over intermediate states
\be
2\text{Im} {\cal M}^{\textrm{1-loop}} = \sum_{\mathrm{states}}  
{\cal M}^{\mathrm{tree}}
{\cal M}^{\mathrm{tree}*}\,,
\ee
which can be two scalars or two gravitons. Their corresponding contributions in the $s$-channel are depicted in Figure \ref{fig:cuts} and given by
\be
2\text{Im} {\cal M}^{\rm 1-loop}|_{\s\s} &= \int d{\rm LIPS}_{\ell_1,\ell_2}{\cal M}^{\mathrm{tree}}_{\s\s}(1^\s, 2^{\s},-\ell_1^\gr, -\ell_2^\gr){\cal M}^{\mathrm{tree}}_{\s\s}(\ell_1^\gr,\ell_2^\gr,3^\s,4^{\s})^*\,, \\
2\text{Im} {\cal M}^{\rm 1-loop}|_{\gr\gr} &= \int d{\rm LIPS}_{\ell_1,\ell_2}\sum_{\mathrm{pols.}}{\cal M}^{\mathrm{tree}}_{\gr\gr}(1^\s, 2^{\s},-\ell_1^\gr, -\ell_2^\gr){\cal M}^{\mathrm{tree}}_{\gr\gr}(\ell_1^\gr,\ell_2^\gr,3^\s,4^{\s})^*\,,
\ee
where $\ell_1, \ell_2$ are the momenta of the intermediate states which satisfy momentum conservation $\ell_1+\ell_2 = p_1+p_2$. The integral is over the Lorentz-invariant phase space for two identical particles, with measure 
\be
d{\rm LIPS}_{\ell_1,\ell_2} = \frac12  (2\pi)^D\delta^{(D)}(\ell_1+\ell_2 - p_1-p_2) \prod_{i=1,2} \frac{d^D\ell_i}{(2\pi)^{D-1}} \theta(\ell_i^0) \delta(\ell_i^2)\,,
\ee
where the $\frac{1}{2}$ is a symmetry factor because the particles crossing the cut are identical.
The amplitudes on the left and the right of the cut depend on the corresponding Lorentz products defined as
\be\label{eq:studefinition_sixmomenta}
&t_1=(p_2-\ell_1)^2,\quad u_1=(p_1-\ell_1)^2\,,\quad t_2=(p_4+\ell_1)^2\,,\quad u_2=(p_3+\ell_1)^2\,,
\ee
as well as the usual Mandelstam variable $s = (p_1+p_2)^2=(p_3+p_4)^2$, 
which satisfy the relations
\be\label{eq:sturelation_6pt}
s+t_1+u_1=s+t_2+u_2=0\,.
\ee
The all-scalar amplitudes ${\cal M}^{\mathrm{tree}}_{\s\s}$ entering the cut with intermediate scalars are 
\be\label{eq:M4s_tree}
{\cal M}^{\rm tree}_{\s\s} &= 8\pi G\p{\frac{st}{u}+\frac{su}{t}+\frac{tu}{s}} + g_2(s^2+t^2+u^2)+ g_3 stu + g_4(s^2+t^2+u^2)^2\nn \\
&+g_5(s^2+t^2+u^2)(stu)+g_6(s^2+t^2+u^2)^3+g'_6(stu)^2  \nn \\
&+ g_7(s^2+t^2+u^2)^2stu+g_8(s^2+t^2+u^2)^4 + g'_8(s^2+t^2+u^2)(stu)^2 \nn \\
&+ g_9(s^2+t^2+u^2)^3stu + g'_9(stu)^3 +\cdots\,.
\ee
with the replacements $t,u\to t_i,u_i$. Here we give them up to dimension ${\cal O}(s^9)$, and the dots denote contributions from higher dimension operators.

In order to describe the two-scalar-two-graviton amplitudes, ${\cal M}^{\mathrm{tree}}_{\gr\gr}$ entering the other cut, it is convenient to decompose them into a basis of gauge-invariant tensor structures \cite{Glover:2003cm, Bern:2017tuc, Boels:2018nrr, Chowdhury:2019kaq, Peraro:2019cjj, Peraro:2020sfm},
\be
{\cal M}_{\gr\gr}(1^\s, 2^{\s},3^\gr, 4^\gr) = {\cal M}^{(1)}_{\gr\gr}T_{\gr\gr}^{(1)} + {\cal M}^{(2)}_{\gr\gr}T_{\gr\gr}^{(2)} + {\cal M}^{(3)}_{\gr\gr}T_{\gr\gr}^{(3)}\,,
\ee
which capture the dependence on the polarization as follows
\be\label{eq:2s2gstruct_def0}
{T}_{\gr\gr}^{(1)}(p_i,\e_i)&= (2G_{1|34|2}+2G_{1|43|2} - (p_1\cdot p_2) H_{34})^2 \,,\nn \\
{ T}_{\gr\gr}^{(2)}(p_i,\e_i)&=(2G_{1|34|2}+2G_{1|43|2} - (p_1\cdot p_2) H_{34})H_{34}\,, \nn \\
{T}_{\gr\gr}^{(3)}(p_i,\e_i)&= H_{34}^2\,,
\ee
with
$G_{1|34|2}=p_{1\mu} F^\mu_{3\nu} F_{4\alpha}^\nu p_2^\alpha$, $ H_{34} = F^\mu_{3\nu} F_{4\mu}^\nu$ and $F_{i}^{\mu\nu} = p_i^\mu \epsilon_i^\nu - p_i^\nu \epsilon_i^\mu$, and $\epsilon_i$ are the polarization vectors.
The coefficients ${\cal M}^{(i)}_{\gr\gr}$ only depend on Lorentz-invariant products such as those in Eq.~\eqref{eq:studefinition_sixmomenta}, and are $(t,u)$-crossing symmetric.
For instance, their tree-level expressions at low energies are
\be\label{eq:2s2g_tree}
{\cal M}^{(1)}_{\gr\gr} &= -\frac{8\pi G}{stu} + \beta^{(1)}_0 + \beta^{(1)}_1 s + \beta^{(1)}_2 s^2 + \beta^{'(1)}_2 tu + \ldots, \nn \\ 
{\cal M}^{(2)}_{\gr\gr} &= -\frac{8\pi G}{2s}\alpha_2+ \beta^{(2)}_0 + \beta^{(2)}_1 s + \beta^{(2)}_2 s^2 + \beta^{'(2)}_2 tu + \beta^{(2)}_3 s^3 + \beta^{'(2)}_3 stu + \ldots, \nn \\
{\cal M}^{(3)}_{\gr\gr} &= -8\pi G\frac{tu}{4s}\alpha_4 + \beta^{(3)}_1 s + \beta^{(3)}_2 s^2 + \beta^{'(3)}_2 tu + \beta^{(3)}_3 s^3 + \beta^{'(3)}_3 stu \nn \\
&\quad + \beta^{(3)}_4 s^4+ \beta^{'(3)}_4 s^2 tu+ \beta^{''(3)}_4 t^2u^2 + \cdots\,.
\ee
Here, we have kept all the terms up to dimension ${\cal O}(s^6)$ in the full amplitude. (Note that the tensor structures have different power counting.) The leading terms in ${\cal M}^{(1)}_{\gr\gr}, {\cal M}^{(2)}_{\gr\gr}$ and ${\cal M}^{(3)}_{\gr\gr}$ correspond to the graviton minimal coupling, curvature squared and curvature cubed interactions respectively. We have normalized these couplings $\alpha_2,\alpha_4$ such that the three-graviton amplitude is
\be
{\cal M}_{hhh} = \sqrt{32\pi G}({\cal A}_1^2+\alpha_2 {\cal A}_1{\cal A}_2 + \alpha_4 {\cal A}_2^2)\,,
\ee
where ${\cal A}_1 = p_1\cdot \epsilon_3 \epsilon_1 \cdot \epsilon_2 +p_3\cdot \epsilon_2 \epsilon_1 \cdot \epsilon_3 + p_2\cdot \epsilon_1 \epsilon_2 \cdot \epsilon_3$ and ${\cal A}_2 = p_1 \cdot \epsilon_3 p_2\cdot \epsilon_1 p_3\cdot \epsilon_2$.  We use $\beta^{(i)}_n$ to denote the couplings of the two-scalar-two-graviton contact interactions. Since we assume the scalar has a shift symmetry, the constant part of ${\cal M}^{(3)}_{\gr\gr}$ vanishes.

The tensor decomposition also facilitates the sum over polarizations in the unitarity cut. All the tensors are manifestly gauge invariant in each leg, without having to use any on-shell condition on other legs. Thus the sum can be carried out by using the simplified completeness relation \cite{Kosmopoulos:2020pcd}
\be
\sum_{\mathrm{pol}}\e_{\mu}(\ell_i)\e_{\nu}(\ell_i) \e^*_{\alpha}(\ell_i)\e^*_{\beta}(\ell_i) \to \frac12 \left(\eta_{\mu\alpha} \eta_{\nu\beta}+\eta_{\mu\beta} \eta_{\nu\alpha}-\frac{2}{D_s-2}\eta_{\mu\nu} \eta_{\alpha\beta} \right),
\ee
where $D_s=\eta^\mu_\mu$ is a state counting parameter which lets us keep track of the counting of states in the loops.\footnote{In intermediate steps we keep this separate from the dimension $D$ arising from the dimensionally-regulated loop integrals.} The sum over states can then be conveniently encoded in the  matrix
\be\label{eq:sewingmat_def}
{\cal T}_{(i,j)} = \sum_{\rm pols.} T_{\gr\gr}^{(i)} T_{\gr\gr}^{(j)*},
\ee
which satisfies ${\cal T}_{(j,i)}(t_1,u_1,t_2,u_2) = {\cal T}_{(i,j)}(t_2,u_2,t_1,u_1)$ and is given by
{\allowdisplaybreaks
\begin{align}
{\cal T}_{(1,1)} &= 
   \frac{1}{2}s^8- 2 s^6 \left( t u + t_1 u_1+t_2 u_2\right) +\frac{1}{2}s^5(t-u)(t_1-u_1)(t_2-u_2)
   \nn \\
&+  \frac{1}{2}s^4 \left(\left(D_s+8\right) t_1 t_2 u_1 u_2 +2 t^2 u^2 +2 \left(t_1 u_1+t_2 u_2\right)\left(6 t u +t_1 u_1+t_2 u_2\right)\right)
\nn \\
&-s^3(t-u)(t_1-u_1)(t_2-u_2)(t u+t_1 u_1+t_2 u_2) \nn \\
&- s^2 \left( D_s\,  t_1 t_2 u_1 u_2 \left(t u+t_1 u_1+t_2 u_2\right) + 16 t u t_1 t_2  u_1 u_2  \right)  \nn \\
&+\frac{D_s}{2}s (t-u)(t_1-u_1)(t_2-u_2)t_1 u_1 t_2 u_2+\frac{1}{2} D_s \left(D_s-3\right)  t_1^2 t_2^2 u_1^2 u_2^2,
\label{eq:t11}
\\ 
{\cal T}_{(2,2)} &= 
\frac{ \left(D_s-4\right) }{8
   \left(D_s-2\right){}^2} s^2 \bigg(4 \left(D_s-3\right) \left(D_s^2-2 D_s+4\right) t_1 t_2
   u_1 u_2 
    \\
   &\qquad+ \left(D_s^2-4\right) \left(s^4-2 s^2 \left(t u+t_1
   u_1+t_2 u_2\right)+s(t-u) \left(t_1-u_1\right) \left(t_2-u_2\right)\right)\bigg),
 \nn \\
{\cal T}_{(3,3)} &= \frac{1}{2} \left(D_s-3\right) D_s  s^4 \\
\
{\cal T}_{(1,2)} &=
\frac{\left(D_s-4\right)}{4 \left(D_s-2\right)} s t_1 u_1  \bigg(  2 D_s \left(D_s-3\right)  t_1 t_2
   u_1 u_2 
    \\
   &
   \qquad+ \left(D_s-2\right) \left(s^4-2 s^2 \left(t u+t_1
   u_1+t_2 u_2\right)+s(t-u) \left(t_1-u_1\right) \left(t_2-u_2\right)\right) 
\bigg),
\nn \\
{\cal T}_{(1,3)} &= \frac{1}{2}(D_s-3)(D_s-4)s^2 t_1^2 u_1^2\,,
\\ 
{\cal T}_{(2,3)} &=\frac{D_s(D_s-3)(D_s-4)}{2(D_s-2)} s^3 t_1 u_1 \,.
\end{align}
}%
In terms of this matrix, one has
\be
\sum_{\mathrm{pols.}}{\cal M}^{\mathrm{tree}}_{\gr\gr}(1^\s, 2^{\s},-\ell_1^\gr, -\ell_2^\gr){\cal M}^{\mathrm{tree}}_{\gr\gr}(\ell_1^\gr,\ell_2^\gr,3^\s,4^{\s})^* = {\cal M}^{(i)}_{\gr\gr}(s,t_1,u_1) {\cal T}_{(i,j)} {\cal M}^{(j)}_{\gr\gr}(s,t_2,u_2)\,,
\ee
where the sum over $i,j$ is implied.

Using integration by parts (IBP) identities \cite{Tkachov:1981wb,Chetyrkin:1981qh} and the reverse unitarity method \cite{Anastasiou:2002yz,Anastasiou:2002qz,Anastasiou:2003yy,Anastasiou:2015yha}, the resulting phase space integral can be decomposed in a basis of cut box, triangle and bubble scalar integrals\footnote{Technically, the triangle integral can be reduced to the bubble using the IBP relation $I_{\itri}(s) = \frac{2(D-3)}{(D-4)s}I_{\ibub}(s)$, but we keep them separate in order to cleanly separate IR divergences, captured in the box and triangle, from UV divergences, captured in the bubble. This can be conveniently done by performing the IBP reduction with massive propagators and sending $m\to 0$ at the end.}
\be \label{eq:cutafterIBP}
\text{Im} {\cal M}^{\textrm{1-loop}} &= H_{\ibox}(s,t,u)I_{\ibox}^{\rm cut}(s,t) + H_{\ibox}(s,u,t)I_{\ibox}^{\rm cut}(s,u) \nn\\
&\hspace{1.5cm}+ H_{\itri}(s,t,u)I_{\itri}^{\rm cut}(s) +H_{\ibub}(s,t,u)I_{\ibub}^{\rm cut}(s)\,,
\ee
where the cut integrals are 
\be
\label{eq:Imbox}
I_{\ibox}^{\rm cut}(s,t) &= \!\int  \frac{d{\rm LIPS}_{\ell_1,\ell_2}}{(p_1-\ell_1)^2 (p_4+\ell_1)^2 } = - r_{\ibox}^{(D)} 2\sin\p{\pi\tfrac{6-D}{2}} \tfrac{s^{\frac{D-6}{2}}}{t}  
  {}_2F_1\left(1, \tfrac{D-4}{2}, \tfrac{D-2}{2} ; \tfrac{s+t}{t} \right) \,,\\
  \label{eq:Imtri}
I_{\itri}^{\rm cut}(s) &=\! \int  \frac{d{\rm LIPS}_{\ell_1,\ell_2}}{(p_1-\ell_1)^2 } = r_{\itri}^{(D)} \sin\p{\pi\tfrac{6-D}{2}} s^{\frac{D-6}{2}} \,,\\
\label{eq:Imbub}
I_{\ibub}^{\rm cut}(s) &=\! \int d{\rm LIPS}_{\ell_1,\ell_2} = r_{\ibub}^{(D)}  \sin\p{\pi\tfrac{4-D}{2}}s^{\frac{D-4}{2}}\,,
\ee
where
\be
\label{eq:rds}
r_{\ibox}^{(D)}
= r_{\itri}^{(D)}  =
\frac{\Gamma\left(\frac{6-D}{2}\right)\Gamma^2\left(\frac{D-4}{2}\right)}{(4\pi)^{D-2}\Gamma(D-3)}\,, 
\quad 
r_{\ibub}^{(D)} 
=\frac{\Gamma\left(\frac{4-D}{2}\right)\Gamma^2\left(\frac{D-2}{2}\right)}{(4\pi)^{D-2}\Gamma(D-2)}\,.
\ee

Finally, from Eq.~\eqref{eq:cutafterIBP} one can read off the coefficient of each integral and easily infer that the full one-loop amplitude is
\be\label{eq:ampdecomp}
i{\cal M}^{\textrm{1-loop}} &= H_{\ibox}(s,t,u)I_{\ibox}(s,t) + H_{\ibox}(t,u,s)I_{\ibox}(t,u) + H_{\ibox}(s,u,t)I_{\ibox}(s,u)  \nn \\
&+ H_{\itri}(s,t,u)I_{\itri}(s) + H_{\itri}(t,s,u)I_{\itri}(t) + H_{\itri}(u,s,t)I_{\itri}(u) \nn \\
&+ H_{\ibub}(s,t,u)I_{\ibub}(s) +  H_{\ibub}(t,s,u)I_{\ibub}(t) +  H_{\ibub}(u,s,t)I_{\ibub}(u) \,,
\ee
where the full off-shell integrals are
\be
\label{eq:Ibox}
I_{\ibox}(s,t) &= \int \frac{d^D\ell_1}{(2\pi)^D} \frac{1}{\ell_1^2 (p_1-\ell_1)^2 (\ell_1-p_1-p_2)^2 (p_4+\ell_1)^2 }  \\
&= i r_{\ibox}^{(D)}\, \frac{2}{st}  
\left( 
   s^{\frac{D-4}{2}} {}_2F_1\left(\tfrac{D-4}{2}, \tfrac{D-4}{2}, \tfrac{D-2}{2} ; \tfrac{s+t}{t} \right) 
+  (-s)^{\frac{D-4}{2}} {}_2F_1\left(1, \tfrac{D-4}{2}, \tfrac{D-2}{2} ; \tfrac{s+t}{t} \right) \right)\,, 
\nn\\
\label{eq:Itri}
I_{\itri}(s) &= \int \frac{d^D\ell_1}{(2\pi)^D} \frac{1}{\ell_1^2 (p_1-\ell_1)^2 (\ell_1-p_1-p_2)^2}  = ir_{\itri}^{(D)} (-s)^{\frac{D-6}{2}} \,,\\
\label{eq:Ibub}
I_{\ibub}(s) &=\int \frac{d^D\ell_1}{(2\pi)^D} \frac{1}{\ell_1^2 (\ell_1-p_1-p_2)^2}  = ir_{\ibub}^{(D)}  (-s)^{\frac{D-4}{2}}\,,
\ee
where each integral should be multiplied by a factor of $\mu^{D_0-D}$ in spacetime dimension $D_0=4,6,8,\ldots$. For simplicity, we have written in the Euclidean region $s<0,t<0$ (or $u$-channel physical kinematics), and analytic continuation to physical scattering kinematics in the $s$ channel is given by taking $(-s) \rightarrow e^{-i\pi} |s|$. One can easily check that Eqs.~\eqref{eq:Imbox}-\eqref{eq:Imbub} are the imaginary parts of these integrals in the $s$-channel, as dictated by the cutting rules. 

As an example, let us consider the contributions to the graviton cut from the minimally coupled ${\cal M}^{\mathrm{tree}}_{\gr\gr}$ amplitudes. These are proportional to 
\be
\sum_{\mathrm{pol}}{\cal M}^{\mathrm{tree}}_{\gr\gr}{\cal M}^{\mathrm{tree}*}_{\gr\gr}\Big|_{{\cal O}(G^2)}  &= {\cal M}^{(1)}_{\gr\gr} {\cal T}_{(1,1)} {\cal M}^{(1)}_{\gr\gr}(s, u_2) 
=\frac{(8\pi G)^2 {\cal T}_{(1,1)}}{s^2t_1u_1t_2u_2} \nn\\ & = \frac{(8\pi G)^2}{s^4} \left(\frac{1}{t_1} + \frac{1}{u_1}\right) \left(\frac{1}{t_2} + \frac{1}{u_2}\right) {\cal T}_{(1,1)}.
\ee
where ${\cal T}_{(1,1)}$ is given in Eq.~\eqref{eq:t11}. Upon IBP reduction, we obtain that for minimal coupling,
\be
\int d{\rm LIPS}_{\ell_1,\ell_2}\sum_{\mathrm{pol}}{\cal M}^{\mathrm{tree}}_{\gr\gr}{\cal M}^{\mathrm{tree}*}_{\gr\gr}\Big|_{{\cal O}(G^2)} 
=2(8\pi G)^2 &\left( t^4 I_{\ibox}^{\rm cut}(s,t) + u^4 I_{\ibox}^{\rm cut}(s,u) \right. \\
&\left.+2 (t^3+u^3)I_{\itri}^{\rm cut}(s)+  h_{\ibub}(s,t,u)I_{\ibub}^{\rm cut}(s) \right), \nn
\ee
where the bubble integral coefficient is given by
\be
 h_{\ibub}(s,t,u) = \tfrac{  D_s\left(D_s-3\right)D(D-2) }{64 \left(D^2-1\right)} s^2- \left(\tfrac{D_s\left(D_s-3\right) }{8 \left(D^2-1\right)}+\tfrac{ D_s(D-3)}{2 (D-1)}-\tfrac{4
   (D-3)}{D-2}\right) tu .
\ee

The full expressions for the coefficients $H_{\ibox}$, $H_{\itri}$ and $H_{\ibub}$ including the contributions from other couplings and the scalar cut are a bit lengthy, so  we give them in Appendix \ref{app:one-loop_amp} and the attached ancillary file \texttt{GravitonLoop.m}.

We have performed several checks on our result for the one-loop amplitude, in particular the $G^2$ contribution. Firstly, by expanding ${\cal M}^{\textrm{1-loop}}$ around $D=D_s=4-2\e$, we have verified that our results agree with the one-loop amplitude for minimal-coupling in four dimensions computed in Ref.~\cite{Dunbar:1995ed}. Furthermore, we checked that the $1/\e^2$ poles cancel, as expected by the absence of collinear divergences in gravity \cite{Akhoury:2011kq}, and find that the $G^2$ contributions to the $1/\e$ poles take the form
\be\label{eq:M1loop_4D_divergence}
\left.{\cal M}^{\textrm{1-loop}}\right|_{G^2} &= \frac{1}{\e} \frac{8\pi G}{8\pi^2} (s \log(-s)+t\log(-t)+u\log(-u)) {\cal M}_{\textrm{tree}}
\nn \\
&+\frac{1}{\e} \frac{203}{80} \frac{(8\pi G)^2}{8\pi^2}(s^2+t^2+u^2)+O(\e^{0}).
\ee
The first line arises from the box and triangle integrals and matches the known form of the IR divergence of the one-loop amplitude \cite{Weinberg:1965nx, Akhoury:2011kq}. The second line is a UV divergence contained in the bubble integrals which renormalizes $g_2$, and we have checked that the corresponding $(\nabla \phi)^4$ counterterm matches Ref.~\cite{tHooft:1974toh}. Additionally, we have checked that in the $t \to 0$ limit the amplitude agrees with the prediction from the eikonal formula for generic $D$ \cite{Muzinich:1987in}.

\subsection{Loop contributions to crossing-symmetric sum rules}\label{sec:loop_contribution_fixeda_sumrules}
We now study the 1-loop contributions to the low-energy part of the fixed-$a$ sum rules. To have a better idea of how loop amplitudes contribute to the sum rules, let us consider a toy example,
\be\label{eq:Mtoy_expression}
{\cal M}_{\textrm{toy}}(s,t) = \p{\frac{-s}{\mu^2}}^{\e} + \p{\frac{-t}{\mu^2}}^{\e} + \p{\frac{s+t}{\mu^2}}^{\e},
\ee
where $\e$ will play the role of the dim reg regulator. So, all the statements we make should be understood as analytic continuation in $\e$. We would like to understand whether ${\cal M}_{\textrm{toy}}$ can be part of the low-energy limit of some ${\cal M}$ satisfying \eqref{eq:crossing_symmetric_relation}. In particular, we expect that, similar to the fixed-$t$ case, the non-analyticity of ${\cal M}_{toy}$ should be reproduced by the integral in \eqref{eq:crossing_symmetric_relation} over the low-energy branch cut,\footnote{We stress again that the integral is originally over the low-energy ``arcs". (For the contour in the crossing-symmetric case, see e.g., \cite{Li:2023qzs}.) To compute the integral of the EFT amplitude over the arc, we deform the contour and obtain \eqref{eq:fixedadispersion_lowenergy_onMtoy}.}
\be\label{eq:fixedadispersion_lowenergy_onMtoy}
x\int_0^{M^2} \frac{ds'}{\pi}\mathrm{Im} \frac{{\cal M}_{\textrm{toy}}(s',\tau(s',a))}{(s')^3}\frac{(2s'-3a)(s')^2}{x(a-s')-(s')^3}.
\ee

Now we have to deal with a subtlety regarding the $\frac{1}{s'}$ factor in the integrand. Without massless loops, it is not an issue because the integral starts at some nonzero value of $s'$. However, for massless loops, the branch cut starts at $s'=0$, and the $\frac{1}{s'}$ factor can give nontrivial contributions (which is true for the fixed-$t$ sum rules discussed in section \ref{sec:fixed-t_loop}). To understand this, we note that the toy amplitude with fixed $a$ is written as
\be
{\cal M}_{\textrm{toy}}(s',\tau(s',a)) = \left(\frac{-s'}{\mu^2}\right)^{\e} + \left(\frac{\frac{s'}{2}\left(1-\sqrt{\frac{s'+3a}{s'-a}}\right)}{\mu^2}\right)^{\e} + \left(\frac{\frac{s'}{2}\left(1+\sqrt{\frac{s'+3a}{s'-a}}\right)}{\mu^2}\right)^{\e}.
\ee
The imaginary part of ${\cal M}_{\textrm{toy}}$ simply comes from the first term $\p{\frac{-s'}{\mu^2}}^{\e}$. The sum of the other two terms, despite each having discontinuity for $0<s'<-3a$, has no imaginary part. Furthermore, we see that it satisfies $\mathrm{Im}[\frac{{\cal M}(s',\tau (s',a))}{(s')^n}]= \frac{\mathrm{Im}[{\cal M}(s',\tau(s',a))]}{(s')^n}$ for any positive integer $n$, and we can rewrite \eqref{eq:fixedadispersion_lowenergy_onMtoy} as
\be\label{eq:fixedadispersion_lowenergy_onMtoy_onlyImM}
x\int_0^{M^2} \frac{ds'}{\pi} \frac{\mathrm{Im}[{\cal M}_{\textrm{toy}}(s',\tau(s',a))]}{(s')^3}\frac{(2s'-3a)(s')^2}{x(a-s')-(s')^3}
\ee
Moreover, this turns out to be true for all the 1-loop amplitudes we consider in this work. As mentioned above, this is yet another advantage of the crossing-symmetric dispersion relation: the dispersive integral really is an integral of $\mathrm{Im}{\cal M}$ even with massless loops, unlike the fixed-$t$ case where one has to be careful with contributions at $s=0$.

Plugging the expression of ${\cal M}_{\textrm{toy}}$ into \eqref{eq:fixedadispersion_lowenergy_onMtoy_onlyImM} and evaluating the integral, one can obtain\footnote{If we chose a renormalization scale $\mu\neq M$ the second line would be multiplied by a factor of $\left(\frac{M^2}{\mu^2}\right)^{\e}$}
\be
&x\int_0^{M^2} \frac{ds'}{\pi} \frac{\mathrm{Im}[{\cal M}_{\textrm{toy}}(s',\tau(s',a))]}{(s')^3}\frac{(2s'-3a)(s')^2}{x(a-s')-(s')^3} =\left(\frac{-s}{\mu^2}\right)^{\e}+\left(\frac{-t}{\mu^2}\right)^{\e}+\left(\frac{s+t}{\mu^2}\right)^{\e} \\
&\hspace{3cm}+\frac{\sin{\pi \e}}{\pi \e}\left[\left(1-{}_2F_1\left(-\e,1,1-\e,\tfrac{s}{M^2}\right)\right)+(s\to t)+(s\to -s-t)\right], \nn
\ee
where $t=\tau(s',a)$ defined in \eqref{eq:tau_def}. Crucially, we see that the first three terms on the right-hand side exactly reproduce ${\cal M}_{\textrm{toy}}$, and the additional terms are analytic and completely crossing-symmetric.

After studying this toy example, we can now consider our actual target, the 2-to-2 amplitude of a real scalar coupled to gravity. In particular, we want to compute the contributions from the one-loop part of the amplitude. Concretely, they are given by (see \eqref{eq:fixeda_lowenergy_def})
\be\label{eq:fixeda_lowenergy_def_oneloop}
\left.-{\cal C}^{\textrm{low}}_{k}(a)\right|_{\textrm {1-loop}}=- \int_0^{M^2}\frac{ds'}{\pi}\mathrm{Im}{\cal M}_{\textrm{1-loop}}(s',\tau(s',a))\frac{(s'-a)^{\frac{k}{2}-1}(2s'-3a)}{(s')^{\frac{3k}{2}+1}}.
\ee
As described in section \ref{sec:1loop_amplitude}, the 1-loop amplitudes can be reduced to a sum of bubble, triangle, and box master integrals. Below we discuss how to compute their contributions to the crossing-symmetric sum rules.

\paragraph{Bubble and triangle contributions}
We first discuss how to compute the contributions from the bubble integral. The triangle contributions can be treated in the same way. As explained above, the bubble part of the 1-loop amplitude can be written as
\be\label{eq:M1loop_bubblepart_expr0}
i{\cal M}^{\rm 1-loop} \supset H(s,t,u)I_{\ibub}(s) + H(t,s,u)I_{\ibub}(t) + H(u,s,t)I_{\ibub}(u)\,,
\ee
where $H(s,t,u)$ is a polynomial invariant under $t \leftrightarrow u$, and $I_{\ibub}(s)$ is the  bubble master integral \eqref{eq:Ibub}. This expression can be also be written as a linear combination of the crossing-symmetric combinations,
\be\label{eq:Mpq_def}
i{\cal M}_{p,q} = (t u)^p s^q I_{\ibub}(s) + (s u)^p t^q I_{\ibub}(t) + (s t)^pu^q I_{\ibub}(u),
\ee
which are simple  modifications of the toy example above.
The imaginary part of this amplitude is
\be
\mathrm{Im}{\cal M}_{p,q}(s',\tau(s',a)) &= \p{\frac{a(s')^2}{a-s'}}^p s'^qI_{\ibub}^{\rm cut}(s') 
=r^{(D)}_{\ibub}\p{\frac{a(s')^2}{a-s'}}^p (s')^{q+\frac{D-4}{2}} \sin\p{\pi\tfrac{4-D}{2}}\,,
\ee
where $r_{\ibub}$ given in Eq.~\eqref{eq:rds}. Plugging the imaginary part into \eqref{eq:fixeda_lowenergy_def_oneloop}, we find that the contribution of each ${\cal M}_{p,q}$ is given by
\be\label{eq:bubble_dispersive_Mpq}
&\left.-{\cal C}^{\textrm{low}}_{k}(a)\right|_{{\cal M}_{p,q}}=\frac{r_{\ibub} (-a)^{k/2} \sin \left(\frac{\pi  D}{2}\right) M^{D-4-3k+4 p+2q} }{\pi  \left(\frac{D}{2}-k+p+q-2\right)}\left( 2 \left(\frac{-a+M^2}{-a}\right)^{\frac{k}{2}-p} \right. \\
&\left.+\frac{D-2 p+2 q-4}{D-3 k+4 p+2 q-4} \, _2F_1\left(-\tfrac{k}{2}+p+1,\tfrac{D}{2}+2 p+q-\tfrac{3 k}{2}-2;\tfrac{D}{2}+2 p+q-\tfrac{3 k}{2}-1;\tfrac{M^2}{a}\right)\right) \nn 
\ee
By expressing \eqref{eq:M1loop_bubblepart_expr0} in terms of ${\cal M}_{p,q}$, the above expression then allows us to compute all the contributions of the bubble amplitude to the fixed-$a$ sum rules.

Note that for some bubble amplitudes, their contribution to the sum rule can be divergent. For example,
\be\label{eq:bubble_sumrule_div_example}
\left.-{\cal C}^{\textrm{low}}_{k=2}(a)\right|_{{\cal M}_{1,0}}=\frac{a}{256\pi^4}\frac{1}{D-6} + (\textrm{finite at $D=6$}).
\ee
This simply comes from the fact that the bubble amplitudes renormalize some EFT couplings. In the above example, ${\cal M}_{1,0}$ renormalizes $g_3$ in $D=6$ (see \eqref{eq:fixeda_EFT_spin246}).
Thus, these divergences are removed by renormalization.

\paragraph{Box contribution}
Now we consider the box amplitudes. In section \ref{sec:1loop_amplitude}, we show that the box part of the 1-loop amplitude is given by 
\be\label{eq:Mbox_def}
i{\cal M}_{\textrm{box}}=(8\pi G)^2\p{(s^4 + t^4) I_{\ibox}(s,t) + (s^4 + u^4) I_{\ibox}(s,u) + (t^4 + u^4) I_{\ibox}(t,u)}.
\ee
where $I_{\ibox}$ is the box master integral given by \eqref{eq:Ibox}. The imaginary part of ${\cal M_{\textrm{box}}}$ can be written as
\be
&\mathrm{Im}{\cal M}_{\textrm{box}}(s',\tau(s',a))  \\
&=2(8\pi G)^2 r^{(D)}_{\ibox}\sin\p{\pi\tfrac{4-D}{2}}s'^{\frac{D-6}{2}}\left(\frac{s'^4+\tau^4}{\tau} {}_2F_1\p{1,\tfrac{D-4}{2},\tfrac{D-2}{2},\tfrac{s'+\tau}{\tau}}+(\tau \to -s'-\tau)\right)\,. \nn
\ee
Therefore, its contribution to the sum rule is
\be\label{eq:box_dispersive_intexpression}
\left.-{\cal C}^{\textrm{low}}_{k}(a)\right|_{{\cal M}_{\textrm{box}}}=&- \int_0^{M^2}\frac{ds'}{\pi}\mathrm{Im}{\cal M}_{\textrm{box}}(s',\tau(s',a))\frac{(s'-a)^{\frac{k}{2}-1}(2s'-3a)}{(s')^{\frac{3k}{2}+1}}  \\
=&2r_{\ibox}(8\pi G)^2  \sin(\tfrac{\pi D}{2})\int_0^{M^2}\frac{ds'}{\pi}(s')^{\frac{D-8-3k}{2}} (s'-a)^{\frac{k}{2}-1}(2s'-3a) \nn \\
& \left(\tfrac{(s')^4+\tau^4}{\tau} {}_2F_1\p{1,\tfrac{D-4}{2},\tfrac{D-2}{2},\tfrac{s'+\tau}{\tau}}+\tfrac{(s')^4+(-s'-\tau)^4}{-s'-\tau} {}_2F_1\p{1,\tfrac{D-4}{2},\tfrac{D-2}{2},\tfrac{\tau}{s'+\tau}}\right)\,, \nn
\ee
where $\tau=-\frac{s'}{2}\p{1-\sqrt{\frac{s'+3a}{s'-a}}}$. This integral can be evaluated in the same way as the fixed-$t$ case, which we explain in Appendix \ref{app:dispersive_int}. As an example, in $D=6$, the result for the $k=2$ sum rule is
\be\label{eq:box_dispersive_D6k2}
&-{\cal C}^{\textrm{low}}_{k=2}(a)\Big|^{D=6}_{{\cal M}_{\textrm{box}}}=\frac{(8\pi G)^2}{3072 \pi ^4 a}\left(36 a^2 \mathrm{Li}_2\left(\tfrac{a}{M^2}\right)-15 a^2 \log ^2\left(\tfrac{-a+M^2}{-a}\right) +27 a^2 \log ^2\left(\tfrac{1+\sqrt{\tfrac{M^2+3 a}{M^2-a}}}{1-\sqrt{\tfrac{M^2+3a}{M^2-a}}}\right) \right. \nn \\
&\left.+18 a^2 \log ^2\left(\tfrac{-a}{M^2}\right)\!+\!12 M^4 \log \left(\tfrac{M^2-a}{-a}\right) \!-\!12 M^4 \sqrt{\tfrac{M^2+3a}{M^2-a}} \log \left(\tfrac{1+\sqrt{\tfrac{M^2+3a}{M^2-a}}}{1-\sqrt{\tfrac{M^2+3 a}{M^2-a}}}\right)\!+\!20 \pi ^2 a^2\right),
\ee
Note that the range of $a$ is $-\frac{M^2}{3}<a<0$, and thus the square root above is always real.

\paragraph{Total}
Combining the bubble contribution \eqref{eq:bubble_dispersive_Mpq} and the box contribution \eqref{eq:box_dispersive_intexpression}, one can then explicitly compute the contribution of the 1-loop amplitude to each sum rule. For example, for the $k=2$ fixed-$a$ sum rule in $D=6$, we obtain\footnote{As discussed above, the couplings featuring in the sum rules are renormalized couplings at the scale $M$. If we were to choose a different scale $\mu\neq M$ this sum rule would contain an additional $a \log\p{\frac{M^2}{\mu^2}}$ to account for the RG running of $g_3$ due to graviton loops. 
}
\be\label{eq:G2_sumrule_spin2_D6}
&\left.-{\cal C}^{\textrm{low}}_{k=2}(a)\right|^{D=6}_{\textrm {1-loop}} =\frac{(8\pi G)^2}{107520 \pi ^4 a}\Bigg(
1260 a^2 \mathrm{Li}_2\left(\tfrac{a}{M^2}\right)-525 a^2 \log ^2\left(\tfrac{M^2-a}{-a}\right)+630 a^2 \log ^2\left(\tfrac{-a}{M^2}\right) \nn \\
&\hspace{3cm}+(420 M^4-1409 a^2) \log \left(\tfrac{M^2-a}{-a}\right)  -420 M^4 \sqrt{\tfrac{M^2+3a}{M^2-a}} \log \left(\tfrac{1+\sqrt{\tfrac{M^2+3a}{M^2-a}}}{1-\sqrt{\tfrac{M^2+3a}{M^2-a}}}\right) 
 \nn \\
& \hspace{3cm}+700 \pi ^2 a^2+2126 a M^2\Bigg)+ \cdots\,,
\ee
where $\cdots$ are contributions from the higher couplings $G g_2,g_2^2,$etc. They can also be systematically computed using \eqref{eq:bubble_dispersive_Mpq}, and we give some of their expressions in Appendix \ref{app:dispersive_int}. A more complete list of $\left.-{\cal C}^{\textrm{low}}_{k}(a)\right|^{D}_{\textrm {1-loop}}$ for $k=2,4,6,8,10$ and $D=4,6,8,10,12$ is given in the ancillary file \texttt{GravitonLoop.m}.

It is important to check that we can apply the functional e.g., $\int dp f(p)$ to the 1-loop contributions. Taking the small-$a$ limit, \eqref{eq:G2_sumrule_spin2_D6} becomes
\be\label{eq:G2_sumrule_spin2_D6_smalla}
&\left.-{\cal C}^{\textrm{low}}_{k=2}(a)\right|_{\textrm {1-loop}}= \frac{(8\pi G)^2M^2}{128\pi^4}\log\p{\frac{-a}{M^2}} + \cdots,
\ee
where $\cdots$ are terms that are not singular as $a\to 0$. The behavior is similar for $D=6$ sum rules with other values of $k$. Since the most singular term is $\log(-a)$, the functionals used to obtain the tree-level bounds also have finite actions on the 1-loop corrections (see \eqref{eq:functional_pexpansion}).

More generally, we find that for other dimensions $D$, the leading singularity of the 1-loop correction in the small-$a$ limit is
\be\label{eq:fixeda_smalla_loop}
&\left.-{\cal C}^{\textrm{low}}_{k}(a)\right|_{\textrm {1-loop}}\sim (-a)^{\frac{D-6}{2}}\log(-a).
\ee
Thus, for $D>4$, the corrections do not lead to any additional divergences when applying the functionals. We leave the discussion for $D=4$ to section \ref{sec:4Dbounds}.

\subsection{Loop corrections to bounds on $g_2,g_3$ with gravity}
\begin{figure}[t]
\centering
\includegraphics[width=12cm]{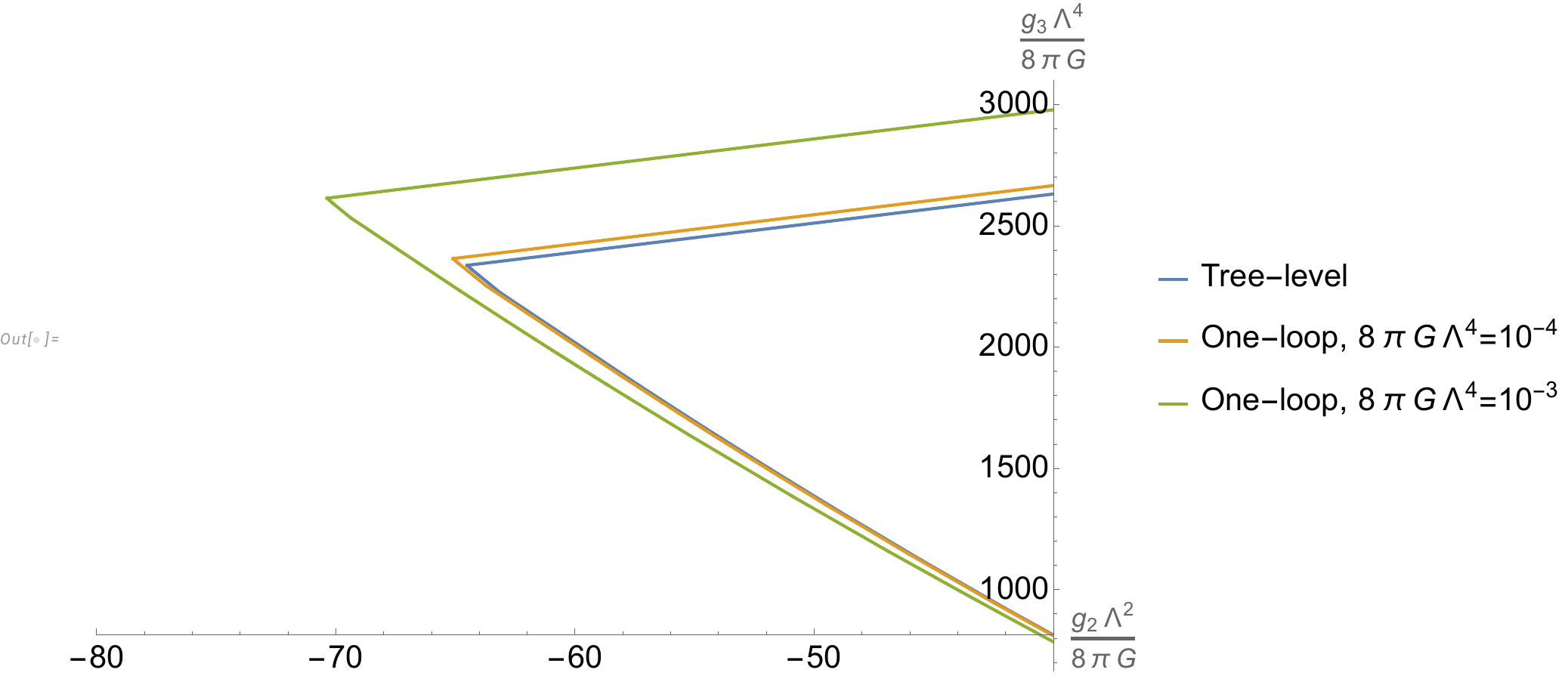}
\caption{Comparison between tree-level bound and one-loop bound on $g_2$ and $g_3$ in $D=6$ for $8\pi G \Lambda^4=\{10^{-4},10^{-3}\}$. The allowed regions are to the right of the curves. We have chosen $M=\frac{\Lambda}{2}$ to suppress the contributions from higher couplings.}
\label{fig:g2-g3-loop}
\end{figure}

We now use the results in the previous subsection to compute 1-loop corrections to the tree-level bounds obtained in section \ref{sec:fixed-a_tree}. In particular, we would like to compute the corrections to the tree-level fixed-$a$ bound in Figure \ref{fig:g2-g3-tree}. The approach is straightforward: we take the functionals that lead to the bounds in Figure \ref{fig:g2-g3-tree}, and apply them to the loop contributions $\left.-{\cal C}^{\textrm{low}}_{k}(a)\right|_{\textrm {1-loop}}$. For instance, the correction of the spin-2 sum rule can be obtained by applying $\int dp f(p)$ to \eqref{eq:G2_sumrule_spin2_D6}. Doing this allows us to obtain a two-sided bound given by
\be\label{eq:g2g3_bound_loop_example}
&-192.7 -21 \frac{g_2 M^2}{8\pi G} - 2.15 \times 10^5 \times (8\pi G M^4) + \ldots \nn \\
&\qquad\qquad  \leq \frac{g_3 M^4}{8\pi G} \leq 194.4 + 3\frac{g_2 M^2}{8\pi G} + 3.47 \times 10^5 \times (8\pi G M^4) + \cdots,\quad (D=6).
\ee

The $\cdots$ in \eqref{eq:g2g3_bound_loop_example} include the contributions from higher couplings such as $G g_2,g_2^2, G g_3,\ldots$. As we discuss earlier, to suppress their contributions, we have to choose $M$ to be below the EFT cutoff $\Lambda$ so that we can truncate the EFT series (with mild assumptions on the asymptotics of these couplings). Additionally, in practice we find that the numerical coefficients of the higher couplings all seem to be much smaller than the coefficient of the $G^2$ term. For example, in \eqref{eq:g2g3_bound_loop_example}, the coefficients of $G g_2, g_2^2$ are all smaller than $10^{-3}$.

In Figure \ref{fig:g2-g3-loop}, we show the allowed region of the dimensionless couplings $\frac{g_2 \Lambda^2}{8\pi G}$ and $\frac{g_3 \Lambda^4}{8\pi G}$ after including the one-loop contributions, choosing two different values ($10^{-4},10^{-3}$) for the dimensionless quantity $8\pi G\Lambda^4$ which controls the loop expansion. The one-loop corrections mainly come from the $G^2$ term in the amplitude, since the higher couplings all have much smaller numerical coefficients, and are further suppressed by choosing $M=\frac{\Lambda}{2}$. This also makes the tree-level bound in Figure \ref{fig:g2-g3-loop} weaker than the one shown in Figure \ref{fig:g2-g3-tree}. 

To recap, the result shown in Figure \ref{fig:g2-g3-loop} is an explicit example which demonstrates that crossing-symmetric dispersive sum rules allow us to compute the one-loop corrections to tree-level dispersive bounds. This approach leads to a weaker tree-level bound, but has finite loop corrections. On the other hand, the fixed-$t$ approach gives stronger tree-level bound, but its loop corrections generally have divergences.

In the above analysis, the loop corrections to the bounds are obtained by evaluating the action of the tree-level functional on the 1-loop part of the dispersive sum rules. In the limit where $8\pi G \Lambda^{D-2} \ll 1$, this is guaranteed to be the optimal bound. However, if we imagine $8\pi G \Lambda^{D-2}$ is a finite but small quantity, then one should adjust to optimization problem \eqref{eq:optimization_tree} in order to get optimal bounds. One way to see this is by plugging $8\pi G \Lambda^4=10^{-2}$ into \eqref{eq:g2g3_bound_loop_example}. Then one can see that the loop correction $3.47\times 10^{5}\times (8\pi G M^4)$ and the tree-level part become comparable. This means that if we were to plot the bounds with $8\pi G \Lambda^4=10^{-2}$ on Figure \ref{fig:g2-g3-loop}, we will find the bounds are significantly weaker than the tree-level bounds.

However, one can in fact obtain bounds that are almost as strong as the tree-level bounds for $8\pi G \Lambda^4=10^{-2}$. The idea is to change the optimization condition in \eqref{eq:optimization_tree} to
\be\label{eq:optimization_loop}
\textrm{maximize}\ &\int_0^{\frac{M}{\sqrt{3}}}dp\ f(p)\left(-\frac{8\pi G}{p^2}-2g_2\right) \nn \\
+\g_0 &\int_0^{\frac{M}{\sqrt{3}}}dp\ \left(f(p)\left.{\cal C}^{\textrm{low}}_{2}(-p^2)\right|_{\textrm {1-loop}}+h_4(p)\left.{\cal C}^{\textrm{low}}_{4}(-p^2)\right|_{\textrm {1-loop}}+h_6(p)\left.{\cal C}^{\textrm{low}}_{6}(-p^2)\right|_{\textrm {1-loop}}\right),
\ee
where $\g_0$ should be thought of as the size of $8\pi G\Lambda^4$. Surprisingly, we find that using the new optimization conditions with $\g_0=10^{-2}$, the new bounds become
\be
&-192.7 -21 \frac{g_2 M^2}{8\pi G} - 3.7 \times 10^{-33} \times (8\pi G M^4) + \ldots \nn \\
&\qquad\qquad  \leq \frac{g_3 M^4}{8\pi G} \leq 194.4 + 3\frac{g_2 M^2}{8\pi G} + 1.4 \times 10^{-35} \times (8\pi G M^4) + \ldots,\quad (D=6),
\ee
where the tree-level part of the bounds only become weaker by $\sim 10^{-25}$ compared to \eqref{eq:g2g3_bound_loop_example}.

This shows that by adjusting the optimization condition, we can make the loop corrections much smaller at the cost of slightly weakening the tree-level bound. Moreover, we also observe similar phenomena when studying the loop corrections to the lower bound of $\frac{g_2 \Lambda^2}{8\pi G}$ (i.e., position of the kink in Figure \ref{fig:g2-g3-loop}). This seems to suggest that in the original tree-level problem, there is a nearly-flat direction in the functional space in which the tree-level bound remains almost the same as the extremal value. More precisely, we find that the extremal functionals for the two different optimization conditions \eqref{eq:optimization_tree}, \eqref{eq:optimization_loop} have almost the same $f(p)$, but have very different $h_4(p), h_6(p)$. Interestingly, their extremal spectra look qualitatively similar. The number of states are small at low $J$ and increase as $J$ gets larger, just like the extremal spectra of the fixed-$t$ sum rules found in \cite{Caron-Huot:2021rmr}. It would be very interesting to study if there is a different optimization condition that can change the qualitative behavior of the extremal spectrum.

\subsection{Loop corrections to positivity bounds}\label{sec:positivity_oneloop}
In this section, we study the positivity bounds and their corrections at one-loop. For weakly-coupled theories without gravity, fixed-$t$ sum rules at forward limit imply that $g_{2n} \geq 0$, where $g_{2n}$ is the coefficient of $(s^2+t^2+u^2)^n$ in the tree-level amplitude. The authors of \cite{Caron-Huot:2021rmr} showed that for weakly-coupled theories with gravity, $g_2$ is allowed to be negative, but all the higher $g_{2n}$'s are still positive, since the tree-level graviton exchange does not contribute to the higher-spin sum rules. We now show that after including one-loop diagrams with gravitons, almost all the $g_{2n}$'s are allowed to be negative.

In section \ref{sec:fixed-a_tree_bounds}, we show that using fixed-$a$ sum rules and impact parameter functionals, one can reproduce the positivity bounds on $g_{2n}$. The method for computing the loop corrections is thus straightforward: we take the functionals and act it on the one-loop contribution to the sum rules. As an example, to derive positivity bounds on $g_4$, we can take the spin-4, spin-6, and spin-8 sum rules, and look for functionals such that
\be\label{eq:pols_g4positivity}
\int_{0}^{\frac{M}{\sqrt{3}}}dp\ h_4(p){\cal C}_{4,-p^2}[m^2,J] + \sum_{k=6,8}X_{k;h_k}[m^2,J] \geq 0,\quad \forall m>M,J=0,2,4,\ldots,
\ee
where $h_6(p), h_8(p)$ satisfy the constraints \eqref{eq:hk_nullconstraints}, but we allow $\int dp\ h_4(p)$ to be nonzero. Moreover, since the spin-4 sum rule is the one that dominates \eqref{eq:pols_g4positivity} at fixed impact parameter, we have to choose $h_4(p)$ to be $h_4(p)=\sum_{n=0}^{N_4}a_n p^{n_0+n}$, where the choice of $n_0$ is described below \eqref{eq:functional_pexpansion}. 

On the low-energy side, the functionals satisfying \eqref{eq:pols_g4positivity} then lead to a bound given by
\be
\int_{0}^{\frac{M}{\sqrt{3}}} dp\ h_4(p) g_4 \geq \int_{0}^{\frac{M}{\sqrt{3}}} dp\ \sum_{k,4,6,8}h_k(p)\left.{\cal C}^{\textrm{low}}_{k}(a)\right|_{\textrm {1-loop}}.
\ee
Therefore, the optimization problem can be set up as
\be
\textrm{maximize}\ \int_{0}^{\frac{M}{\sqrt{3}}} dp\ \sum_{k,4,6,8}h_k(p)\left.{\cal C}^{\textrm{low}}_{k}(a)\right|_{\textrm {1-loop}},\qquad \int_{0}^{\frac{M}{\sqrt{3}}} dp\ h_4(p)=1.
\ee

We carry out this numerical computation in various spacetime dimensions, focusing on the $G^2$ term at one loop. Using a $14$-dimensional space of functionals (with $\{N_4,N_6,N_8\}=\{9,8,8\}$), we obtain\footnote{We also recompute some of the bounds with a functional space of $25$ dimensions. The resulting improvement on the bounds are less than $1\%$.}
\be\label{eq:g4_bound_loop}
g^{R}_4 \geq  \begin{cases} 
 -\frac{(8\pi G)^2}{M^2}\times 8.99\times 10^{-5} + \cdots, & (D=6), \\
 (8\pi G)^2\times 1.18 \times 10^{-8} + \cdots, & (D=8),  \\
(8\pi G)^2 M^2 \times 5.63\times 10^{-12} + \ldots, & (D=10), \\
(8\pi G)^2 M^4 \times 9.34\times 10^{-16} + \ldots,& (D=12).
\end{cases}
\ee

Similarly, we can use the spin-6, spin-8,spin-10 sum rules to compute the one-loop corrections to the positivity bound of $g_6$. With $\{N_6,N_8,N_{10}\}=\{9,9,9\}$, we find
\be\label{eq:g6_bound_loop}
g^{R}_6 \geq  \begin{cases} 
 -\frac{(8\pi G)^2}{M^6}\times 1.47\times 10^{-5} + \cdots, & (D=6), \\
 -\frac{(8\pi G)^2}{M^4}\times 4.15 \times 10^{-9} + \cdots, & (D=8), \\
-\frac{(8\pi G)^2}{M^2} \times 2.77\times 10^{-12} + \ldots, & (D=10), \\
-(8\pi G)^2 \times 6.19\times 10^{-15} + \ldots,& (D=12).
\end{cases}
\ee
The bounds shown above are written in terms of the renormalized couplings $g_{2n}^R$ at scale $\mu=M$, and the scheme is chosen such that $g_{2n}^R$ is the coefficient of $(s^2+t^2+u^2)^n$ in the one-loop amplitude, as described in section \ref{sec:loop_contribution_fixeda_sumrules}.

These results show that $g_4,g_6$ are allowed to be negative after including the graviton loops. More generally, the sign of the graviton loop correction depends on the power counting of $g_{2n}$ and $G^2$. We expect that as we lower the scale $M$, the resulting dispersive bounds should become weaker. Hence, if the $g_{2n}$ coupling has higher power counting than $G^2$ (i.e., $G^2$ is more important than $g_{2n}$ at low energy), then one gets $g_{2n} \geq \textrm{(negative number)}\times \frac{(8\pi G)^2}{M^{\#}}$. On the other hand, if $g_{2n}$ has lower power counting, then one gets $g_{2n} \geq \textrm{(positive number)}\times (8\pi G)^2 M^{\#}$. Note that for a fixed spacetime dimension, only finitely many $g_{2n}$'s have lower power counting than $G^2$. Thus, almost all the $g_{2n}$'s get negativity from graviton loops. 

In the case where $g_{2n}$ and $G^2$ have the same power counting, such as $g_4$ in $D=8$ and $g_6$ in $D=12$, the sign is not fixed and one has to do the computation to find the sign of bound. Indeed, from \eqref{eq:g4_bound_loop} and \eqref{eq:g6_bound_loop}, we see that one of them ($g_4$ in $D=8$) is positive, but the other one ($g_6$ in $D=12$) is negative.

Note that some of the bounds on $g_4,g_6$ can also be obtained using the forward limit. In particular, from \eqref{eq:fixeda_smalla_loop} we know that the one-loop contribution to the sum rule is finite at $a=0$ for $D>6$.\footnote{\eqref{eq:fixeda_smalla_loop} only contains the box contribution, but the bubble contribution is always finite at $a=0$.} Then, one can consider ${\cal C}_{4}(0)$ and ${\cal C}_{6}(0)$ to obtain bounds on $g_4,g_6$. The correction to the positivity bound will come from the finite contribution from the one-loop amplitude at $a=0$. We find that bounds obtained from the forward limit all seem to be slightly weaker than the bounds above from the impact parameter functionals. For example, for $g_6$ in $D=8$, the forward limit ${\cal C}_{6}(0)$ gives a bound $g_6 \geq -\frac{(8\pi G)^2}{M^4}\times 3.42 \times 10^{-8}$, which is approximately an order of magnitude weaker than the bound in \eqref{eq:g6_bound_loop}.

Let us also make a quick comment on how higher-loop corrections can modify the story. By simple power-counting, the box diagram with graviton in the loop goes like $G^2M^{D}$, and the two-loop diagram goes like $G^3M^{2D-2}$. Therefore, the two-loop or higher-loop contributions in \eqref{eq:g4_bound_loop} or \eqref{eq:g6_bound_loop} will be less important than the leading one-loop term at low energies.

So far, we have only included the $G^2$ contribution, which we expect to be dominant when $M$ is sufficiently lower than the EFT cutoff $\Lambda$. One can also include corrections from higher coupling, but as we discuss in at the beginning of Sec.~\ref{sec:fixed-a_loop}, we need to choose $M$ to be lower than $\Lambda$ to make sure their contributions are suppressed. As an example, we consider $g_4$ and $g_6$ in $D=6$. If we set $M=\frac{\Lambda}{10}$ and write the bounds in terms of dimensionless quantities $\frac{g_k \Lambda^{2(k-1)}}{8\pi G}$ and $8\pi G \Lambda^{D-2}$, we obtain
\be\label{eq:g4g6_bounds_withhighercouplings}
\frac{g^R_4 \Lambda^6}{8\pi G} \geq& -0.00899(8\pi G \Lambda^4)\left(1- \frac{g_3 \Lambda^4}{8\pi G} \!\times\! 2.39 \!\times\! 10^{-6} - \p{\frac{g_2 \Lambda^2}{8\pi G}}^2 \!\times\! 4.78 \!\times\! 10^{-6} + O(10^{-6}) \!\right), \nn \\
\frac{g^R_6 \Lambda^{10}}{8\pi G}\geq& -14.7(8\pi G \Lambda^4)\left(1+ \frac{g_2 \Lambda^2}{8\pi G} \times 0.004+ \frac{g_3 \Lambda^4}{8\pi G} \times 7.31\times 10^{-6} \right. \nn \\
&\qquad\qquad\qquad\qquad \left.+ \p{\frac{g_2 \Lambda^2}{8\pi G}}^2 \times 1.46\times 10^{-5} + O(10^{-8})\right).
\ee
For simplicity, here we have assumed the scalar-graviton and graviton-graviton interactions are minimally-coupled. The corrections from including non-minimal couplings can be obtained straightforwardly using the expressions given in Appendix \ref{app:dispersive_int}. To estimate the errors for the bounds in \eqref{eq:g4g6_bounds_withhighercouplings}, we have made the assumption that all the dimensionless quantities $\frac{g_k \Lambda^{2(k-1)}}{8\pi G}$ are $O(1)$ numbers, which are established at tree-level in \cite{Caron-Huot:2021rmr} for small values of $k$. Thus, the estimated error $O(10^{-6})$ for the $g_4$ bound (and similarly for the $g_6$ bound) comes from the fact that the next term has a $\p{\frac{M}{\Lambda}}^6$ factor from the EFT counting. 

One might also notice that the $\p{\frac{M}{\Lambda}}^2$ term in the $g_4$ bound and the $\p{\frac{M}{\Lambda}}^6$ term in the $g_6$ bound are vanishing. This is because they have the same power counting as $g_4$, $g_6$, and hence renormalize the couplings. In particular, their one-loop contributions all come from the bubble integral. With the renormalization scheme we choose, they are only proportional to non-analytic terms. Therefore, these terms control the RG running of $g_4,g_6$, and in the sum rules they become $\log\p{\frac{M}{\mu}}$, which can be removed by setting $\mu=M$. This is in contrast to the case of the box integral for $G^2$, which has nonvanishing contributions to the dispersive sum rules even after choosing $\mu=M$, as shown in \eqref{eq:g4_bound_loop}, \eqref{eq:g6_bound_loop}.

\section{$D=4$ dispersive bounds}\label{sec:4Dbounds}
In this section, we study the dispersive bounds for $D=4$. As pointed out in \cite{Caron-Huot:2021rmr,Caron-Huot:2022ugt}, there is a tension between having finite low-energy contribution and imposing positivity in the impact parameter space in $D=4$. This is why we have been restricting to $D>4$ in the previous sections. 

By introducing an IR regulator in the functional, one can still get regulated bounds in $D=4$. We review this approach in section \ref{sec:4D_IRcutoff} and consider the 1-loop corrections to these bounds. Interestingly, we find that with a better understanding of the structures of the bounds at higher loops, one can in fact perform a resummation for all loops and obtain a finite bound in $D=4$ even when the IR regulator is set to zero. This is due to a universal behavior of the gravitational amplitude in the small $t/s$ limit, which we assume is described by the eikonal formula. We give the argument in section \ref{sec:4D_eikonal}.

\subsection{Bounds with IR regulators}\label{sec:4D_IRcutoff}
Let us first consider the $D=4$ bounds with IR regulators at tree-level and one-loop. More concretely, in this subsection we study the lower bound on $g_2$ in terms of the Newton's constant $G$. We use the crossing-symmetric sum rules to avoid the issues with fixed-$t$ sum rules we mention in section \ref{sec:fixed-t_loop}. 

Consider the spin-2 fixed-$a$ sum rule. At tree-level, it is given by
\be
-\frac{8\pi G}{a} + 2g_2 + g_3 a = \left\langle \frac{{\cal P}_J(\sqrt{\frac{m^2+3a}{m^2-a}})}{m^6}(2m^2-3a) \right\rangle.
\ee
To obtain a bound on $g_2$, we set $a=-p^2$ and apply a functional $\int_0^{\frac{M}{\sqrt{3}}}dp\ f(p)$ with a constraint $\int_0^{\frac{M}{\sqrt{3}}}dp\ f(p)p^2=0$ so that we remove the $g_3$ contribution. This gives
\be\label{eq:4D_functionalonspin2}
8\pi G \int_0^{\frac{M}{\sqrt{3}}}dp\ \frac{f(p)}{p^2} + 2g_2 \int_0^{\frac{M}{\sqrt{3}}}dp\ f(p) = \left\langle \int_0^{\frac{M}{\sqrt{3}}}dp\ f(p)\frac{{\cal P}_J(\sqrt{\frac{m^2-3p^2}{m^2+p^2}})}{m^6}(2m^2+3p^2) \right\rangle.
\ee
To have positive contribution for all $m>M$ and $J$, it is necessary that the integral on right-hand side is positive in the impact parameter space ($m\to \infty$ with fixed $b=\frac{2J}{m}$) for all $b$. It turns out that in $D=4$, this forces (see \cite{Caron-Huot:2021rmr} for more details)
\be
\lim_{p\to 0}\frac{f(p)}{p} >0.
\ee
We can normalize $f(p)$ so that this number is $1$. Namely, at small $p$ we have $f(p)=p + \ldots$. However, this implies that graviton pole contribution $\int_0^{\frac{M}{\sqrt{3}}}dp\ \frac{f(p)}{p^2}$ on the left-hand side of \eqref{eq:4D_functionalonspin2} is divergent.

One way to regulate this divergence is to follow \cite{Caron-Huot:2021rmr,Caron-Huot:2021enk} and consider a regulated functional with an IR cutoff $\int_{p_{\textrm{IR}}}^{\frac{M}{\sqrt{3}}}dp f(p)$. Then, the graviton pole contribution becomes
\be
8\pi G \int_{p_{\textrm{IR}}}^{\frac{M}{\sqrt{3}}}dp \frac{f(p)}{p^2} = 8\pi G \log\p{\frac{M}{\sqrt{3}p_{\textrm{IR}}}} + \ldots,
\ee
where $\ldots$ are finite at $p_{\textrm{IR}} \to 0$. In this work, we will instead use dimensional regularization and introduce a $p^{D-4}$ factor to the integrand when applying the functional. The contribution from the graviton pole can then be written as a pole in $D$,
\be
8\pi G \int_0^{\frac{M}{\sqrt{3}}}dp\ p^{D-4}\frac{f(p)}{p^2} = 8\pi G \frac{1}{D-4} + \ldots.
\ee
The advantage of using dimensional regularization is that it nicely captures all the IR divergences, including the ones from the loop integral and the ones from applying the functional near $p=0$. Since the divergences always come from the $p\sim 0$ region, in what follows we will often suppress the integration range of $p$ when discussing divergences.

Using the regulated functional, one can obtain a lower bound on $g_2$ by solving the positivity conditions given by \eqref{eq:pols_fixeda}, along with the optimization conditions
\be
\textrm{maximize}~ -\int_0^{\frac{M}{\sqrt{3}}}dp\ f(p),\quad \lim_{p\to 0}\frac{f(p)}{p}=1.
\ee
In $D=4$, we choose the basis function for $f(p), h_k(p)$ to be
\be\label{eq:fixeda_4D_functional}
f(p)=&(1-\sqrt{3}p)^2(p+c_2 p^2 + c_3 p^3 + \ldots), \nn \\
h_k(p)=& b_{k,2}p^2 + b_{k,3}p^3 + \ldots.
\ee
Let us explain why we need the $(1-\sqrt{3}p)^2$ factor for $f(p)$. First of all, the additional $\sqrt{3}$ is simply because we are integrating $p$ up to $\frac{M^2}{\sqrt{3}}$ instead of $M^2$ as in the fixed-$t$ case. Without the $(1-\sqrt{3}p)^2$ factor, the leading behavior of the high-energy part at large impact parameter $b=\frac{2J}{m}$ is given by
\be
&\int dp\ (p+c_2 p^2 + \ldots) \left\langle \frac{P_J(\sqrt{\tfrac{m^2-3p^2}{m^2+p^2}})}{m^6}(2m^2+3p^2)\right\rangle \nn \\
&\sim -\frac{c_2}{b^3} + \frac{\sqrt{2}}{3^{\frac{1}{4}}b^{\frac{3}{2}}\sqrt{\pi}}\cos\left(\frac{b}{\sqrt{3}}-\frac{3\pi}{4}\right) + \ldots,
\ee
where $\ldots$ are subleading terms at large $b$. So, the dominant contribution is oscillatory and it is impossible to make it positive for all $b$. If we now introduce the $(1-\sqrt{3}p)^2$ factor, we get
\be
&\int dp\ (1-\sqrt{3}p)^2(p+c_2 p^2 + \ldots) \left\langle \frac{P_J(\sqrt{\tfrac{m^2-3p^2}{m^2+p^2}})}{m^6}(2m^2+3p^2)\right\rangle \nn \\
&\sim -\frac{(c_2-2\sqrt{3})}{b^3} + \frac{2\sqrt{2}3^{\frac{3}{4}}}{b^{\frac{7}{2}}\sqrt{\pi}}\cos\left(\frac{b}{\sqrt{3}}+\frac{\pi}{4}\right) + \ldots.
\ee
We see that in this case the leading term is the non-oscillatory term $\frac{1}{b^3}$, and therefore it is possible to get a positive functional. For the higher-spin sum rules/null constraints, we also choose their corresponding functions $h_k(p)$ to start at $p^2$. This is because even though the sum rules do not have divergence near $p=0$ at tree-level, the one-loop contribution from $G^2$ goes like $\frac{\log(p)}{p^2}$ (see \eqref{eq:fixeda_smalla_loop}), and thus $h_k(p)$ has to go like $\sim p^2$ to make the integral finite. Using the choice of basis function given by \eqref{eq:fixeda_4D_functional}, we find that the tree-level numerical bound for $g_2$ in $D=4$ is given by\footnote{We use a $20$-dimensional functional space. The result is weaker than the bound from the fixed-$t$ sum rules obtained in \cite{Caron-Huot:2021rmr,Caron-Huot:2021enk}. This is again because we use the fixed-$a$ sum rules and different basis functions for $h_k(p)$ so that we can compute the one-loop correction.}
\be\label{eq:g2_bound_D4_tree}
g_2 \geq -32.4 \times \frac{8\pi G}{M^2}\frac{1}{D-4}.
\ee

We now consider the one-loop correction to this bound, and focus on the leading IR divergences. So, we would like to compute\footnote{Thanks to our choice of $h_k(p)$, the contributions from higher-spin sum rules with $k=4,6$ do not have IR divergences from the integral near the small $p$ region. They can still contain divergences from the loop integral, but they are subleading contributions.} 
\be
-\int dp\ f(p) \left.{\cal C}^{\textrm{low}}_2(a)\right|_{\textrm{1-loop}},
\ee
where the one-loop term $\left.-{\cal C}^{\textrm{low}}_2(a)\right|_{\textrm{1-loop}}$ is given by \eqref{eq:fixeda_lowenergy_def_oneloop}. IR divergences can either come from the loop amplitude itself, or from applying the functional near $p\sim 0$. In $\left.-{\cal C}^{\textrm{low}}_2(a)\right|_{\textrm{1-loop}}$, the former will appear as a dim reg pole, and the latter will appear as singularity near $a=0$. To understand how to include both of them, let us consider the action of the functional to a term of the form $\frac{1}{(D-4)^{n_1}}\frac{\log^{n_2}(-a)}{-a}$. Then, we get
\be
\int dp\ p^{D-4} f(p)\frac{1}{(D-4)^{n_1}}\frac{2^{n_2}\log^{n_2}(p)}{p^2} = \frac{(-1)^{n_2}n_2!}{(D-4)^{n_1+n_2+1}} + \ldots.
\ee
From the observation that each power of $\log(-a)$ gives a $\frac{1}{D-4}$ after applying the functional, we conclude that the IR divergences can be captured by the following limit of $\left.-{\cal C}^{\textrm{low}}_2(a)\right|_{\textrm{1-loop}}$:
\be\label{eq:eikonal_limit_sumrule}
\lim_{\substack{a\to 0,\ D\to 4 \\ (-a)^{D-4}\ \textrm{fixed}}} \left.-{\cal C}^{\textrm{low}}_2(a)\right|_{\textrm{1-loop}}.
\ee

Moreover, since $\left.-{\cal C}^{\textrm{low}}_2(a)\right|_{\textrm{1-loop}}$ is given by \eqref{eq:fixeda_lowenergy_def_oneloop} and ${\cal M}_{\textrm{1-loop}}$ is explicitly known, we can exchange the limit and the integral. (In other words, the contribution from $s'=0$ in \eqref{eq:fixeda_lowenergy_def_oneloop} does not contribute to the leading divergences.) After plugging in the expression of ${\cal M}_{\textrm{1-loop}}$ and taking the limit \eqref{eq:eikonal_limit_sumrule}, we obtain
\be
\left.-{\cal C}^{\textrm{low}}_2(a)\right|_{\textrm{1-loop}} = \frac{(8\pi G)^2M^2}{2\pi^2}\frac{1}{D-4}\frac{(-a)^{\frac{D-4}{2}}}{a} + \ldots,
\ee
where $\ldots$ are subleading in the limit \eqref{eq:eikonal_limit_sumrule}. Adding this contribution to \eqref{eq:4D_functionalonspin2}, we then obtain that the one-loop correction to the lower bound on $g_2$ is given by
\be\label{eq:g2_bound_D4_loop}
g_2 \geq -32.4\times \p{\frac{8\pi G}{M^2}\frac{1}{D-4}-(8\pi G)^2\frac{1}{4\pi^2 (D-4)^2}+ \ldots},
\ee
where $\ldots$ include terms that are less divergent than $\frac{1}{D-4}$ at tree-level or $\frac{1}{(D-4)^2}$ at one-loop.

For higher couplings such as $g_4$, we can perform similar analysis to study the one-loop corrections to their lower bound. At tree-level, it obeys a positivity bound $g_4\geq 0$. After including the graviton loop contributions, the behavior at small $a$ will again lead to a divergent bound that contains a $G^2/(D-4)^2$ term as opposed to the finite bounds in $D>4$ computed in section \ref{sec:positivity_oneloop}.

\subsection{Finite bounds from the eikonal amplitude}\label{sec:4D_eikonal}

An important observation for this problem is that there are two small parameters: Newton's constant $G$ and $D-4$ (or the IR cutoff $p_{\mathrm{IR}}$). In the above analysis, we are taking the weak coupling limit $G\to 0$ first (so that we focus on tree-level and one-loop), and then we take $D \to 4$. The one-loop bound \eqref{eq:g2_bound_D4_loop} shows that the leading divergence of the $G^2$ correction is a double pole $\frac{1}{(D-4)^2}$, as opposed to the tree-level bound which only has a $\frac{1}{D-4}$ pole. In fact, as one goes to higher order in the loop expansion, one gets even higher powers of $\frac{1}{D-4}$. This is a clear sign that the $G \to 0$ limit and the $D \to 4$ limit do not commute with each other.

However, it seems reasonable to also consider the other limit, where we first take $D \to 4$ (i.e., take the IR cutoff to zero) while keeping $G$ finite. It is in this limit that we will see that under certain assumptions we can resum all the IR divergences at each order in $G$ and get a finite quantity.

As we argue above, the leading IR divergences is given by \eqref{eq:eikonal_limit_sumrule}. By \eqref{eq:fixeda_lowenergy_def}, it can be written as
\be\label{eq:4D_sumrule_IRdiv}
&\lim_{\substack{a\to 0,\ D\to 4 \\ (-a)^{D-4}\ \textrm{fixed}}} -{\cal C}^{\textrm{low}}_2(a) \nn \\ &=-\lim_{\substack{a\to 0,\ D\to 4 \\ (-a)^{D-4}\ \textrm{fixed}}}\partial_x{\cal M}(x,a)|_{x=0}-\lim_{\substack{a\to 0,\ D\to 4 \\ (-a)^{D-4}\ \textrm{fixed}}} \int_0^{M^2}\frac{ds'}{\pi}\frac{2}{(s')^3}\mathrm{Im}{\cal M}(s',a), \nn \\
&=-\lim_{\substack{a\to 0,\ D\to 4 \\ (-a)^{D-4}\ \textrm{fixed}}}\partial_x{\cal M}(x,a)|_{x=0} -\int_0^{M^2}\frac{ds'}{\pi}\frac{2}{(s')^3}\mathrm{Im}\left[\lim_{\substack{a\to 0,\ D\to 4 \\ (-a)^{D-4}\ \textrm{fixed}}}{\cal M}(s',a)\right],
\ee
where we have taken the small-$a$ limit of \eqref{eq:fixeda_lowenergy_def_oneloop}, and used the fact that $\tau(s',a)=a+O(a^2)$.\footnote{In the third line of \eqref{eq:4D_sumrule_IRdiv}, we have exchanged the integral over $s'$ and the limit. For one-loop amplitudes, one can explicitly check that the $s'=0$ region does not contribute to the leading divergences, and thus it is safe to take the limit inside the integral. More generally, since the integral can always be defined in terms of the low-energy arcs that does not touch $s'=0$, we expect the leading divergence can always be computed by the last line of \eqref{eq:4D_sumrule_IRdiv}.}

\subsubsection{Eikonal amplitude}
Our goal now is to understand the behavior of the amplitude at $a \to 0, D\to 4$ with fixed $(-a)^{D-4}$, at all orders in $G$. It turns out that gravitational amplitudes have universal behavior in this limit due to eikonalization.

Since $a$ and $t$ are identical at small $a$, in this subsection we will switch back to the $t$ variable. The behavior of the amplitudes at large $G s$ and small $t/s$ is captured by the eikonal formula \cite{Muzinich:1987in, Amati:1987wq, Amati:1987uf}:
\be
i{\cal M}_{eik}(s,t=-\vec q_\perp^{\,2}) = 2s \int d^{D-2} \vec x_\perp e^{i\vec q_{\perp}\cdot \vec x_{\perp}}(e^{i\chi(b)}-1),
\ee
where $b=|\vec x_\perp|$, and $\chi(b)$ is the eikonal phase coming from the inverse Fourier transform of the graviton propagator,
\be
i\chi(b)=2i G s \Gamma\left(\frac{D-2}{2}\right)\frac{(\pi b^2)^{\frac{4-D}{2}}}{D-4}.
\ee

While the eikonal formula is technically only valid for large $G s\gg 1$, one might expect that its validity can be extended for finite $G s$, as long as $\frac{s}{-t} \gg 1$ (and $D \to 4$ such that $(-t)^{D-4}$ is fixed). One check we can do is to show that the formula is consistent with the twice-subtracted dispersion relation,
\be\label{eq:2SDR_expr}
&\frac{{\cal M}_{\mathrm{low}}(s,t)}{s(s+t)} + \mathrm{Res}_{s'=0}\left[\left(\frac{1}{s'-s}+\frac{1}{s'+s+t}\right)\frac{{\cal M}_{\mathrm{low}}(s',t)}{s'(s'+t)}\right] \nn \\
&=\int \frac{ds'}{\pi}\left(\frac{1}{s'-s}+\frac{1}{s'+s+t}\right)\mathrm{Im}\left[\frac{{\cal M}_{\mathrm{high}}(s',t)}{s'(s'+t)}\right].
\ee
Indeed, it has been recently checked in \cite{Haring:2024wyz} that the twice-subtracted dispersion relation correctly reproduces the graviton pole at low energy from the eikonal amplitude. (See also \cite{Caron-Huot:2021rmr,Caron-Huot:2021enk} for similar checks using the spin-2 dispersive sum rule.)

For our purpose, we are interested in the limit $t \to 0,D \to 4$ with $(-t)^{D-4}$ fixed. In this limit, the eikonal amplitude takes an even simpler form given by (see e.g., eq. (3.4) of \cite{DiVecchia:2019myk}) 
\be\label{eq:Meikonal_smalleps}
{\cal M}(s,t) =&{\cal M}_{tree}\exp\left(\frac{2i G s}{D-4}\left(\frac{-t}{\mu^2}\right)^{\frac{D-4}{2}} \right)+\ldots \nn \\
=&-\frac{8\pi G s^2}{t}\exp\left(\frac{2i G s}{D-4}\left(\frac{-t}{\mu^2}\right)^{\frac{D-4}{2}}\right)+\ldots,
\ee
where $\ldots$ are subleading terms in the $t \to 0,D \to 4$, fixed $(-t)^{D-4}$ limit. After plugging the formula into the twice-subtracted dispersion relation, the left-hand side of \eqref{eq:2SDR_expr} gives
\be
&\frac{{\cal M}_{\mathrm{low}}(s,t)}{s(s+t)} + \mathrm{Res}_{s'=0}\left[\left(\frac{1}{s'-s}+\frac{1}{s'+s+t}\right)\frac{{\cal M}_{\mathrm{low}}(s',t)}{s'(s'+t)}\right] \nn \\
&= -\frac{8\pi G}{t}\exp\left(\frac{2i G s}{D-4}\left(\frac{-t}{\mu^2}\right)^{\frac{D-4}{2}}\right) + \ldots,
\ee
where $\ldots$ are subleading at small $t$. On the other hand, the heavy contribution in the right-hand side of \eqref{eq:2SDR_expr} is given by
\be
&\int_0^{\infty} \frac{ds'}{\pi}\left(\frac{1}{s'-s}+\frac{1}{s'+s+t}\right)\mathrm{Im}\left[\frac{{\cal M}_{\mathrm{high}}(s',t)}{s'(s'+t)}\right] \nn \\
&=\int_0^{\infty} \frac{ds'}{\pi}\left(\frac{1}{s'-s}+\frac{1}{s'+s}\right)\frac{1}{(s')^2}\left(-\frac{8\pi G (s')^2}{t}\right)\sin\left(\frac{2 G s'}{D-4}\left(\frac{-t}{\mu^2}\right)^{\frac{D-4}{2}}\right) +\ldots \nn \\
&=-\frac{8\pi G}{t}\exp\left(\frac{2i G s}{D-4}\left(\frac{-t}{\mu^2}\right)^{\frac{D-4}{2}}\right) + \ldots.
\ee
Therefore, we see that in this particular limit the entire eikonal amplitude is exactly reproduced by its imaginary part from the dispersion relation (not just the graviton pole). One can also perform the same check using the crossing-symmetric dispersion relation, but in the forward limit it is essentially the same as the fixed-$t$ dispersion relation, so the check is identical.

We emphasize that the correct way of interpreting the check above is that one should imagine there is an underlying nonperturbative S-matrix that satisfies the twice-subtracted dispersion relation. The check we did shows that it is consistent that the S-matrix is described by the eikonal formula \eqref{eq:Meikonal_smalleps} in the limit $t \to 0, D\to 4$ with fixed $(-t)^{D-4}$.

An independent check that Eq. \eqref{eq:Meikonal_smalleps} captures the correct small $t$ behavior of the amplitude, is that it is also reproduced by the exponentiation of gravitation infrared divergences in momentum space, which takes the form \cite{Weinberg:1965nx}
\be
{\cal M}_{\rm IR} = {\cal M}_{\rm tree} e^{ {\cal I}},
\ee
where ${\cal I}$ is the IR divergence of the one-loop amplitude
\be
{\cal I}=  \frac{{\cal M}_{\rm 1-loop}^{\rm div}}{{\cal M}_{\rm tree}} =- \frac{4G}{\pi} \frac{1}{(D-4)^2} \left( s \left(\frac{s}{\mu^2}\right)^{\frac{D-4}{2}}\!\! e^{-i\pi \frac{D-4}{2}} + t\left(\frac{-t}{\mu^2}\right)^{\frac{D-4}{2}}\!\! + u\left(\frac{-u}{\mu^2}\right)^{\frac{D-4}{2}} \right).
\ee

In light of the discussion above, from now on we will assume that the eikonal formula describes the amplitude at small $t$ at all orders in $G$, and, in particular  limit $t \to 0, D\to 4$ with $(-t)^{D-4}$ and $Gs$ fixed.

\subsubsection{Finite $D=4$ bounds}

We are now ready to resum the divergences of the dispersive bound. We would like to compute
\be\label{eq:FD_definition}
F(D)\equiv M^2\int dp\ f(p) \lim_{\substack{a\to 0,\ D\to 4 \\ (-a)^{D-4}\ \textrm{fixed}}} -{\cal C}^{\textrm{low}}_2(a)\Big|_{a=-p^2},
\ee
and study the behavior of $F(D)$ near $D=4$ while keeping $G$ finite. The additional $M^2$ factor is introduced to make $F(D)$ dimensionless. Plugging in \eqref{eq:4D_sumrule_IRdiv} and the eikonal formula \eqref{eq:Meikonal_smalleps}, we obtain that the leading divergence of $F(D)$ near $D=4$ is given by
\be
F(D)=&\int dp\ p^{D-3}\frac{8\pi G M^2}{p^2} - \int dp\ p^{D-3}\int_0^{M^2}\frac{ds'}{\pi}\frac{2}{(s')^3}\frac{8\pi GM^2 (s')^2}{p^2}\sin\p{\frac{2G s'}{D-4}p^{D-4}\mu^{4-D}},
\ee
where we have normalized $f(p)$ such that $f(p)=p+O(p^2)$ at small $p$. Evaluating the above integral, we have
\be\label{eq:FD_resummed}
F(D) =  \frac{8\pi G M^2}{D-4} + 16\sin^2\p{\frac{GM^2}{D-4}} -\frac{16 GM^2}{D-4}\mathrm{Si}\p{\frac{2GM^2}{D-4}},
\ee
where $\mathrm{Si}(x)=\int_0^x \frac{\sin t}{t} dt$.

\begin{figure}[t]
\centering
\includegraphics[width=12cm]{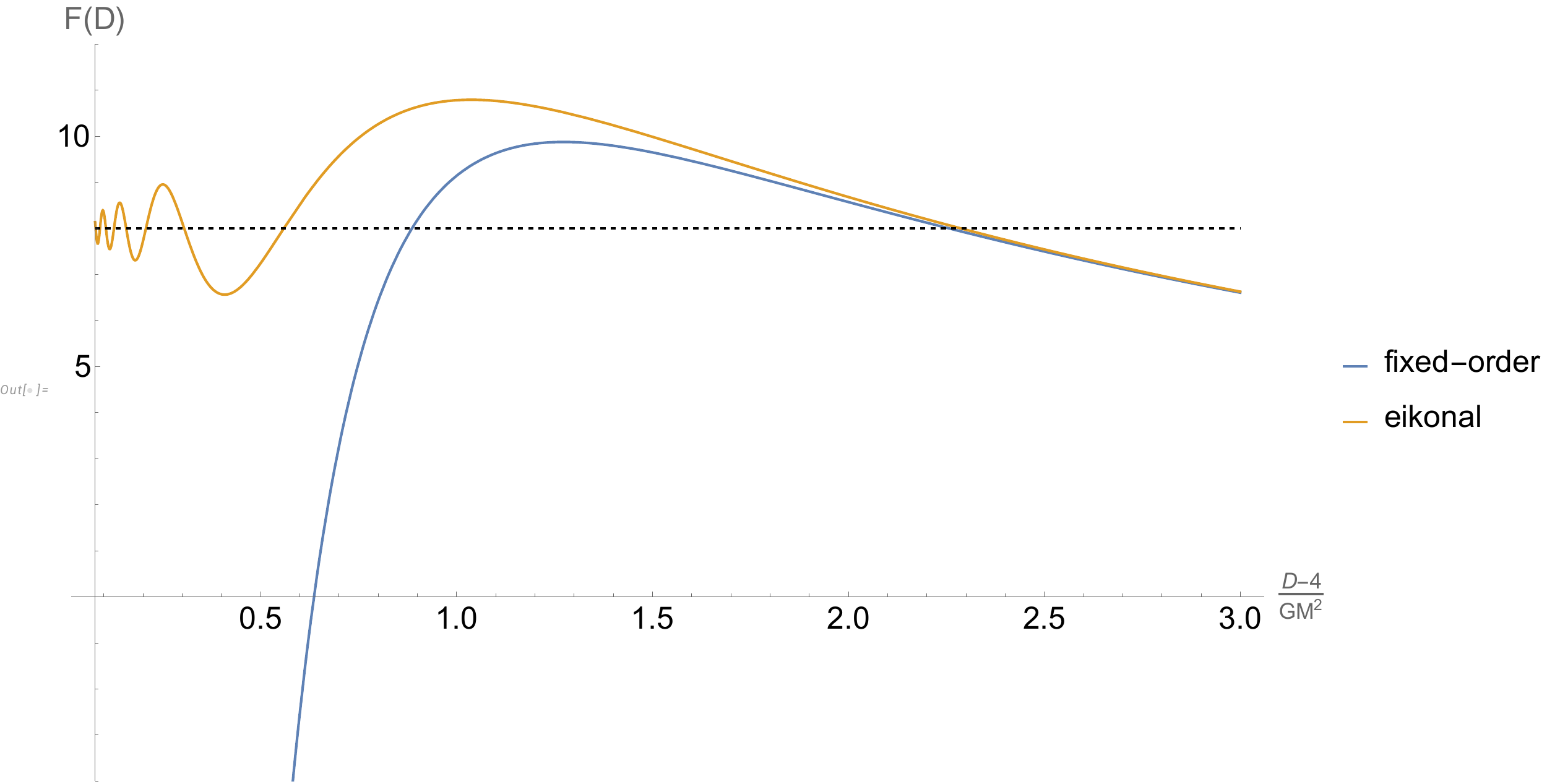}
\caption{Plot for the leading IR divergences of the functional action on the spin-2 sum rule $F(D)$ (defined by \eqref{eq:FD_definition}) near $D=4$. The blue curve represents the result from fixed-order (up to one-loop) calculation \eqref{eq:FD_fixedorder}, and orange curve is from resummation using the eikonal formula \eqref{eq:FD_resummed}. Away from $D=4$, the two curves agree with each other. As $D \to 4$, the fixed-order result diverges, but the eikonal result approaches a finite value represented by the black dashed line.}
\label{fig:eikonal}
\end{figure}

As a check, we can further expand the the above expression at small $G$, and we find
\be\label{eq:FD_fixedorder}
F_{\textrm{fixed-order}}(D)=\frac{8\pi GM^2}{D-4}-\frac{(8\pi GM^2)^2}{4\pi^2(D-4)^2},
\ee
which indeed agrees with the one-loop result \eqref{eq:g2_bound_D4_loop}. The resummed result also clearly shows that the two limits $G \to 0$ and $D \to 4$ do not commute. In fact, \eqref{eq:FD_resummed} only depends on the dimensionless quantity $\frac{G M^2}{D-4}$. The weak coupling limit $G \to 0$ corresponds to $\frac{G M^2}{D-4} \to 0$, but taking $D \to 4$ first corresponds to taking $\frac{G M^2}{D-4} \to \infty$. In the latter limit, we find
\be\label{eq:FD_4D_limit}
\lim_{D\to 4} F(D) = 8.
\ee
Surprisingly, $F(D)$ becomes finite at $D=4$. Another way of seeing why one should expect to get a finite quantity in this limit is to look at \eqref{eq:FD_definition} and remember the fact that the eikonal amplitude itself satisfies the dispersion relation. Therefore, when we take $M\to \infty$ (which also corresponds to $\frac{G M^2}{D-4} \to \infty$), the entire dispersive integral should vanish, and thus $\frac{F(D)}{M^2} \to 0$. Assuming $F(D)$ has a Laurent series expansion in $\frac{D-4}{GM^2}$, this implies $\lim_{D\to 4}F(D) = \lim_{M\to \infty} F(D)$ must be finite.

In Figure \ref{fig:eikonal}, we show the plots for $F(D)$ obtained from the one-loop calculation \eqref{eq:FD_fixedorder} and the resummation \eqref{eq:FD_resummed} as a function of the dimensionless quantity $\frac{D-4}{GM^2}$. As expected, when $\frac{GM^2}{D-4}$ is small, the weak coupling expansion dominates, and the one-loop result agrees nicely with the resummed result. However, near $D=4$, the one-loop result is divergent, but the resummed result approaches the finite value given by \eqref{eq:FD_4D_limit}.

Recall that $F(D)$ appears in the sum rule as
\be\label{eq:4D_functionalonspin2'}
F(D) + 2g_2 \int_0^{\frac{M}{\sqrt{3}}}dp\ f(p) = \left\langle \int_0^{\frac{M}{\sqrt{3}}}dp\ f(p)\frac{{\cal P}_J(\sqrt{\frac{m^2-3p^2}{m^2+p^2}})}{m^6}(2m^2+3p^2) \right\rangle.
\ee
Repeating the same numerical computation done in section \ref{sec:4D_IRcutoff}, we obtain a new lower bound on $g_2$ given by
\be\label{eq:g2_bound_D4_resummed}
g_2 \geq -\frac{259}{M^4}+ \ldots.
\ee
We expect this bound can have subleading corrections (denoted as $\ldots$ above) coming from resumming the subleading IR divergences at all orders in $G$ and other loop contributions such as $O(g_2^2)$.

Similarly, we can perform the same resummation analysis for $g_4$. We find the leading IR divergences at all orders in $G$ again resum into a finite quantity, giving
\be
\frac{8}{3M^6} + 4g_4 \int dp\ h_4(p) + \ldots = \int dp\ h_4(p)\left\langle \frac{P_J(\sqrt{\tfrac{m^2-3p^2}{m^2+p^2}})}{m^{12}}(2m^2+3p^2)(m^2+p^2)\right\rangle.
\ee
This yields a finite bound $g_4 \geq -\frac{\#}{M^8}$.

By resumming the leading IR divergences at all orders in $G$, we have managed to get a somehow better bound than the one before (eq. \eqref{eq:g2_bound_D4_tree}), in the sense that it turns a logarithmically divergent bound into a finite bound. However, this bound has another problem: it is not suppressed by $G$. This means the higher-loop contributions like $O(g_2^2)$ in the $\ldots$ terms don't have to be suppressed even if we assume $G M^2$ is small. So, we might have to be more careful interpreting this bound. Most likely, this bound should not be interpreted as a bound on the Wilson coefficient $g_2$, but as a bound on an observable that reduces to $g_2$ at weak coupling.

The resummed finite bound \eqref{eq:g2_bound_D4_resummed} is expected to be valid in the limit where the IR cutoff is strictly much smaller than all the other parameters of the theory (including $GM^2$). In more realistic situations, the suppression from $GM^2$ is often a much bigger effect. For example, let us take the bound with IR regulator \eqref{eq:g2_bound_D4_tree}, and replace $\frac{1}{D-4}$ with $\log\p{\frac{M}{m_{\textrm{IR}}}}$. If we choose $m_{\textrm{IR}}$ to be the Hubble scale, then for $M=10~\textrm{TeV}$ the logarithmic divergence becomes $\log\p{\frac{M}{m_{\textrm{IR}}}}\sim 100$. On other hand, the suppression from $GM^2$ is of the size $(M/M_{\textrm{planck}})^2 \sim 10^{-30}$. This makes the bound with IR regulator become $g_2 \gtrsim -\frac{10^{-27}}{M^4}$,
which is much stronger than \eqref{eq:g2_bound_D4_resummed}.

In the analysis that leads to the bound \eqref{eq:g2_bound_D4_resummed}, we have only used the eikonal formula in the low-energy region (i.e., the left-hand side of the sum rule \eqref{eq:4D_functionalonspin2'}). On the other hand, the $p \to 0$ limit of the high-energy contribution should also be described by the eikonal formula. If we also assume eikonal for these high-energy contributions, it seems likely that we can recover a $G$-dependent bound, possibly with an IR logarithm where the IR cutoff is replaced by a physical scale.\footnote{We thank Brando Bellazzini for discussions on this point.}

Another interesting place to explore the comparison between these two types of bounds is in AdS/CFT, where the logarithmic divergences are regulated by $R_{\textrm{AdS}}$. In particular, in \cite{Caron-Huot:2021enk} it was shown that the $g_2$ coupling in $\textrm{AdS}_4/\textrm{CFT}_3$ has a lower bound of the form\footnote{Strictly speaking, the bound in \cite{Caron-Huot:2021enk} was obtained using fixed-$t$ dispersive CFT sum rules, which can be problematic when including loop corrections as discussed in section \ref{sec:fixed-t_loop}. A crossing-symmetric version of dispersive CFT sum rules was studied in \cite{Bissi:2022fmj}, but to our knowledge there has not been a similar analysis that derives a lower bound on $g_2$. Here, we assume such bound exists.}
\be\label{eq:g2_CFT_fixedorder}
g_2 \geq -\# \frac{G}{M^2} \log\p{\# MR_{\textrm{AdS}}} + O\p{G^2M^4\log^2\p{\# MR_{\textrm{AdS}}},\frac{1}{MR_{\textrm{AdS}}}},
\ee
where $\#$ are computable $O(1)$ constants. This bound was obtained assuming the hierarchy $1\ll MR_{\textrm{AdS}} \ll \frac{1}{GM^2}$. From the CFT perspective, the limit is equivalent to $1\ll \Delta_{\textrm{gap}} \ll c_T$, where $c_T$ is the coefficient of the $\langle TT \rangle$ two-point function, and $\Delta_{\textrm{gap}}$ is the gap of the single-trace higher-spin spectrum of the CFT.

Our calculation here suggests that one should be able to resum the leading logarithms $\log(M R_{\textrm{AdS}})$ in \eqref{eq:g2_CFT_fixedorder} at all orders in $G$ (or equivalently $\frac{1}{c_T}$) using the eikonal method in AdS/CFT \cite{Cornalba:2006xk,Cornalba:2006xm,Cornalba:2007zb,Meltzer:2019pyl}. We expect to get an expression analogous to \eqref{eq:FD_resummed}. It would be nice to verify this from an explicit CFT calculation.

\section{Conclusions}\label{sec:conclusion}

In this paper, we have studied the effects of including EFT loops to the dispersive bounds on gravitational EFTs. In particular, we focus on the $2\to 2$ scattering amplitude of a real scalar coupled to gravity considered in \cite{Caron-Huot:2021rmr}. The low-energy EFT is assumed to have a weak coupling expansion, and we included the contributions at one-loop level. We found that the usual approach for deriving tree-level bounds using the fixed-$t$ dispersion relation encounters several issues due to the forward limit divergences of the one-loop amplitude with graviton loops. By introducing the crossing-symmetric dispersion relation, which does not require taking any forward limit, we were able to obtain dispersive sum rules whose one-loop contribution at low energy is finite.

As an application, we computed the bounds on higher-derivative contact interactions using the crossing-symmetric dispersive sum rules. The resulting bounds at tree-level are qualitatively the same as the bounds obtained from fixed-$t$ sum rules, but importantly, the bounds also have finite one-loop corrections, whereas the fixed-$t$ bounds become problematic at loop level.

At tree-level, dispersive sum rules in the forward limit imply that a family of Wilson coefficients $g_{2n}$ must be positive. We computed the one-loop corrections to these bounds and showed that generically, $g_{2n}$'s are allowed to be negative at loop level. The loss of positivity is dominated by the graviton loops, and hence is of the size $O(G^2)$.

Moreover, we revisited the problem in $D=4$, where the tree-level dispersive bounds for gravitational EFTs involve logarithmic divergences. By assuming that the eikonal formula resums the leading logarithmic divergences at all loops (including for $s\sim M^2$), we found that the resulting resummed bound is finite even when taking the IR cutoff to zero. However, the bound is not suppressed by $G$, and thus one should interpret it as part of the nonperturbative S-matrix that reduces to the coupling at low energies.

There is still room for improvement on the results we presented here. For instance, in order to truncate the EFT series at one-loop, we had choose the size of the arc $M$ to be smaller than the EFT cutoff $\Lambda$. In other words, even though at tree-level one can show that all the Wilson coefficients have to be suppressed by $G$ at the cutoff, at loop-level we can only show weak coupling at a scale $M$ smaller than the cutoff. For smaller $M$, the EFT expansion has better convergence and we can include just a few leading terms at one-loop. However, this also makes the resulting bounds weaker. Ideally, one would like to choose an intermediate $M$ so that the bound is reasonably strong, but at the same time does not involve too many couplings at one-loop. It would also be interesting to ``smear over $M$" by integrating the sum rules over $M$ against a function $f(M)$. Also, when applying functionals to the crossing-symmetric dispersion relations, we restricted the range of the variable $a$ to be $a \in [-\frac{M^2}{3},0]$. If we allow for a larger region of validity of our assumptions (listed in the beginning of section \ref{sec:fixed-t_tree}) on the complex $t$ plane, we should be able to apply functionals with support on a larger domain of $a$, and they could lead to stronger bounds.

A natural generalization of our work is to consider the case of four-graviton scattering, and obtain one-loop corrections to the tree-level bounds in \cite{Caron-Huot:2022jli,Caron-Huot:2022ugt}. The four-graviton amplitudes admit superconvergence sum rules which do not require improvement. Hence, it is possible to include the one-loop contribution using the fixed-$t$ sum rules and this was explored recently in \cite{Caron-Huot:2024lbf}. On the other hand, for the higher-derivative couplings which do not appear in the superconvergence sum rules, we expect one would need to use the crossing-symmetric sum rules to obtain one-loop corrections to their bounds. It would also be interesting to revisit the eikonal argument in section \ref{sec:4D_eikonal} and apply it to four-graviton scattering in $D=4$. One important ingredient for this calculation would be the one-loop four-graviton amplitudes in general dimensions, which could be useful for the primal S-matrix bootstrap \cite{Guerrieri:2020bto, EliasMiro:2021nul, Haring:2022sdp,EliasMiro:2022xaa, Guerrieri:2021ivu, Guerrieri:2022sod,EliasMiro:2023fqi}.

The one-loop contributions to the sum rules also involve non-minimal two-scalar-two-graviton interactions, which we did not study in this work. By including their contributions, one can derive bounds involving these couplings just by studying the four-scalar amplitude. Alternatively, one could also set up a mixed system involving four-scalar, two-scalar-two-graviton, and four-graviton amplitudes and derive tree-level bounds on those couplings \cite{Hong:2023zgm, Xu:2024iao}. It would be interesting to compare the bounds obtained by these two different methods.

Another direction is to study similar problems in AdS/CFT for the four-point functions of scalar \cite{Caron-Huot:2021enk} and stress tensor \cite{Chang:2023szz}. In particular, one would need to compute the contribution of the bulk one-loop diagrams to the dispersive CFT sum rules. As shown in \cite{Caron-Huot:2021enk,Chang:2023szz}, dispersive CFT sum rules are equivalent to flat space fixed-$t$ sum rules in the appropriate flat-space limit. Thus, we expect to encounter the same issues discussed in section \ref{sec:fixed-t_loop}.\footnote{Although it is true that the divergences will all be regulated by $R_{\mathrm{AdS}}$ in AdS/CFT, if the divergences are power divergences, then the resulting bound would not be consistent with EFT scaling. In other words, the bounds will have unexpected powers of $M R_{\textrm{AdS}}$ that do not agree with dimensional analysis.} Therefore, in order to obtain finite one-loop corrections (modulo factors of $\log(M R_{\textrm{AdS}})$), one would either need to focus on the superconvergence sum rules for the stress tensor four-point function, or develop a crossing-symmetric version of dispersive CFT sum rules. (See \cite{Bissi:2022fmj} for a recent study of position-space crossing-symmetric dispersion relation in CFT.) Furthermore, as discussed at the end of section \ref{sec:4D_eikonal}, we expect that one could resum the logarithmic divergences of the bounds in $\textrm{AdS}_4/\textrm{CFT}_3$ using eikonal method.

Finally, it would be very interesting to understand whether the eikonal formula really captures the forward behavior of gravitational amplitudes at finite $G$, and the possible implications of this fact for the infrared safety of more general observables in four-dimensional quantum gravity. 

\section*{Acknowledgements}
We thank Brando Bellazzini, Simon Caron-Huot, Subham Dutta Chowdhury, Clay C\'ordova, Yue-Zhou Li, David Simmons-Duffin, and Sasha Zhiboedov for helpful discussions, and the authors of \cite{Beadle2025} for coordinating submission. We also thank the Burke Institute for Theoretical Physics for support during the initial stages of this work. This research was enabled in part by support provided by the BC DRI Group and the Digital Research Alliance of Canada (Cedar cluster). CHC is supported by a Kadanoff fellowship at the University of Chicago.

\appendix

\section{One-loop EFT amplitudes}\label{app:one-loop_amp}
In this appendix, we record the expression of the four-scalar amplitude at one loop up to ${\cal O}(s^4)$ order in the EFT expansion, and the full expression up to ${\cal O}(s^6)$ order is given in the ancillary file. As reviewed in section \ref{sec:1loop_amplitude}, the amplitude can be written as a sum of scalar box, triangle and bubble integrals as in Eq.~\eqref{eq:ampdecomp}.
The box coefficient and triangle coefficients are
\be
H_{\ibox}(s,t,u) &= (8\pi G)^2(s^4+t^4)\,,\\
H_{\itri}(s,t,u) &= -2(8\pi G)^2 s(3s^2-4tu) + 4(8\pi G)g_2 s^2(s^2-tu)+2(8\pi G)g_3 s^3 tu \nn \\
&+ 8(8\pi G)g_4 s^2(s^2-tu)^2 + 4(8\pi G)g_5s^3tu(s^2-tu) + \cdots\,,
\ee
whereas the bubble coefficients are
\begin{align}
&H_{\ibub}(s,t,u) \nn \\
&=\frac{(8\pi G)^2}{64 (D-2) (D^2-1)}\left((D-2) s^2 \left(D^2 \left(D_s^2-3 D_s+162\right)+D \left(-2 D_s^2+6 D_s+28\right)-128\right) \right. \nn \\
&\qquad\qquad\qquad\qquad\left.-8 t u \left(4 D^3 D_s-16 D^2 (D_s-4)+D \left(D_s^2+D_s+2\right)-2 \left(D_s^2-15 D_s+34\right)\right)\right) \nn \\
&+ \frac{(8\pi G)^2\alpha_2}{64 \left(D^2-1\right) (D_s-2)}\left((D_s-4) s \left(D^2 \left(D_s(D_s -3)s^2-16 (D_s-2) t u\right)\right.\right.\nn \\
&\qquad\qquad\qquad\qquad\qquad \left.\left.+D \left(-2 D_s^2 s^2+6 D_s s^2+32 D_s t u-64 t u\right)-8 \left(D_s^2-9 D_s+12\right) t u\right)\right) \nn \\
&+\frac{(8\pi G)g_2}{8 (D-2) (D^2-1)}\left(\left(-37 D^3+64 D^2+44 D-48\right) s^3+8 \left(8 D^3-8 D^2-7 D+6\right) s t u\right) \nn \\ 
&+\frac{(8\pi G)^2 \alpha_4}{128 \left(D^2-1\right)}\left(D_s^2-7 D_s+12\right) s^2 \left(D(D-2) s^2-8 t u\right) \nn \\
&+\frac{(8\pi G)^2\alpha^2_2(D_s-4)s^2}{256 \left(D^2-1\right) (D_s-2)^2}\left(D(D-2) \left(\left(D_s^3-5 D_s^2+10 D_s-12\right) s^2-8 \left(D_s^2-4\right) t u\right)\right. \nn \\
&\qquad\qquad\qquad\qquad\qquad\quad\left. -8 D_s \left(D_s^2-8 D_s+10\right) t u\right) \nn \\
&+ \frac{(8\pi G)g_3 }{16 (D-2) \left(D^2-1\right)}s^2 \left(D^3 \left(9 s^2-64 t u\right)+D^2 \left(64 t u-12 s^2\right)+D \left(56 t u-12 s^2\right)-48 t u\right)
 \nn \\
&- \frac{8\pi G \beta^{(2)}_0(D_s-4) s^2}{32 \left(D^2-1\right) (D_s-2)}\left(D^2 \left(D_s^2 s^2-3 D_s s^2-16 D_s t u+32 t u\right)\right. \nn \\
&\qquad\qquad\qquad\qquad\qquad\left.+D \left(-2 D_s^2 s^2+6 D_s s^2+32 D_s t u-64 t u\right)-8 \left(D_s^2-9 D_s+12\right) t u\right) \nn \\
&-8\pi G \beta^{(3)}_1 \frac{(D-2) \left(D_s^2-7 D_s+12\right) s^4}{8 (D-1)}+g_2^2 \frac{9 D^2 s^4+6 D s^4-8 s^2 t u}{8 \left(D^2-1\right)} + \cdots,
\end{align}
The expressions of $H_{\ibox}$, $H_{\itri}$, and $H_{\ibub}$ up to ${\cal O}(s^6)$ order are given by \texttt{Hbox}, \texttt{Htriangle}, and \texttt{Hbubble} respectively in the ancillary file.

\section{Dispersive integrals}\label{app:dispersive_int}
\subsection{Crossing-symmetric sum rules}
Let us first explain how to compute the box diagram dispersive integral. We will consider the integral in the fixed-$a$ sum rule given by \eqref{eq:box_dispersive_intexpression}. The integral for the fixed-$t$ sum rules can be computed in the same way. The integral we would like to evaluate is
\be\label{eq:box_dispersive_intexpression_inAppendix}
\left.-{\cal C}^{\textrm{low}}_{k}(a)\right|_{{\cal M}_{\textrm{box}}} 
 &=2r_{\ibox}(8\pi G)^2  \sin(\tfrac{\pi D}{2})\int_0^{M^2}\frac{ds'}{\pi}(s')^{\frac{D-6}{2}}\left(\frac{((s')^4+\tau^4){}_2F_1\p{1,\frac{D-4}{2},\frac{D-2}{2},\frac{s'+\tau}{\tau}}}{\tau}\right. \nn \\
& \left.+\frac{((s')^4+(-s'-\tau)^4){}_2F_1\p{1,\frac{D-4}{2},\frac{D-2}{2},\frac{\tau}{s'+\tau}}}{(-s'-\tau)}\right)\frac{(s'-a)^{\frac{k}{2}-1}(2s'-3a)}{(s')^{\frac{3k}{2}+1}}.
\ee

More precisely, we want to compute the analytic continuation of the above integral at fixed $D=D_0$, where $D_0=4,6,8,10,\ldots$. If one na\"ively sets $D=D_0$, one will see that the integral can have divergences due to the $s'=0$ endpoint. The idea is to separately consider the singular part that can lead to divergences and do the integral analytically to find its analytic continuation. For the remaining part, we can simply expand around $D=D_0$ since the integral is finite. More concretely, we expand the integrand of \eqref{eq:box_dispersive_intexpression_inAppendix} around $s'=0$ and rewrite it as\footnote{The $\log(s')$ terms are in fact absent in \eqref{eq:box_dispersive_intexpression_singularandregular} for the fixed-$a$ sum rules. However, if one studies the analogous integral for the fixed-$t$ sum rules, the $\log(s')$ terms will appear in general. So, we include them in \eqref{eq:box_dispersive_intexpression_singularandregular} since they can be treated in the same way.}
\be\label{eq:box_dispersive_intexpression_singularandregular}
\left.-{\cal C}^{\textrm{low}}_{k}(a)\right|_{{\cal M}_{\textrm{box}}} 
 &=2r_{\ibox}(8\pi G)^2  \sin(\tfrac{\pi D}{2})\int_0^{M^2}\frac{ds'}{\pi}(s')^{\frac{D-D_0}{2}}\nn \\
 &\times\p{\frac{g_{\frac{3k+2-D_0}{2}}(a)+h_{\frac{3k+2-D_0}{2}}(a)\log(s')}{(s')^{\frac{3k+2-D_0}{2}}} + \cdots + \frac{g_{1}(a)+h_{1}(a)\log(s')}{s'} + \textrm{regular}}.
\ee
After expanding the ``regular" piece in the parentheses around $D=D_0$, we can obtain the closed-form expression of the integral at $D=D_0$. The result can have poles at $D=D_0$, but they will be removed after renormalization, as discussed in \eqref{eq:bubble_sumrule_div_example}. For example, performing this analysis for $D=6, k=2$, we obtain \eqref{eq:box_dispersive_D6k2}.

Having explained how to compute the one-loop contributions to the dispersive sum rules, we now record some of the explicit expressions of the fixed-$a$ sum rules. We will give the explicit expressions of the $D=6$ sum rules with $k=2$ and $k=10$, and include the all couplings up to the same order as $g_2^2$. We write the expressions after doing renormalization (see discussions below \eqref{eq:G2_sumrule_spin2_D6}) and setting $\mu=M$. Let us first define
\be
h_1(a) = \log ^2\left(\tfrac{1+\sqrt{\frac{M^2+3a}{M^2-a}}}{1-\sqrt{\frac{M^2+3a}{M^2-a}}}\right),\quad h_2(a) = \sqrt{\tfrac{M^2+3a}{M^2-a}} \log \left(\tfrac{1+\sqrt{\frac{M^2+3a}{M^2-a}}}{1-\sqrt{\frac{M^2+3a}{M^2-a}}}\right).
\ee
The results are given by
{\allowdisplaybreaks
\be
&\left.-{\cal C}^{\textrm{low}}_{k=2}(a)\right|_{\textrm {1-loop}} \nn \\
&=\frac{(8\pi G)^2}{107520 \pi ^4 a}\bigg(1260 a^2 \mathrm{Li}_2\left(\tfrac{a}{M^2}\right)-525 a^2 \log ^2\left(\tfrac{M^2-a}{-a}\right)+630 a^2 \log ^2\left(\tfrac{-a}{M^2}\right) \nn \\
&\qquad\qquad\qquad+945 a^2 h_1(a)-1409 a^2 \log \left(\tfrac{M^2-a}{-a}\right)+420 M^4 \log \left(\tfrac{M^2-a}{-a}\right) \nn \\
&\qquad\qquad\qquad-420 M^4  h_2(a) +700 \pi ^2 a^2+2126 a M^2\bigg)\nn \\
&+ \frac{8 \pi G}{143360 \pi ^4}\left(8 \pi G \alpha_2 \left(31 a^2 \log \left(\tfrac{M^2-a}{-a}\right) + M^2 (-35 a - 9 M^2)\right)\right. \nn \\
&\qquad\qquad\qquad \left.- 8 g_2 \left(M^2 (-203 a + 83 M^2) - 23 a^2 \log \left(\tfrac{M^2-a}{-a}\right)\right)\right) \nn \\
&+\frac{8\pi G}{430080 \pi ^4}\left(-6 g_3 \left(M^2 \left(46 a^2 - 91 a M^2 + 20 M^4\right) + 46 a^3 \log \left(\tfrac{-a + M^2}{-a}\right)\right) + 28 \beta^{(3)}_1 M^4 (-9a + 4 M^2) \right. \nn \\
&\qquad\qquad\qquad\left.+ 3 \beta^{(2)}_0 \left(-62 a^3 \log \left(\tfrac{-a + M^2}{-a}\right) - 62 a^2 M^2 + 35 a M^4 + 12 M^6\right) \right) \nn \\
&+g_2^2\frac{2a^3 \log \left(\tfrac{-a + M^2}{-a}\right) + 2 a^2 M^2 + 133 a M^4 - 60 M^6}{107520 \pi ^4} + \ldots
,\quad (D=6),
\ee
}%
{\allowdisplaybreaks
\be
&\left.-{\cal C}^{\textrm{low}}_{k=10}(a)\right|_{\textrm {1-loop}} \nn \\
&=-\frac{(8\pi G)^2}{2980454400 \pi ^4 a^7 M^{24}} \nn \\
&\times \bigg(1050 M^2 \left(1512 a^7-3276 a^6 M^2+105 a^4 M^6+30 a^3 M^8+30 a^2 M^{10}+10 a M^{12}+5 M^{14}\right) \left(a-M^2\right)^4 h_2(a) \nn \\
&\quad +1050 \left(2772 a^{12}-9576 a^{11} M^2+13860 a^{10} M^4-15400 a^9 M^6+17325 a^8 M^8 \right. \nn \\
&\qquad\qquad\quad\left.-12672 a^7 M^{10}+3696 a^6 M^{12}-5 M^{24}\right) \log\left(\tfrac{-a+M^2}{-a}\right) \nn \\
&\quad +a \left(2397780 a^{11}-1819440 a^{10} M^2-17470656 a^9 M^4+39607680 a^8 M^6 \right. \nn \\
&\qquad\quad-31542000 a^7 M^8+8757060 a^6 M^{10}+122500 a^5 M^{12}-54600 a^4 M^{14} \nn \\
&\qquad\quad\left.+23625 a^3 M^{16}-14000 a^2 M^{18}+5250 a M^{20}-10500 M^{22}\right)\bigg) \nn \\
&+\frac{8\pi G}{3973939200 \pi ^4 M^{22}} \bigg(165 \alpha_2 (8\pi G) \left(1008 a^5-3612 a^4 M^2+4312 a^3 M^4-1323 a^2 M^6-888 a M^8+504 M^{10}\right)\nn \\
&-8 g_2 \left(453600 a^5-2519748 a^4 M^2+5584040 a^3 M^4-6185025 a^2 M^6+3433320 a M^8-766920 M^{10}\right)\bigg) \nn \\
&+\frac{8\pi G}{25804800 \pi ^4 M^{20}}\bigg(g_3 \left(-8208 a^5+37040 a^4 M^2-66150 a^3 M^4+58320 a^2 M^6-25320 a M^8+4320 M^{10}\right) \nn \\
&\qquad\qquad\qquad+\beta^{(2)}_0 \left(-2376 a^5+8600 a^4 M^2-10395 a^3 M^4+3240 a^2 M^6+2220 a M^8-1296 M^{10}\right)\nn \\
&\qquad\qquad\qquad+8 \beta^{(3)}_1 \left(378 a^5-1960 a^4 M^2+4095 a^3 M^4-4320 a^2 M^6+2310 a M^8-504 M^{10}\right)\bigg) \nn \\
&+g_2^2 \frac{-4752 a^5+24760 a^4 M^2-51975 a^3 M^4+55080 a^2 M^6-29580 a M^8+6480 M^{10}}{19353600 \pi ^4 M^{20}}+ \ldots
,\quad (D=6).
\ee
}%

In the ancillary file, we give the expressions of \texttt{CSDROneLoop[k,d]}, which is $\left.-{\cal C}^{\textrm{low}}_{k}(a)\right|_{\textrm {1-loop}}$ in dimension $d$, for $k=2,4,6,8,10$ and $d=4,6,8,10,12$. They are sufficient for deriving all the bounds we present in the main text.

\subsection{Fixed-$t$ sum rules}
In this appendix, we also give the one-loop contributions to several fixed-$t$ dispersive sum rules. Even though they are not explicitly used in the main text, we hope that the explicit expressions can make the issues discussed in section \ref{sec:fixed-t_loop} clearer, and potentially be useful for future applications.

We consider the bubble and box contributions separately. As mentioned in the main text, and bubble amplitude can be written in terms of ${\cal M}_{p,q}$ defined in \eqref{eq:Mpq_def}. For a spin-$k$ fixed-$t$ sum rule in dimension $D$, the contribution from ${\cal M}_{p,q}$ with $p\neq \tfrac{k}{2}$ is given by
\be
&-\int_0^{M^2} \frac{ds}{\pi}\textrm{Im} \left[\p{\frac{1}{s}+\frac{1}{s+t}}\frac{{\cal M}_{p,q}}{[s(s+t)]^{\frac{k}{2}}}\right] \nn \\
&=r_{\ibub}^{(D)}\frac{(-1)^{p+1}\sin(\pi \tfrac{D-4}{2})}{(\tfrac{k}{2}-p)\pi}t^{p}M^{D-4+2q+2p-4n}\left(1+\frac{t}{M^2}\right)^{p-\tfrac{k}{2}}\nn \\
&\qquad \times \left(1+\frac{\tfrac{D-4}{2}+q-p}{k-p-\tfrac{D-4}{2}-q}{}_2F_1(\tfrac{k}{2}-p,1,1+k-p-\tfrac{D-4}{2}-q,\tfrac{t}{M^2+t})\right),
\ee
and for the special case $p=\tfrac{k}{2}$, we find
\be
&-\int_0^{M^2} \frac{ds}{\pi}\textrm{Im} \left[\p{\frac{1}{s}+\frac{1}{s+t}}\frac{{\cal M}_{p,q}}{[s(s+t)]^p}\right] \nn \\
&=-r_{\ibub}^{(D)}\frac{\sin(\pi(\tfrac{D-4}{2}-p))t^{p}M^{D-4+2q-2p}}{\pi} \nn \\
&\qquad \times\left(\frac{2}{p-\tfrac{D-4}{2}-q}-\frac{t}{(1+p-\tfrac{D-4}{2}-q)M^2}{}_2F_1(1,1+p-\tfrac{D-4}{2}-q,2+p-\tfrac{D-4}{2}-q,-\tfrac{t}{M^2})\right) \nn \\
&+(-1)^{p+q}r_{\ibub}^{(D)}\left(\frac{-t}{\mu^2}\right)^{\tfrac{D-4}{2}+q}.
\ee

For the box contribution, we will focus on the one appearing in our one-loop amplitude, namely ${\cal M}_{\textrm{box}}$ defined in \eqref{eq:Mbox_def}. Similar to the fixed-$a$ case, we will compute its contribution to a spin-$k$ sum rule for fixed $D$ and $k$. More precisely, let us define
\be
-B^{\textrm{box}}_{k;D_0} \equiv - \int_0^{M^2} \frac{ds}{\pi}\textrm{Im} \left[\p{\frac{1}{s}+\frac{1}{s+t}}\frac{{\cal M}_{\textrm{box}}}{[s(s+t)]^{\frac{k}{2}}}\right],
\ee
where $D_0=4,6,8,\ldots$ is the spacetime dimension. As an example, let us give here the explicit expressions of $-B^{\textrm{box}}_{2;6}$ and $-B^{\textrm{box}}_{6;10}$. For $k=2,D_0=6$, we have
\be
&-B^{\textrm{box}}_{2;6} \nn \\
&= \frac{(8\pi G))^2}{1536 \pi ^4 M^2 t \left(M^2+t\right)^2}\Bigg(72 M^2 t^2 \left(M^2+t\right)^2 \textrm{Li}_2\left(\tfrac{t}{M^2+t}\right)+15 M^2 t^2 \left(M^2+t\right)^2 \log ^2\left(-\tfrac{t}{M^2}\right) \nn \\
&\qquad\qquad -6 t \log \left(-\tfrac{t}{M^2}\right) \left(-2 M^8-4 M^6 t+3 M^4 t^2+5 M^2 t^3+4 M^2 t \left(M^2+t\right)^2 \log \left(\tfrac{M^2+t}{M^2}\right)+t^4\right) \nn \\
& \qquad \qquad -2 \left(M^2+t\right) \left(-5 \pi ^2 M^2 t^2 \left(M^2+t\right)-12 M^2 t^2 \left(M^2+t\right) \log ^2\left(\tfrac{M^2+t}{M^2}\right)\right. \nn \\
&\qquad\qquad\qquad\qquad\qquad \left.+3 \left(2 M^8+4 M^6 t-3 M^4 t^2-5 M^2 t^3-t^4\right) \log \left(\tfrac{M^2+t}{M^2}\right)\right) \Bigg),
\ee
and for $k=6,D_0=10$, we find
\be
&-B^{\textrm{box}}_{6;10} \nn \\
&=-\frac{(8\pi G)^2}{353894400 \pi ^8 M^2 t^3 \left(M^2+t\right)^6}\Bigg(-270 M^2 t^2 \left(M^2+t\right)^6 \textrm{Li}_2\left(\tfrac{t}{M^2+t}\right) \nn \\
&\qquad +\left(M^2+t\right) \left(18 \left(10 M^8+20 M^6 t+16 M^4 t^2+6 M^2 t^3-5 t^4\right) \left(M^2+t\right)^4 \log \left(\tfrac{M^2+t}{M^2}\right) \right. \nn \\
&\qquad\qquad\quad-M^2 t \left(180 M^{12}+990 M^{10} t+2268 M^8 t^2+2781 M^6 t^3+1906 M^4 t^4+728 M^2 t^5+115 t^6\right) \nn \\
&\qquad\qquad\quad\left.+45 M^2 t^2 \left(M^2+t\right)^5 \log ^2\left(\tfrac{M^2+t}{M^2}\right)\right) \nn \\
&\qquad+3 t^2 \log \left(-\tfrac{t}{M^2}\right) \left(t \left(150 M^{12}+825 M^{10} t+1870 M^8 t^2+2160 M^6 t^3+1314 M^4 t^4+369 M^2 t^5+30 t^6\right)\right. \nn \\
&\qquad\qquad\quad \left.-150 M^2 \left(M^2+t\right)^6 \log \left(\tfrac{M^2+t}{M^2}\right)\right)\Bigg).
\ee

Combining bubble and box contribution, one can then calculate the full one-loop contribution to the fixed-$t$ sum rule,
\be
\left.-B^{\textrm{low}}_{k;D_0}\right|_{\textrm{1-loop}} \equiv - \int_0^{M^2} \frac{ds}{\pi}\textrm{Im} \left[\p{\frac{1}{s}+\frac{1}{s+t}}\frac{{\cal M}^{\textrm{1-loop}}}{[s(s+t)]^{\frac{k}{2}}}\right].
\ee
In the ancillary file, we give the full expressions of $\left.-B^{\textrm{low}}_{k;d}\right|_{\textrm{1-loop}}$ \texttt{FixedtOneLoop[k,d]} for $k=2,4,6,8,10$ and $d=4,6,8,10,12$.

\section{Details on numerics}\label{app:numerics}
In this appendix, we give more details on our numerical computation performed in sections \ref{sec:fixed-a_tree}, \ref{sec:fixed-a_loop}, and \ref{sec:4Dbounds}. We closely follow the method described in \cite{Caron-Huot:2021rmr}. The main difference for us is that we are using the fixed-$a$ sum rules. As a worked example, we will explain the approach for obtaining bounds on $g_2,g_3$, namely Figure \ref{fig:g2-g3-tree} and \ref{fig:g2-g3-loop}. Other bounds can be obtained using similar method.

As explained in the main text, we would like to look for functions $f(p),h_k(p)$ such that
\be\label{eq:pols_fixeda_inapp}
&\int_0^{\frac{1}{\sqrt{3}}}dp\ f(p) {\cal C}_{2,-p^2}[m^2,J] \nn \\
+& \sum_{k=4,6,\ldots} \int_0^{\frac{1}{\sqrt{3}}}dp\ h_k(p) {\cal C}_{k,-p^2}[m^2,J] \geq 0,\quad \forall m>1, J=0,2,4,\ldots,
\ee
where ${\cal C}_{k,-p^2}[m^2,J]$ is defined by \eqref{eq:fixeda_highenergy_def}. We also impose additional constraints,
\be\label{eq:hk_nullconstraints_inapp}
\int_0^{\frac{1}{\sqrt{3}}}dp\ h_k(p) p^{2n} =0,\qquad n=0,1,\ldots,\tfrac{k}{2}.
\ee
We have set $M=1$ since one can use dimensional analysis to restore the dependence on $M$. For any $f(p),h_k(p)$ satisfying \eqref{eq:pols_fixeda_inapp} and \eqref{eq:hk_nullconstraints_inapp}, we have
\be\label{eq:lowenergybound_inapp}
&\int_0^{\frac{1}{\sqrt{3}}}dp\ f(p)\left(\frac{8\pi G}{p^2}+2g_2+g_3 p^2\right) \nn \\
-& \int_0^{\frac{1}{\sqrt{3}}} dp\ f(p) \left.{\cal C}^{\textrm{low}}_{2}(a)\right|_{\textrm {1-loop}} - \sum_{k=4,6,\ldots} \int_0^{\frac{1}{\sqrt{3}}} dp\ h_k(p) \left.{\cal C}^{\textrm{low}}_{k}(a)\right|_{\textrm {1-loop}} \geq 0.
\ee
The plots in Figure \ref{fig:g2-g3-tree} and \ref{fig:g2-g3-loop} are obtained by plotting different bounds of the form \eqref{eq:lowenergybound_inapp} with different optimization conditions. In practice, we follow \cite{Caron-Huot:2021rmr} and choose a point $(x_0,y_0)$ on the $(\tfrac{g_2 \Lambda^2}{8\pi G},\tfrac{g_3 \Lambda^4}{8\pi G})$ plane, and extremize the distance in the allowed region from the point along a ray of angle $\theta$. We obtain Figure \ref{fig:g2-g3-tree} by picking $(x_0,y_0)$ close to the kink and scanning over different values of $\theta$.

Now, it remains to impose the positivity conditions \eqref{eq:pols_fixeda_inapp}. We choose the functions $f(p),h_k(p)$ to be linear combinations of powers of $p$ (see \eqref{eq:functional_pexpansion}). Integrating ${\cal C}_{2,-p^2}[m^2,J]$ against $p^n$, we get (for even integer $J$)
\be\label{eq:fixedmfixedJint_k2_expr}
&\int_0^{\frac{1}{\sqrt{3}}}dp\ p^n {\cal C}_{2,-p^2}[m^2,J] \nn \\
&= \sum_{j=0}^{\frac{J}{2}}(-1)^j \frac{2^{2 j} m^{-2 (j+3)} 3^{\frac{1}{2} (-2 j-n-1)} \Gamma \left(\tfrac{J}{2}+1\right) \left(\tfrac{D+J-3}{2}\right)_j}{\Gamma (j+1) \left(\tfrac{D-2}{2}\right)_j \Gamma \left(-j+\frac{J}{2}+1\right)} \nn \\
&\quad \times \left(\frac{2 m^2 \, {}_2F_1\left(j,\tfrac{2 j+n+1}{2};\tfrac{2 j+n+3}{2};-\frac{1}{3 m^2}\right)}{2 j+n+1}+\frac{{}_2F_1\left(j,\tfrac{2 j+n+3}{2};\tfrac{2 j+n+5}{2};-\frac{1}{3 m^2}\right)}{2 j+n+3}\right),
\ee
and the integrals of $p^n {\cal C}_{k,-p^2}[m^2,J]$ for $k=4,6,\ldots$ take similar forms. Ideally, one would like to impose \eqref{eq:pols_fixeda_inapp} for all $m>1$. This can be achieved using $\tt{SDPB}$ if one could approximately write \eqref{eq:fixedmfixedJint_k2_expr} as a polynomial in $m$ times a positive function. Although it would be interesting to see if there is such an approximation, here we will simply use the discretization/refinement procedure explained in \cite{Caron-Huot:2021rmr}. In other words, we impose
\be\label{eq:pols_discrete_fixeda_inapp}
&\sum_{n=0}^{N_2}a_n\int_0^{\frac{1}{\sqrt{3}}}dp\ p^{n_0+n} {\cal C}_{2,-p^2}[m^2,J] + \sum_{k=4,6,\ldots} \sum_{i_k=0}^{N_k}\beta_{i_k,k} \int_0^{\frac{1}{\sqrt{3}}}dp\ p^{i_k} {\cal C}_{k,-p^2}[m^2,J] \geq 0, \nn \\
&\forall m=m_1,\ldots,m_n ,\quad J=0,2,4,\ldots, J_{\textrm{max}},
\ee
where $m_1,\ldots,m_n$ is a set of discrete points. We can choose them by first doing a change of variable $m^2=\frac{1}{1-x}$, and then evenly sample in the $x$-space for $x\in [0,1]$. This does not guarantee positivity for all $m>1$, but if there is any negative region, one can add more sample points there to the positive condition \eqref{eq:pols_discrete_fixeda_inapp} and repeat the process.

It is also important to impose \eqref{eq:pols_fixeda_inapp} in the limit where $m\to \infty, J\to \infty$ with fixed impact parameter $b=\frac{2J}{m}$. In this limit, the $k=2$ sum rule in \eqref{eq:pols_fixeda_inapp} gives the leading contribution, and it becomes
\be\label{eq:pols_inb_fixeda_inapp}
\sum_{n=0}^{N_2} \frac{a_n}{3^{\frac{n+1}{2}}(n+1)}{}_1F_2\p{\tfrac{n+1}{2};\tfrac{D-2}{2},\tfrac{n+3}{2};-\tfrac{b^2}{12}} \geq 0.
\ee
To impose the above condition for all $b$, we choose a $B_{\mathrm{max}}$ and use the same discretization approach as for the $m$-space positivity condition for $b\leq B_{\mathrm{max}}$. For $b>B_{\mathrm{max}}$, we take the $b \to \infty$ limit of \eqref{eq:pols_inb_fixeda_inapp} and rewrite it as the positive semidefiniteness of a $2\times 2$ matrix whose entries are polynomials in $b$ (see \cite{Caron-Huot:2021rmr} for more details).

We use $\tt{SDPB}$ to find the coefficients $a_n, \beta_{i_k,k}$ satisfying \eqref{eq:pols_discrete_fixeda_inapp} and \eqref{eq:pols_inb_fixeda_inapp}. The implementation depends on several parameters, including the dimension of the functional space, number of discretized sample points in the $m$-space and $b$-space, $J_{\textrm{max}}$, $B_{\textrm{max}}$, and the order of polynomial approximation in the $b\to \infty$ limit of \eqref{eq:pols_inb_fixeda_inapp}. The dimension of the functional space used to obtain the bounds in this work are all given in the main text. For the other parameters, we choose them such that the bounds remain the same (within desired precision) when the parameters are slightly increased. For the bounds in section \ref{sec:fixed-a_tree}, we use the same parameters as \cite{Caron-Huot:2021rmr}. For the $g_4$ and $g_6$ bounds in section \ref{sec:positivity_oneloop}, we choose $J_{\textrm{max}}=74$, and the other parameters are the same.

\bibliographystyle{JHEP}
\bibliography{LoopDispersive}

\end{document}